\documentclass[draft]{agujournal2019}
\usepackage{url} 
\usepackage{lineno}
\usepackage[inline]{trackchanges} 
\usepackage{soul}
\usepackage{amsmath}
\usepackage{amssymb}
\usepackage{graphicx}
\usepackage{xcolor}
\usepackage{tcolorbox}
\usepackage{enumitem}

\newcounter{obscount}
\newenvironment{obsquestion}[1]{%
    \stepcounter{obscount}%
    \begin{tcolorbox}[
        colback=gray!5,
        colframe=gray!30,
        boxrule=0.5pt,
        left=10pt,
        right=10pt,
        top=8pt,
        bottom=8pt,
        title={\textbf{#1}},
        fonttitle=\bfseries,
        coltitle=black,
        colbacktitle=gray!15
    ]
    \begin{description}[leftmargin=0pt, style=nextline] 
}{%
    \end{description} 
    \end{tcolorbox}
    \vspace{0.5em}
}
\draftfalse

\journalname{Journal of Advances in Modeling Earth Systems (JAMES)}

\begin{document}

%
%


\title{Crowdsourcing the Frontier: Advancing Hybrid Physics-ML Climate Simulation via a \$50,000 Kaggle Competition}

%
%



\authors{Jerry Lin$^{1,2}$, Zeyuan Hu$^{3}$, Tom Beucler$^{4,5}$, Katherine Frields$^{1}$, Hannah Christensen$^{6}$, Walter Hannah$^{7}$, Helge Heuer$^{8}$, Peter Ukkonen$^{6}$, Laura A. Mansfield$^{6}$, Tian Zheng$^{9}$, Liran Peng$^{1}$, Ritwik Gupta$^{10,11}$, Pierre Gentine$^{12}$, Yusef Al-Naher, Mingjiang Duan$^{13}$, Kyo Hattori$^{14}$, Weiliang Ji$^{13}$, Chunhan Li$^{13}$, Kippei Matsuda$^{15}$, Naoki Murakami$^{16}$, Shlomo Ron, Marec Serlin$^{17}$, Hongjian Song$^{13}$, Yuma Tanabe, Daisuke Yamamoto, Jianyao Zhou, Mike Pritchard$^{1,3}$}

\affiliation{1}{Department of Earth System Sciences, University of California at Irvine, Irvine, CA, USA}
\affiliation{2}{Department of Computing \& Data Sciences, Boston University, Boston, MA, USA}
\affiliation{3}{NVIDIA Research}
\affiliation{4}{Faculty of Geosciences and Environment, University of Lausanne, Lausanne, VD, Switzerland}
\affiliation{5}{Expertise Center for Climate Extremes, University of Lausanne, Lausanne, VD, Switzerland}
\affiliation{6}{Department of Physics, University of Oxford, Oxford, United Kingdom}
\affiliation{7}{Lawrence Livermore National Laboratory}
\affiliation{8}{Deutsches Zentrum f{\"u}r Luft- und Raumfahrt, Institut f{\"u}r Physik der Atmosph{\"a}re, Oberpfaffenhofen, Germany}
\affiliation{9}{Department of Statistics, Columbia University, New York, NY, USA}
\affiliation{10}{University of California, Berkeley, Berkeley, CA, USA}
\affiliation{11}{University of Maryland, College Park, MD, USA}
\affiliation{12}{LEAP Science and Technology Center, School of Engineering and Applied Sciences, Climate School, Columbia University}
\affiliation{13}{Z Lab, China}
\affiliation{14}{ABEJA Inc., Japan}
\affiliation{15}{Kawasaki Heavy Industries, Ltd., Japan}
\affiliation{16}{DeNA Co., Ltd}
\affiliation{17}{Uber Technologies, Inc.}




\correspondingauthor{Jerry Lin}{jlin404@bu.edu}



\begin{keypoints}
\item Online stability in the low-resolution real-geography setting is reproducibly achievable across diverse architectures.
\item Offline and online zonal mean biases are near-identical across architectures; online runs underestimate tropical precipitable water.
\item An expanded variable list is universally beneficial offline but has diverging, architecture-dependent effects online.
\end{keypoints}

%
%

%
%


\begin{abstract}
\\
Subgrid machine-learning (ML) parameterizations have the potential to introduce a new generation of climate models that incorporate the effects of higher-resolution physics without incurring the prohibitive computational cost associated with more explicit physics‑based simulations. However, important issues, ranging from online instability to inconsistent online performance, have limited their operational use for long-term climate projections. To more rapidly drive progress in solving these issues, domain scientists and machine learning researchers opened up the offline aspect of this problem to the broader machine learning and data science community with the release of ClimSim, a NeurIPS Datasets and Benchmarks publication, and an associated Kaggle competition. This paper reports on the downstream results of the Kaggle competition by coupling emulators inspired by the winning teams' architectures to an interactive climate model (including full cloud microphysics, a regime historically prone to online instability) and systematically evaluating their online performance. Our results demonstrate that online stability in the low-resolution real-geography setting is reproducible across multiple diverse architectures, which we consider a key milestone. All tested architectures exhibit strikingly similar offline and online biases, though their responses to architecture-agnostic design choices (e.g., expanding the list of input variables) can differ significantly. Multiple Kaggle-inspired architectures achieve state-of-the-art (SOTA) results on certain metrics such as zonal mean bias patterns and global RMSE, indicating that crowdsourcing the essence of the offline problem is one path to improving online performance in hybrid physics-AI climate simulation.

\end{abstract}

\section*{Plain Language Summary}

Future climate models may use machine learning (ML) to replace small-scale physical processes that are otherwise too costly to simulate directly over long timescales. Such ``hybrid'' physics–ML models could improve predictions by reducing uncertainties from current approximations. But making them run reliably in full climate simulations has been a major challenge. To speed progress, scientists created an open dataset, benchmarking framework, and global competition to drive improvement for these ML components. This paper follows up on that competition by testing ideas from the winning teams within hybrid climate models. For the first time, we show that stable hybrid simulation is now reproducible across a range of diverse ML architectures. We find that different architectures share similar patterns of errors both before and after coupling, although their responses to added training inputs can differ. Finally, some competition-inspired designs achieve state-of-the-art scores on individual performance measures, but no single approach beats the previous benchmark (Hu et al. 2025) on every metric.

%
%

\section{Introduction}

Earth system models (or simpler climate models) used for contemporary long-term projections operate at a coarse resolution ($>25$ km along the horizontal) and rely on hand-tuned subgrid parameterizations that imperfectly represent the effects of subgrid processes like convection, radiation, and turbulence, resulting in systematic biases and large error bars in the amount of expected warming \cite{O-Gorman2012-gc, Webb2013-gz, Sherwood2014-si, Ceppi2016-hs, Schneider2017-od}. To improve the representation of deep convection and clouds, the climate modeling community is actively exploring the use of computationally expensive kilometer-scale global simulations \cite{Hohenegger2023-iq, Taylor2023-pm}. However, even at these resolutions, convection is only partially resolved and the need for parameterizations for processes like turbulence or microphysics is never fully eliminated \cite{Schneider2017-od, Taylor2023-pm}. In addition, this approach becomes intractable when considering the fact that multiple realizations of future climates over long time horizons and emissions scenarios are required for proper uncertainty quantification, depiction of extremes, and actionable adaptation \cite{Kay2015-ja}.

An alternative approach for more explicitly resolving some subgrid processes without relying on a natively high-resolution discretization is Multiscale Modeling Framework (MMF), also known as superparameterization \cite{Randall2003-db, Khairoutdinov2005-fk, Randall2013-hl}. In MMF, a Cloud Resolving Model (CRM) with periodic boundary conditions and higher spatial and temporal resolution than the host climate model is embedded inside each coarse-model grid cell. The effective kilometer scale resolution in the MMF approach yields several tangible benefits, ranging from improvements in the representation of precipitation extremes to more faithfully reproducing the periodicity of the El Niño Southern Oscillation \cite{Stan2010-id, Li2012-yl, Randall2013-hl}. Ultimately, this approach has not been adopted for long-term climate projections as the computational cost remains prohibitively high \cite{Gentine2018-ux}. Additionally, the artificial scale separation that allows for massively parallel computation in MMF introduces unnatural artifacts and its own set of compromises for the sake of computational efficiency \cite{Jansson2022-vv, Hannah2022-im}. Because of this and other factors, more effort has been devoted to advancing Global Cloud Resolving Models (GCRMs) for operational use in recent years.

Interestingly, the clean spatial and temporal scale separation between the embedded CRMs and GCM in the MMF approach has conveniently (albeit unintentionally) given birth to a new means for bringing kilometer scale fidelity at a fraction of the computational cost. This is because the MMF approach integrates CRMs within the GCM framework at the GCM's native spatial and temporal resolution. This process inherently summarizes (``coarse-grains'') the CRM behavior to the GCM scale at each timestep, making the CRM output a natural target for emulation by machine learning (ML) models, particularly neural networks (NNs) \cite{Rasp2018-fk, Gentine2018-ux, Mooers2021-sh, Han2023-nm}. Unfortunately, this promise remains unrealized despite years of interest because of emergent failure modes when deploying these ML models in an ``online'' setting (i.e. when dynamically coupled to the GCM) \cite{Brenowitz2020-bv, Lin2025-ya, Chen2025-jv}. Online emulation has proved particularly challenging in the limit of full complexity emulation of the CRMs used in MMFs, such as when difficult-to-emulate microphysical tendencies are included in the scale coupling. To engage the machine learning research community on these issues, a group of climate scientists and machine learning researchers released ClimSim, a NeurIPS Datasets and Benchmarks publication, software repository, and collection of datasets consisting of multivariate CRM input-output pairs generated using the Energy Exascale Earth System Model-Multiscale Modeling Framework (E3SM-MMF) (See Section 2.1 in Methods for more details) \cite{Yu2023-on}. This was followed up by \citeA{Hu2025-mf}, which showcased stable five-year online simulations with explicit cloud condensate coupling for the first time, and ClimSim-Online, which democratized online testing via a containerized version of E3SM-MMF \cite{Yu2025-cl}. In parallel, the official ClimSim Kaggle competition ran for three months during the summer of 2024, attracting approximately 700 teams from around the world who submitted over 10,000 predictions in pursuit of a \$50,000 prize pool \cite{leap-atmospheric-physics-ai-climsim}.
 
Our hypothesis is that by expanding both awareness and accessibility of this research problem to those in the data science and machine learning community, we can facilitate advancements that would not have happened had this research problem stayed confined to domain scientists limited by time, personnel, and funding constraints. Thanks to the visibility of Kaggle, which is widely considered the leading platform for data science and machine learning competitions, the CRM emulation problem attracted the talents of some of the world's best data scientists and machine learning engineers in a short timeframe. The leaderboard revealed that hundreds of teams were able to surpass the strongest offline $R^2$ baseline set using the U-Net architecture from Hu et al. 2025, raising the question---do the same innovations that led to improvements in offline skill also yield benefits online?

This study approaches this question systematically, evaluating both architecture-specific and architecture-agnostic design decisions inspired by the top of the Kaggle leaderboard online. The following results will demonstrate their methods' potential to achieve state-of-the-art hybrid simulation results and will reveal aspects of the hybrid physics-ML simulation problem that remain resistant to Kaggle-inspired innovations. In the low-resolution real-geography ClimSim limit, we can report that online stability when including the complexity of full microphysical coupling is reproducibly achievable across diverse architectures going forward. However, different architectures can respond differently to architecture-agnostic design decisions in ways that can cause one architecture to experience less online drift and another to become completely unstable online. This is particularly evident when expanding the input variable list to include convective memory, large-scale forcings, and latitudinal information. Nevertheless, both offline and online biases are structurally similar across all architectures and architecture-agnostic design decisions, and offline biases across multiple variables change systematically as a function of precipitation percentile. In the following sections, we showcase how Kaggle-inspired innovations have pushed the frontier of hybrid physics-ML climate simulation while isolating the remaining challenges that lie ahead.

\section{Methods}

Designing a reasonably controlled and scientific intercomparison inspired by top Kaggle users' decisions naturally required several trade-offs. In order to make our post-competition analysis tractable, we do not exhaustively investigate every architecture and architecture-agnostic design decision utilized in the winning solutions from the Kaggle competition. This would be impractical as winning solutions were often composed of a weighted average of vastly different architectures highly optimized for strong performance on a singular offline scalar metric (averaged $R^2$ across all target variables). Instead, we intercompare online performance of five architectures inspired by the methods of the five winning teams alongside the U-Net architecture of \citeA{Hu2025-mf}, a strong baseline. We evaluate both offline and online performance using held-out simulation data from MMF as ground truth (see Section 2.1 for details). Additional details regarding calculation of $R^2$, Root Mean Squared Error (RMSE), and zonal averaging can be found in Texts S1-S3 in the SI. With the exception of architectures in the multirepresentation configuration, which uses multiple variations of the input using different input normalizations (detailed in Section 2.3.4), all models use the same input normalizations, output normalizations, microphysics constraint, and feature pruning used in \citeA{Hu2025-mf}. Details regarding input and output normalizations as well as feature pruning  are detailed in Tables \ref{tab:input_table}, \ref{tab:output_table}, and \ref{tab:expanded_inputs}. The microphysics constraint, as seen in Figure 2 from \citeA{Hu2025-mf}, mirrors the one-moment microphysics scheme in the CRM, using temperature to diagnose the mixing ratio of liquid and ice cloud. While the diagnostic relationship does not hold exactly on the GCM grid scale, the constraint was shown to greatly reduce online biases in \citeA{Hu2025-mf}. Each architecture is trained across five configurations using architecture-agnostic design decisions inspired by the Kaggle competition winners as well as the input variable list expansion developed by \citeA{Hu2025-mf}. We make no modifications to the hyperparameters when adapting architectures from the winning solutions in the Kaggle competition. To gauge variability induced by choice of seed, we train three models for each architecture and configuration pair using different seeds. This yields a total of 90 models that we trained for the entire study (6 architectures $\times$ 5 configurations $\times$ 3 seeds). All models were coupled online to the same MMF used to generate the ClimSim data, using an FTorch binding to facilitate GPU-accelerated hybrid simulations \cite{Atkinson2025-ba}.

For our Kaggle architectures, we chose the \textit{Squeezeformer} from the first place team, the \textit{Pure ResLSTM} from the second place team, a custom architecture called the \textit{Pao Model} from the third place team, a \textit{ConvNeXt} convolutional NN from the fourth place team, and an \textit{Encoder-Decoder LSTM} from the fifth place team. Feature selection, preprocessing, postprocessing, and training settings such as batch size, number of epochs, loss function, learning rate, and learning rate scheduler were unified across all architectures to facilitate fairer comparisons. To understand how different architectures responded to different architecture-agnostic design decisions, we created ``standard," ``confidence loss," ``difference loss," ``multirepresentation," and ``expanded variable list" configurations (detailed in Section 2.2).  Input variables and input normalizations for the ``standard," ``confidence loss," and ``difference loss" configurations are the same and are shown in Table \ref{tab:input_table}. The ``multirepresentation" configuration makes use of the same input variables but across three separate normalizations. The ``expanded variable list" configuration incorporates additional input variables (shown in Table \ref{tab:expanded_inputs}) that matches what was used for the U-Net-expanded model seen in \citeA{Hu2025-mf}. All models across all configurations share the same output variables and output normalizations, which are listed in Table \ref{tab:output_table}.

\subsection{Dataset and Reference Climate Simulation}

ClimSim provides three different 10-year simulations generated by E3SM-MMF of varying complexity: a low-resolution aquaplanet version, a low-resolution real-geography version, and a high-resolution real-geography version. All NNs in this study were trained, validated, and tested (offline) using the low-resolution real-geography version. This version was configured using the ``F2010-MMF1” compset and ``ne4pg2” grid \cite{Yu2025-cl}. The effective horizontal resolution for the GCM is extremely coarse at approximately $11.5^\circ \times 11.5^\circ$ as there are only 384 columns arranged in an unstructured, cubed-sphere grid. Each GCM column is composed of 60 vertical levels extending up to 65 km in altitude, and the model timestep is 20 minutes. The embedded CRMs are 2D and aligned in the north-south direction with 64 columns across their domain and 2 km horizontal grid spacing. The forcing from the GCM on the domain mean of each CRM is 1D, and the CRMs are not designed to have spatial information that correspond to actual locations on the globe. Taking advantage of the clean scale separation provided by MMF, only data at the GCM grid's discretization is used for training, validation, and testing. Sea Surface Temperatures (SSTs) and sea ice coverage are prescribed to be similar to present-day climatology (with SSTs varying according to a fixed annual cycle), and aerosols are transparent to radiation code. The first 7 years and 1st month of the 8th year of the simulation data are used for training, the rest of the 8th year and the first month of the 9th year are used for validation, and rest of the final 2 years are used for offline testing. No subsampling was used to create the training dataset, but the validation dataset subsamples every 7th timestep while the test dataset subsamples every 12th timestep. This yields 69,783,552 samples for the training data (235.2 GB), 1,564,086 samples for the validation data (4.7 GB), and 1,681,920 samples for the test data (5.5 GB). The storage sizes shown here are after preprocessing using the ``standard" variable list, described in Section 2.3.1 and shown in Tables \ref{tab:input_table} and \ref{tab:output_table}. 

For online simulation and testing, both hybrid runs and the MMF reference run used a different initial condition than that used to create the original ClimSim dataset. This was necessary because the original dataset did not provide restart files, requiring us to start from a spun-up state rather than from scratch.

Since our NN parameterizations are trained exclusively on a low-resolution MMF simulation, and our primary goal is to advance fundamental hybrid modeling challenges---such as stability, reliability, and both offline and online accuracy relative to MMF---we omit direct comparisons of our hybrid simulations to ERA5 reanalysis or observations. For similar reasons, we also refrain from comparing against E3SM configurations that use traditional parameterizations (e.g. E3SMv1, E3SMv2, and E3SMv3). These non-MMF versions are fundamentally different models that were not tuned to reproduce E3SM-MMF behavior. Moreover, prior studies have already compared E3SM-MMF to its non-MMF counterparts, in some cases identifying biases or unphysical artifacts unique to E3SM-MMF \cite{Hannah2020-ez, Hannah2025-ah, Xia2025-cn}.

\begin{table}[ht]
 \centering
 \begin{tabular}{l l l l}
 \hline
  \textbf{Variable} & \textbf{Units} & \textbf{Description} & \textbf{Normalization} \\
 \hline
    $T(z)$ & K & Temperature & (x-mean)/(max-min)  \\
    $RH(z)$ &  & Relative humidity   \\
    $q_n(z)$ & kg/kg & Liquid and ice cloud mixing ratio & 1 - exp(-$\lambda x$)  \\
    Liquid partition $(z)$ & & Diagnostic microphysics constraint \\
    $u(z)$ & m/s & Zonal wind & (x-mean)/(max-min)  \\
    $v(z)$ & m/s & Meridional wind & (x-mean)/(max-min)  \\
    O3$(z)$ & mol/mol & Ozone volume mixing ratio & (x-mean)/(max-min)  \\
    CH4$(z)$ & mol/mol & Methane volume mixing ratio & (x-mean)/(max-min)  \\
    N2O$(z)$ & mol/mol & Nitrous volume mixing ratio & (x-mean)/(max-min)  \\
    PS & Pa & Surface pressure & (x-mean)/(max-min)  \\
    SOLIN & W/m\textsuperscript{2} & Solar insolation & x/(max-min) \\
    LHFLX & W/m\textsuperscript{2} & Surface latent heat flux & x/(max-min)  \\
    SHFLX & W/m\textsuperscript{2} & Surface sensible heat flux & x/(max-min)  \\
    TAUX & W/m\textsuperscript{2} & Zonal surface stress &  (x-mean)/(max-min) \\
    TAUY & W/m\textsuperscript{2} & Meridional surface stress & (x-mean)/(max-min)  \\
    COSZRS & & Cosine of solar zenith angle & (x-mean)/(max-min)  \\
    ALDIF & & Albedo for diffuse longwave radiation & (x-mean)/(max-min)  \\
    ALDIR & & Albedo for direct longwave radiation & (x-mean)/(max-min)  \\
    ASDIF & & Albedo for diffuse shortwave radiation & (x-mean)/(max-min)  \\
    ASDIR & & Albedo for direct shortwave radiation & (x-mean)/(max-min)  \\
    LWUP & W/m\textsuperscript{2} & Surface upward longwave flux & (x-mean)/(max-min) \\
    ICEFRAC & & Sea-ice area fraction   \\
    LANDFRAC & & Land area fraction   \\
    OCNFRAC & & Ocean area fraction   \\
    SNOWHICE & m & Snow depth over ice & (x-mean)/(max-min) \\
    SNOWHLAND & m & Snow depth over land & (x-mean)/(max-min) \\
 \hline
 \end{tabular}
  \caption{List of input features used in the standard, confidence loss, and difference loss configurations. Here $\lambda = 1/\overline{q_n^*})$ where $\overline{q_n^*}$ refers to the average liquid and ice cloud mixing ratio for liquid and ice cloud mixing ratios $> 10^{-7}$.\label{tab:input_table}}.
 \end{table}

 \begin{table}[ht]
 \centering
 \begin{tabular}{l l l l}
 \hline
  \textbf{Variable} & \textbf{Units} & \textbf{Description} & \textbf{Normalization} \\
 \hline
    $dT/dt(z,t_{0})$ & K/s & Temperature tendency & x/std  \\
    $dq_v/dt(z,t_{0})$ & kg/kg/s & Water vapor tendency & x/min(std,$\gamma_1$)  \\
    $dq_n/dt(z,t_{0})$ & kg/kg/s & Liquid and ice cloud tendency & x/min(std,$\gamma_1$)  \\
    $du/dt(z,t_{0})$ & m/s\textsuperscript{2} & Zonal wind tendency  & x/min(std,$\gamma_2$) \\
    $dv/dt(z,t_{0})$ & m/s\textsuperscript{2} & Meridional wind tendency  & x/min(std,$\gamma_2$) \\
    NETSW & W/m\textsuperscript{2} & Net shortwave flux at surface & x/std  \\
    FLWDS & W/m\textsuperscript{2} & Downward longwave flux at surface & x/std  \\
    PRECSC & m/s & Snow rate (liquid water equivalent) & x/std  \\
    PRECC & m/s & Rain rate & x/std  \\
    SOLS & W/m\textsuperscript{2} & Downward visible direct solar flux to surface  & x/std \\
    SOLL & W/m\textsuperscript{2} & Downward near-IR direct solar flux to surface & x/std  \\
    SOLSD & W/m\textsuperscript{2} & Downward visible diffuse solar flux to surface & x/std  \\
    SOLLD & W/m\textsuperscript{2} & Downward near-IR diffuse solar flux to surface & x/std  \\
 \hline
 \end{tabular}
  \caption{List of output variables used by all models. $\gamma_1$ and $\gamma_2$ are lower bounds to prevent the output normalizations from creating value that are too large. $\gamma_1 = 3^{-10}$ and $\gamma_2 = 10^{-6}$. The top 15 levels of NN output tendencies for water vapor, total cloud, and both zonal and meridional wind are ``pruned" and set to 0. \label{tab:output_table}}
 \end{table}

\subsection{Architectures}
\begin{figure}[!htbp]
 \centering
 \includegraphics[width=\textwidth]{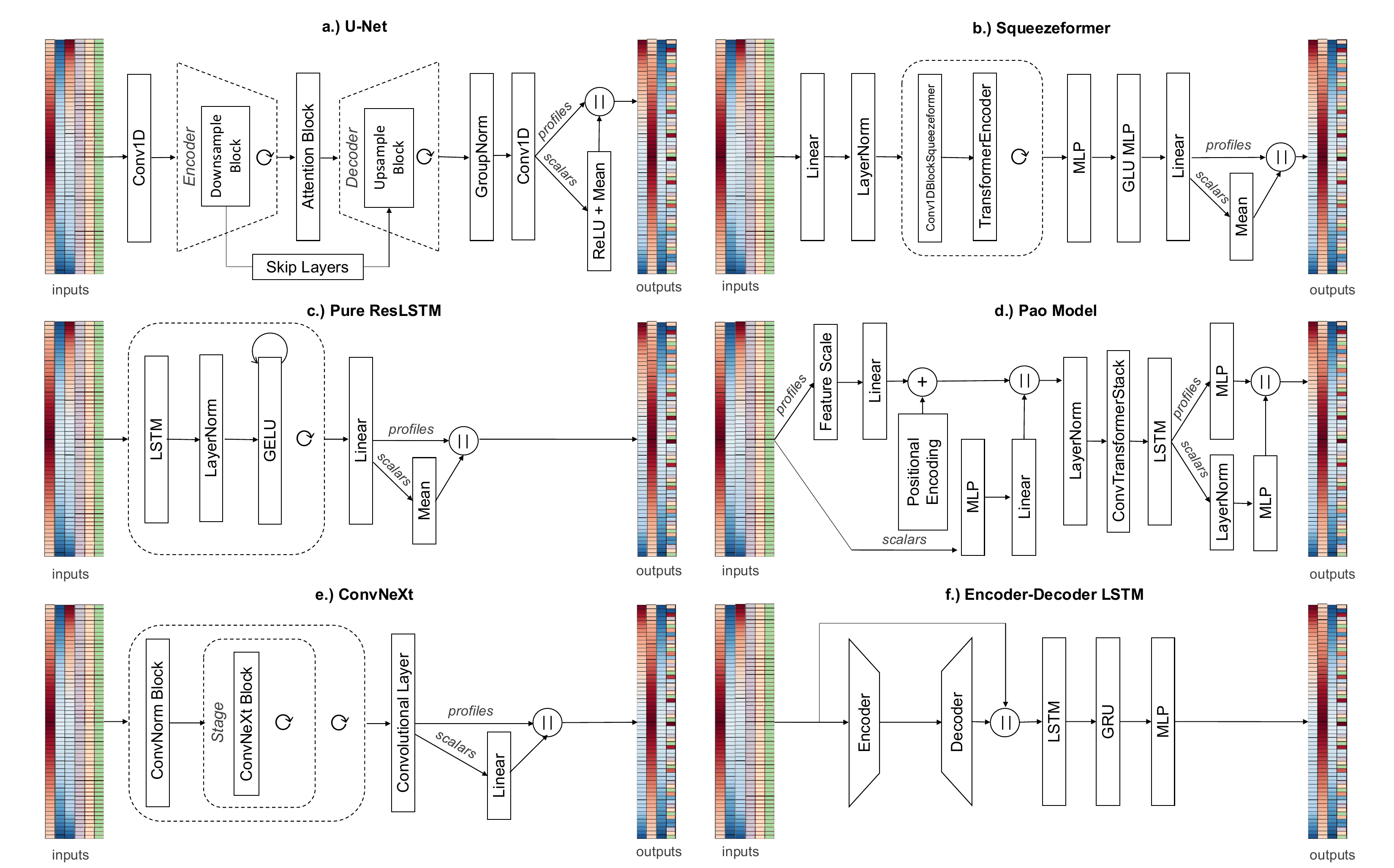}
 \setlength{\belowcaptionskip}{-1em}%

\caption{Architecture diagrams for the U-Net from \cite{Hu2025-mf}, Squeezeformer, Pure ResLSTM, Pao Model, ConvNeXt, and Encoder-Decoder LSTM from the 1st place, 2nd place, 3rd place, 4th place, and 5th place teams in the 2024 LEAP ClimSim Kaggle competition.}
 \label{fig:climsim3_diagrams}
\end{figure}

\subsubsection{U-Net Baseline}

The U-Net architecture is taken from \citeA{Hu2025-mf}, which itself was adapted from a 2-D U-Net created for score-based diffusion modeling in \citeA{song2020score}. In this architecture, shown in Figure \ref{fig:climsim3_diagrams}a, an encoder progressively downsamples the vertical dimension while expanding the feature space (i.e. channels), and the decoder reverses this operation. Skip connections directly link layers of the encoder to corresponding layers of the decoder, preserving fine-grained details that might be lost during the downsampling process. The scalar output variables, which are each represented by a channel of 60 values, are made non-negative via a ReLU function and condensed into scalar values (i.e. $\mathbb{R}^{60} \rightarrow \mathbb{R}^{1}$) via averaging. These scalar values are then concatenated back to the vertically-resolved output variables for the final output vector. This architecture serves as a useful baseline against which to compare winning architectures from the Kaggle competition.

\subsubsection{Squeezeformer}

The Squeezeformer, shown in Figure \ref{fig:climsim3_diagrams}b, is an architecture used by the first place team in the Kaggle competition. Originally invented for automatic speech recognition by \citeA{kim2022squeezeformer}, this architecture has been successful in other contexts, having been used for first and second place finishes in Kaggle competitions for fingerspelling recognition and RNA folding, respectively \cite{stanford-ribonanza-rna-folding, asl-fingerspelling}. This architecture integrates convolutional and transformer components to process sequential data. Initially, input variables are projected into a high-dimensional embedding space of 384 features by a linear layer, followed by layer normalization. The core of the network consists of a series of identical blocks stacked together. Each block first passes the data through a \texttt{Conv1DBlockSqueezeformer} to capture local contextual information as well as relationships between features, and then through a standard \texttt{TransformerEncoder} with multi-head self-attention to model global dependencies across the sequence. The \texttt{Conv1DBlockSqueezeformer} blocks are characterized by gated linear units for selective feature activation, depthwise convolutions for cross-feature mixing, and efficient channel attention to adaptively recalibrate channel-wise features. The result is then passed through an MLP that expands the feature space to 2048 dimensions. This is followed by a Gated Linear Unit Multi-Layer Perceptron (GLUMLP), which projects the features into a lower-dimensional embedding space, splits them into two halves, applies a non-linear activation to one half, multiplies the two halves element-wise, and then linearly transforms the output back to 2048 dimensions. Finally, a linear layer maps the features to the desired number of output variables, and scalar output variables are averaged before being concatenated to the vertically-resolved variables in the final output vector.

\subsubsection{Pure ResLSTM}

The ``Pure ResLSTM", shown in Figure \ref{fig:climsim3_diagrams}c is a multi-layer bidirectional Long Short-Term Memory (LSTM) network adapted from one of the architectures used by the second place team. This architecture consists of multiple LSTM layers, each followed by a layer normalization and Gaussian Error Linear Unit (GELU) activation. The model takes input profiles and scalar features, processes them through 10 blocks each composed of a bidirectional LSTM layer, layer normalization layer, and GELU activation applied to a weighted sum of the current layer and the previous layer. The result is passed through a linear layer, and the channels corresponding to scalar variables are averaged before being concatenated back to the vertically-resolved variables in the final output vector. The use of recurrent neural networks across the vertical dimension can be thought of as embedding a physical prior of locality \cite{Ukkonen2025-jt}.

\subsubsection{Pao Model}

The ``Pao Model", shown in Figure \ref{fig:climsim3_diagrams}d, is adapted from an architecture created from scratch by the third place team in the Kaggle competition. This architecture employs a unique strategy for processing vertically-resolved variables and scalar variables separately before and after using a combined representation that is passed through convolutional, transformer, and LSTM elements. Vertically-resolved variables are first linearly scaled and projected into a high-dimensional embedding space of 160 features before being added to a positional embedding that encodes the vertical discretization. Scalar variables are nonlinearly mapped to a 160 dimensional embedding space via an MLP and linear layer. The transformed tensors are then concatenated and fed into a core sequence of residual blocks (composed of convolutional and transformer encoder elements) followed by a two-layer bidirectional LSTM. Finally, the output is split and processed by separate MLPs to produce vertically-resolved variables and scalar output variables before being concatenated together for the final output vector.

\subsubsection{ConvNeXt}

The ``ConvNeXt" model, shown in Figure \ref{fig:climsim3_diagrams}e, is an architecture used by the fourth place team in the Kaggle competition. It is adapted from work by \citeA{liu2022convnet}, who invented this new generation of convolutional NNs to be competitive with vision transformers, which have become increasingly dominant for vision tasks. ConvNeXt inspired architectures have also made their way into modern ocean emulators and deep learning weather prediction models \cite{Dheeshjith2024-ym, Karlbauer2024-nv, Dheeshjith2025-mw, Duncan2025-ev, Bonev2025-dw}. The 1D ConvNeXt model here employs a stack of ``ConvNorm blocks" and ``stages", alternating between each four times. Each ConvNorm block is composed of a convolutional layer that expands the number of features and a batch normalization layer. The stages are composed of multiple modified ConvNeXt blocks that are each characterized by depthwise convolutions with large kernel sizes for efficient spatial mixing, batch normalization (as opposed to layer normalization), pointwise linear transformations, and a residual connection. The final stage outputs are processed through a convolutional layer that produces combined profile and scalar predictions, with the tensor corresponding to output scalar variables linearly projected into the desired number of output scalar variables.

\subsubsection{Encoder-Decoder LSTM}

The ``Encoder-Decoder LSTM," shown in Figure \ref{fig:climsim3_diagrams}f, is an architecture used by the fifth place team in the Kaggle competition that has already been adopted in follow-up work \cite{Heuer2025-gu}. This architecture first uses an encoder-decoder MLP structure to learn a combined latent representation of all input variables before recurrent processing. The decoded latent representation, which mixes information across variables and levels, is used to create additional input variables for the bidirectional LSTM layer, breaking the locality prior assumed when using vertically recurrent NNs in other ML parameterizations \cite{Ukkonen2025-jt}. The LSTM output is subsequently input into a bidirectional Gated Recurrent Unit (GRU) layer, which refines the sequence representations with a smaller hidden size. Finally, the GRU output is passed through another MLP to generate the final output vector.

\subsection{NN Configurations}
In this section, we compare five architecture-agnostic design decisions, referred to as ``configurations'' for short, that were inspired by design decisions made by winners of the Kaggle competition and the input variable feature expansion in \citeA{Hu2025-mf}. We evaluated these five configurations across all six architectures to detect potential systematic effects. All NNs were trained for 12 epochs with a batch size of 1024, learning rate of 0.0001, AdamW optimizer, Huber loss, and a learning rate scheduler with a ``step" decay strategy in which the learning rate decreases 95\% every 3 epochs.

\subsubsection{Standard Configuration}

The standard configuration serves as a minimal baseline from which other configurations make changes. It can be viewed as identical to the configuration choices of \cite{Hu2025-mf} except by restricting the input variable list to variables available to Kaggle participants. 

\subsubsection{Confidence Loss Configuration}

The confidence loss configuration is motivated by the ``confidence head" used by the first place team and has also recently been adopted by \citeA{Heuer2025-gu} in their development of a ``confidence-guided mixing parameterization'' for transferring knowledge across atmospheric general circulation models. The confidence head mirrors the final prediction layer of the architecture and is used to predict the loss, $L_{\text{pred}}$, for each output in the original prediction layer. The confidence head's prediction for this loss is denoted $\hat{L}_{\text{pred}}$. The loss function for this configuration applies the $\mathrm{Huber}$ loss to the sum of the original loss and the loss for the confidence head prediction, $L_{\text{loss}}$. The combined loss is denoted $L_{\text{conf}}$, and the 50/50 weighting used is justified by the fact that the confidence loss is in loss units. \citeA{Heuer2025-gu} also showed that this framework can be used for uncertainty estimation; however, our implementation detached the prediction loss being used by the confidence head, preventing this application.

\begin{center}
$L_{\text{pred}} = \mathrm{\mathrm{Huber}}(\hat{y}, y)$

$L_{\text{loss}} = \mathrm{\mathrm{Huber}}(\hat{L}_{\text{pred}}, L_{\text{pred}})$

$L_{\text{conf}} = L_{\text{pred}} + L_{\text{loss}}$
\end{center}

\subsubsection{Difference Loss Configuration}

The difference loss configuration, inspired by the second place team and also adopted by \cite{Heuer2025-gu}, is intended to improve the vertical structure of the predicted tendencies and adds an additional term in the loss function that compares vertical differences for the vertically-resolved variables $y_{\text{diff}}$. The combined loss is denoted $L_{\text{total}}$.

\begin{center}
$L_{\text{pred}} = \mathrm{\mathrm{Huber}}(\hat{y}, y)$

$L_{\text{diff}} = \mathrm{\mathrm{Huber}}(\hat{y}_{\text{diff}}, y_{\text{diff}})$

$L_{\text{total}} = L_{\text{pred}} + L_{\text{diff}}$
\end{center}

\subsubsection{Multirepresentation Configuration}

The multirepresentation configuration, inspired by the first place team, makes use of multiple simultaneous ways of encoding vertical profiles to help the model leverage complementary statistical views of the data. This is motivated by the heterogeneous statistical properties of atmospheric vertical profiles, where variables like moisture exhibit strong vertical gradients, skewness, and level-specific variability. Level-wise normalization preserves fine-grained vertical structure, column-wise normalization emphasizes global profile statistics (e.g., total column-integrated quantities), and the log-symmetric transform mitigates skewness in skewed distributions (common in microphysical variables) while preserving symmetry \cite{Hu2025-mf}. This configuration only changes normalization for input variables and distinguishes between vertically-resolved variables $x_{\text{lev}}$ and scalar (i.e., level-invariant) variables $x_{\text{scalar}}$. Scalar variables are normalized using standard z-scores. For vertically-resolved variables, we construct three parallel representations:

\begin{enumerate}
    \item \textbf{Level-wise normalization}: Each feature at each vertical level is normalized using its own level-specific mean and standard deviation.
    \item \textbf{Column-wise normalization}: Each vertically-resolved feature is normalized across all levels using a single global mean and standard deviation for that feature.
    \item \textbf{Level-wise log-symmetric transformation}: A smooth, sign-preserving logarithmic transform (shown below) is applied to the levelwise-normalized features to reduce skewness while handling negative values.
    \[
    x_{\text{log}} = 
    \begin{cases}
    \log{(x_{\text{lev}} - \alpha_{\text{lev}} + 1)} & \text{where } x_{\text{lev}} \geq \alpha_{\text{lev}}, \\
    -[\log{(\alpha_{\text{lev}} - x_{\text{lev}} + 1)}] & \text{where } x_{\text{lev}} < \alpha_{\text{lev}}, \\
    \end{cases}
    \]
    In this case, $\alpha$ is defined separately for each vertically-resolved variable and vertical level. For a given variable $x_{\text{lev}}$ with variable and level mean $\mu_{\text{lev}}$, $\alpha_{\text{lev}}$ is equal to the minimum of $\left( \frac{x_{\text{lev}} - \mu_{\text{lev}}}{\mu_{\text{lev}}} \right)$ across all $x_{\text{lev}}$ in the training data. 
\end{enumerate}





\subsubsection{Expanded Variable List Configuration}

The expanded variable list configuration appends inputs listed in Table \ref{tab:expanded_inputs} to the baseline set of input variables listed in Table \ref{tab:input_table}. This expanded variable list matches the variable list used to achieve State-Of-The-Art (SOTA) results shown in \citeA{Hu2025-mf}. These expanded input variables consist of tendencies and large-scale forcings at two timesteps as well as sin and cosine of latitude.

\begin{table}[ht]

 \centering
 \begin{tabular}{l l l l}
 \hline
  \textbf{Variable} & \textbf{Units} & \textbf{Description} & \textbf{Normalization} \\
 \hline
    $dT_{adv}/dt(z,t_{0})$ & K/s & Large-scale forcing of temperature at (t) & x/(max-min)  \\
    $dq_{T,adv}/dt(z,t_{0})$ & kg/kg/s & Large-scale forcing of total water at (t) & x/(max-min) \\
    $du_{adv}/dt(z,t_{0})$ & m/s\textsuperscript{2} & Large-scale forcing of zonal wind at (t) & x/(max-min)  \\
    $dT_{adv}/dt(z,t_{-1})$ & K/s & Large-scale forcing of temperature at (t-1) & x/(max-min)  \\
    $dq_{T,adv}/dt(z,t_{-1})$ & kg/kg/s & Large-scale forcing of total water at (t-1) & x/(max-min) \\
    $du_{adv}/dt(z,t_{-1})$ & m/s\textsuperscript{2} & Large-scale forcing of zonal wind at (t-1) & x/(max-min)  \\
    $dT/dt(z,t_{-1})$ & K/s & Temperature tendency at (t-1) & x/std \\
    $dq_v/dt(z,t_{-1})$ & kg/kg/s & Water vapor tendency at (t-1) & x/std  \\
    $dq_n/dt(z,t_{-1})$ & kg/kg/s & Total cloud tendency at (t-1) &  x/std  \\
    $du/dt(z,t_{-1})$ & m/s\textsuperscript{2} & Zonal wind tendency at (t-1) & x/std  \\
    $dT/dt(z,t_{-2})$ & K/s & Temperature tendency at (t-2) & x/std \\
    $dq_v/dt(z,t_{-2})$ & kg/kg/s & Water vapor tendency at (t-2) & x/std  \\
    $dq_n/dt(z,t_{-2})$ & kg/kg/s & Total cloud tendency at (t-2) &  x/std  \\
    $du/dt(z,t_{-2})$ & m/s\textsuperscript{2} & Zonal wind tendency at (t-2) & x/std  \\
    cos(lat) & & Cosine of latitude   \\
    sin(lat) & & Sine of latitude   \\
 \hline
 \end{tabular}
  \caption{List of additional input features used in the Expanded Variable List configuration.  \label{tab:expanded_inputs}}
 \end{table}

\section{Results}

\subsection{Offline $R^2$ comparison}

\begin{figure}
    \centering
    \includegraphics[width=\textwidth]{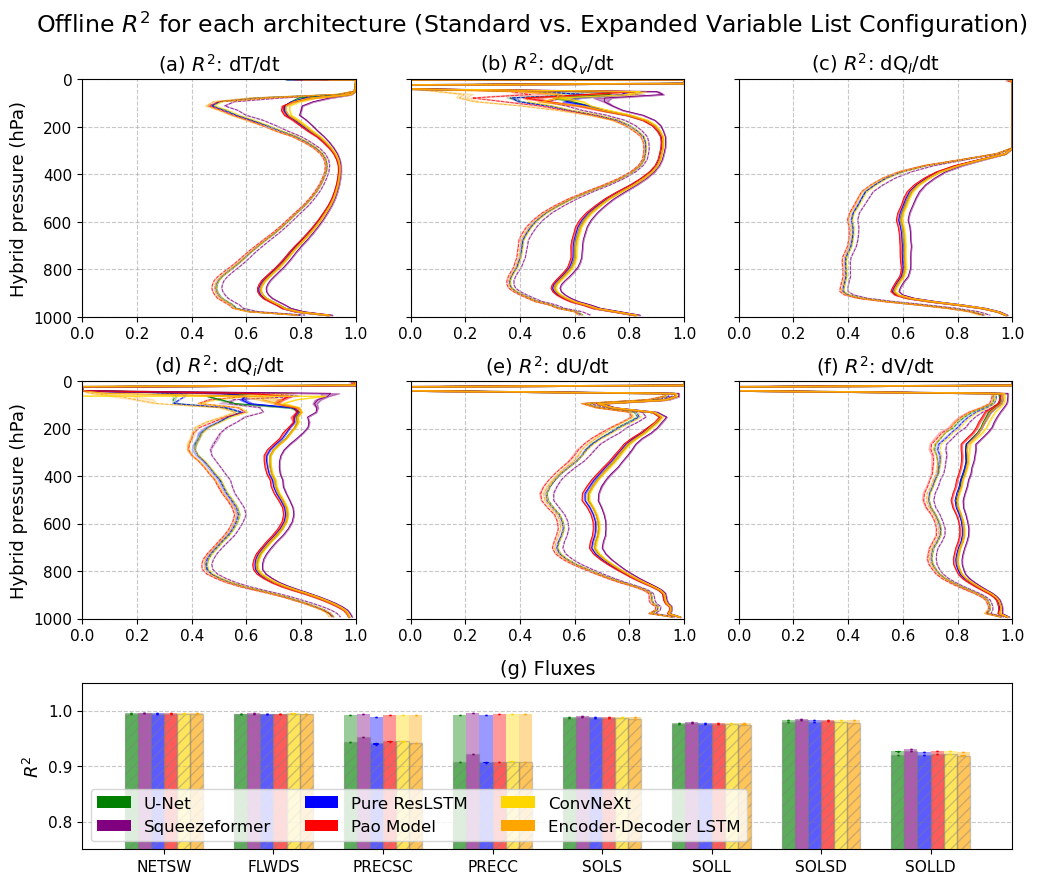}
 \setlength{\belowcaptionskip}{-1em}%

\caption{Offline $R^2$ values for each variable across architectures for the standard configuration (depicted using dashed lines and hatched bar charts) and the expanded variable list configuration (depicted using solid lines and semi-transparent shaded bar charts). For vertically-resolved variables, the colored lines depict the median $R^2$ while the shading shows the min and max across seeds for each architecture. For scalar variables, the bars show the medians while the vertical lines at the top of each bar show the min-max range.}

 \label{fig:offline_r2_lines_standard_vs_v6}
\end{figure}

In the Kaggle competition leaderboard, the winning solutions outperformed the U-Net submission by a considerable margin in offline $R^2$ skill. Surprisingly, this gap nearly vanishes after controlling for feature selection, preprocessing, normalization, training settings, postprocessing, and using individual models as opposed to ensembles. This is shown in Figure \ref{fig:offline_r2_lines_standard_vs_v6}, which displays offline $R^2$ scores for both the standard configuration (depicted with dashed lines and hatched bar charts) and the expanded variable list configuration. A possible explanation for the large gap in the Kaggle competition is that the output normalization used by the U-Net resulted in worse performance (i.e. negative $R^2$) for stratospheric liquid cloud tendencies. While \citeA{Hu2025-mf} and this work make use of a microphysics constraint that mirrors the one-moment microphysics scheme in the CRM to diagnose liquid and ice cloud tendencies as a function of total liquid and ice cloud tendency and temperature, Kaggle competitors coalesced around a ``trick" that set stratospheric liquid cloud tendencies to $-\frac{q_l}{1200}$ (where $q_l$ represents the amount of liquid cloud for a given unpredictable stratospheric level). This ``trick" is based on the assumption that stratospheric moisture would tend to 0 (since there are 1200 seconds in a GCM timestep). Finally, the best offline $R^2$ scores in this paper are not universally superior to those reported in \citeA{Hu2025-mf}. Nevertheless, when controlling for the aforementioned factors, the Squeezeformer, which was used in the first place solution, shows clear positive separation from all architectures on all variables, and the reverse is true for the Pao Model, which was used in the third place solution. For all other architectures, the offline $R^2$ is less differentiated, with results varying based on variable and configuration. There is also very little variation in offline $R^2$ between seeds, as shown by the min-max shading for the vertically-resolved variables and the nearly invisible vertical lines showing min-max range for the scalar variables. Finally, as expected from previous work, there is a universal improvement to offline skill across architectures when expanding the input variable list to include variables like convective memory and large-scale forcings \cite{Lin2025-ya, Hu2025-mf}. The confidence loss, difference loss, and multirepresentation configurations yield smaller benefits compared to the offline $R^2$ of the standard configuration and are omitted from Figure \ref{fig:offline_r2_lines_standard_vs_v6} for visual clarity. However, as shown in Figure S3, the multirepresentation configuration yields large improvements in offline $R^2$, specifically for liquid cloud tendencies above 200 hPa and ice cloud tendencies above 400 hPa. This validates the idea that multiple complementary statistical views of the data can be useful for cloud mixing ratios, which have highly skewed distributions with long tails extending beyond 10 standard deviations \cite{Hu2025-mf}. For completeness, we show how the confidence loss and difference loss configurations compare to the standard configuration in Figures S1 and S2, and we compare the models in the expanded variable list configuration to the models evaluated in \citeA{Hu2025-mf} in Figure S4.

\subsection{Online Stability and Error Growth}
\begin{figure}
    \centering
    \includegraphics[width=\textwidth]{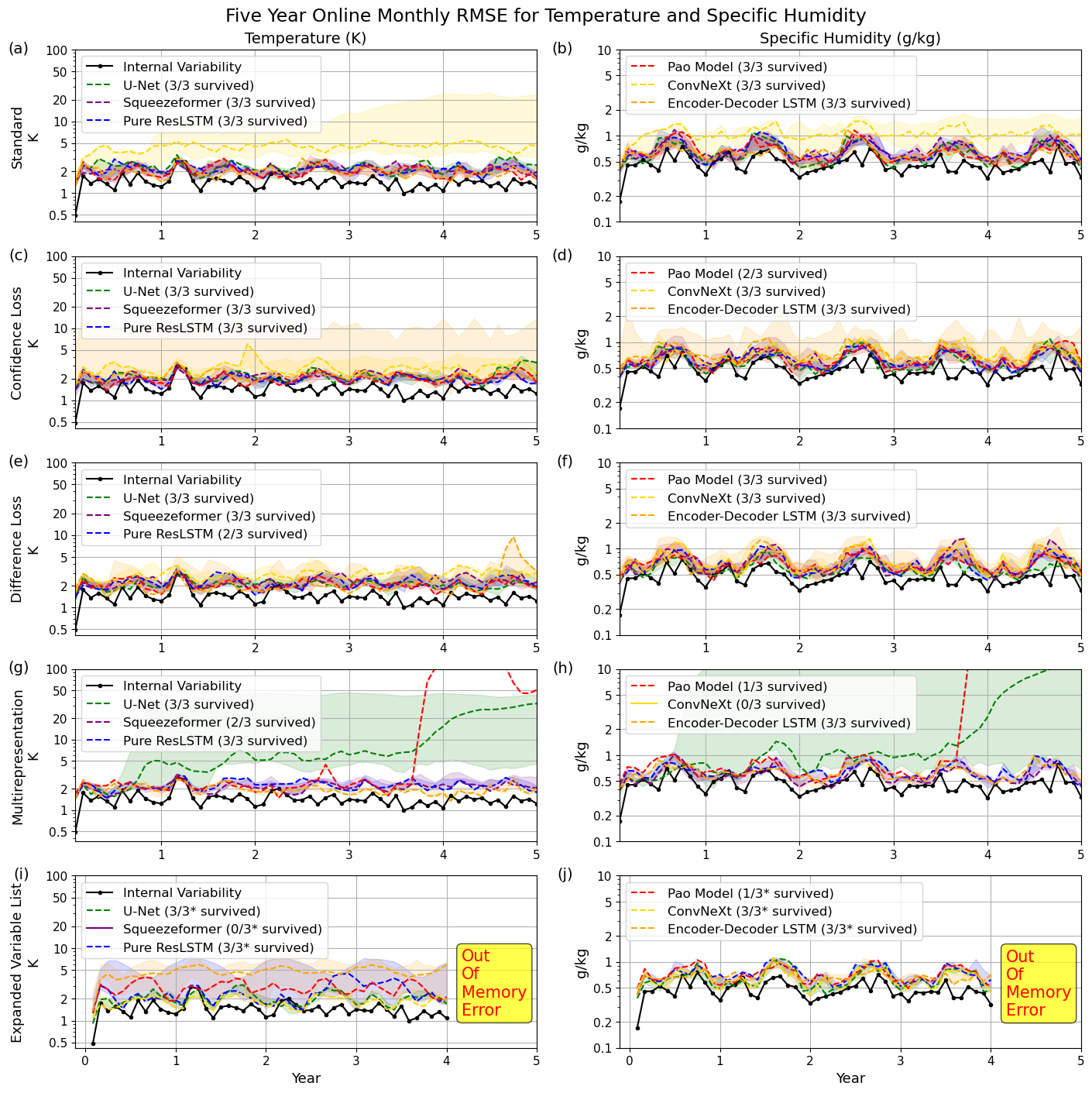}
 \setlength{\belowcaptionskip}{-1em}%

\caption{Online monthly RMSE for temperature and moisture for all architectures in each configuration. Shading indicates the inter-seed range (min to max RMSE across three seeds per month after excluding RMSE from hybrid simulations that crash at any point). Dashed lines show RMSE from the seed whose monthly mean absolute deviation from MMF RMSE is closest to the median absolute deviation across seeds. For visual clarity, RMSE for hybrid simulations that crash due to numerical instability are not shown. Internal variability is approximated via monthly RMSE when compared to another MMF simulation (which in general are not bit-for-bit reproducible). Subplots i and j only show data up to four years because of out-of-memory issues that caused most of the hybrid simulations to terminate in the middle of the fifth year. Asterisk (*) indicates that survival is assessed via integrating for four, and not five, simulation years.}

 \label{fig:five_year_online_monthly_rmse_t_and_q}
\end{figure}

Online, most architectures across configurations exhibit stable performance with minimal drift (i.e. systematic or growing errors), yet with some notable exceptions, as seen in Figure \ref{fig:five_year_online_monthly_rmse_t_and_q}, which shows monthly RMSE for temperature and specific humidity across all configurations for each architecture. Monthly RMSE for other prognostic variables are shown in Figures S5, S6, S7, and S8, and the formula by which we calculate it is shown in Text S2. These results suggest that stability in the low-resolution real-geography ClimSim setting may be the default expectation going forward. We note that these results were obtained using the microphysics constraint by \citeA{Hu2025-mf}, which was shown to improve online stability. Even so, cases of instability and drift still exist, and they arise for different architectures depending on choice of configuration.

In the standard configuration (Figure \ref{fig:five_year_online_monthly_rmse_t_and_q} a,b), all hybrid simulations stably integrate without drift with the exception of those corresponding to the ConvNeXt architecture (yellow). Interestingly, this is achieved without the use of convective memory, inconsistent with previous experience by some of our team that had implicated this as important (albeit in a higher-resolution, aquaplanet setting using a different MMF version) in \citeA{Lin2025-ya}.

In the confidence loss and difference loss configurations, we see the first signs of inter-seed variability in online stability and drift. In the confidence loss configuration, one of the seeds for the Encoder-Decoder LSTM experiences sustained high online temperature RMSE and one of the seeds for the Pao Model crashes due to numerical instability in the middle of the fourth simulation year. In the difference loss configuration, one of the seeds for the Pure ResLSTM architecture crashes in the middle of the fifth simulation year. Impactful inter-seed variability in online performance has been shown in prior work and is confirmed here, emphasizing the ongoing importance of training ensembles of ML parameterizations before drawing design conclusions \cite{Han2023-nm}. Interestingly, the ConvNeXt architecture, which experienced significant online temperature error in the standard configuration, integrates stably without drift across all seeds in the confidence loss and difference loss configurations.

For the multirepresentation configuration (Figure \ref{fig:five_year_online_monthly_rmse_t_and_q} g,h), the two architectures without convolutional elements---the Pure ResLSTM and the Encoder-Decoder LSTM---are the only ones to stably integrate without issue. Across the U-Net, ConvNeXt, and Pao Model architectures, every hybrid simulation experiences catastrophic drift or numerical instability. In the case of the Squeezeformer architecture, one of the hybrid simulations crashes from numerical instability within the first hundred simulation days and an annually recurring liquid cloud RMSE maximum emerges, consistent with systematic biases at a discrete phase of the seasonal cycle (Figure S6).

In the expanded variable list configuration, architectures that make repeated use of attention or LSTM elements are negatively impacted. The Squeezeformer architecture is worst affected, with hybrid simulations crashing in the first ten simulation days. Meanwhile, only one out of three hybrid simulations using the Pao Model runs to completion, and the inter-seed variability in online temperature RMSE for both the Pure ResLSTM and the Encoder-Decoder LSTM is worsened. Conversely, hybrid simulations using the ConvNeXt architecture (which experiences catastrophic drift in the standard configuration) achieve best-in-class online temperature error in this configuration. Finally, the U-Net shows no signs of drift or numerical instability, although this was also the case in the standard configuration.

We excluded the fifth simulation year when assessing the online results for the expanded variable list configuration. This is because 10/18 hybrid simulations using this configuration encountered out-of-memory issues caused by our transition from the PyTorch-Fortran binding developed in \citeA{Hu2025-mf} to an FTorch binding (which makes use of CUDA GPU acceleration, resulting in a 4x speedup). In our implementation of the FTorch binding, we neglected to make appropriate use of \texttt{torch\_delete} (which is now automated in newer versions of FTorch), introducing out-of-memory issues. Nevertheless, based on the self-similarity of most seasonal RMSE variations we do not believe curtailing the simulations at four years significantly impacts our conclusions, and, in general, we use four-year simulations when comparing configurations and five-year simulations when directly comparing our online results to those seen in \citeA{Hu2025-mf}.




\subsection{SOTA Online Results and a Revealed Persistent Secondary Bias}

\begin{figure}
    \centering
    \includegraphics[width=\textwidth]{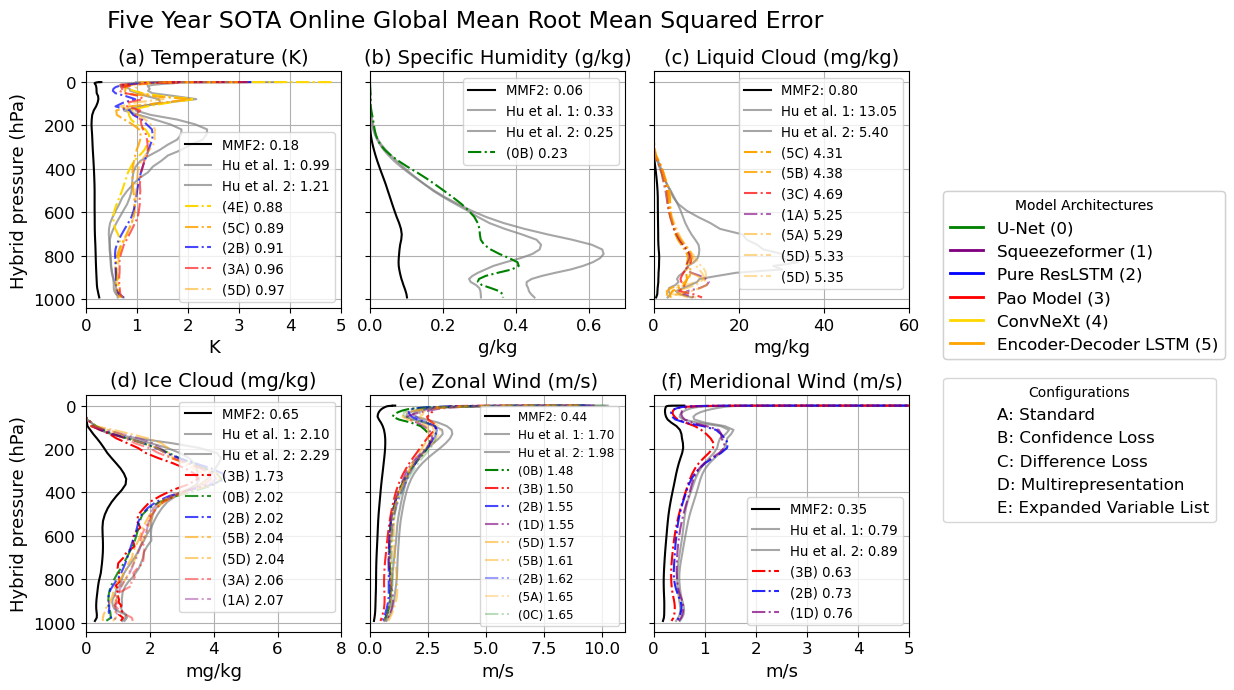}
 \setlength{\belowcaptionskip}{-1em}%

\caption{Vertical profiles of online global RMSE across temperature, specific humidity, liquid cloud, ice cloud, zonal wind, and meridional wind for architectures that surpass those of the U-Net shown in \citeA{Hu2025-mf}. Each subplot contains a legend that uses a number and letter to denote the architecture and configuration corresponding to each vertical profile.}

 \label{fig:five_year_online_sota_results}
\end{figure}

Regarding online five year global RMSE, no single hybrid simulation yields a simultaneous improvement over the best results in \citeA{Hu2025-mf} across all field variables of temperature, liquid cloud, ice cloud, zonal wind, and meridional wind. However, Kaggle-inspired architectures and design decisions did lead to state-of-the-art (SOTA) variable-specific results, as seen in Figure \ref{fig:five_year_online_sota_results}. For the lowest online RMSE achieved by all hybrid simulations in each category, we see an 11.1\% improvement for temperature, 8\% improvement for moisture, 20.2\% improvement for liquid cloud, 17.6\% improvement for ice cloud, 12.9\% improvement for zonal wind, and 20.3\% improvement for meridional wind, relative to results from \citeA{Hu2025-mf}. For online moisture RMSE, only one of 90 hybrid simulations outperforms the best results from \citeA{Hu2025-mf}. This simulation uses the same U-Net architecture but with our confidence loss configuration. For online liquid cloud RMSE, five out of seven hybrid simulations that achieve a lower RMSE than \citeA{Hu2025-mf} make use of the Encoder-Decoder LSTM architecture. These results could suggest that different architectures are better suited at emulating different tendencies; however, confirming this hypothesis and understanding associated mechanisms would require a more targeted study with greater sampling of the online consequences of intrinsic machine learning uncertainties \cite{Lin2025-ya}.

\begin{figure}
    \centering
    \includegraphics[width=\textwidth]{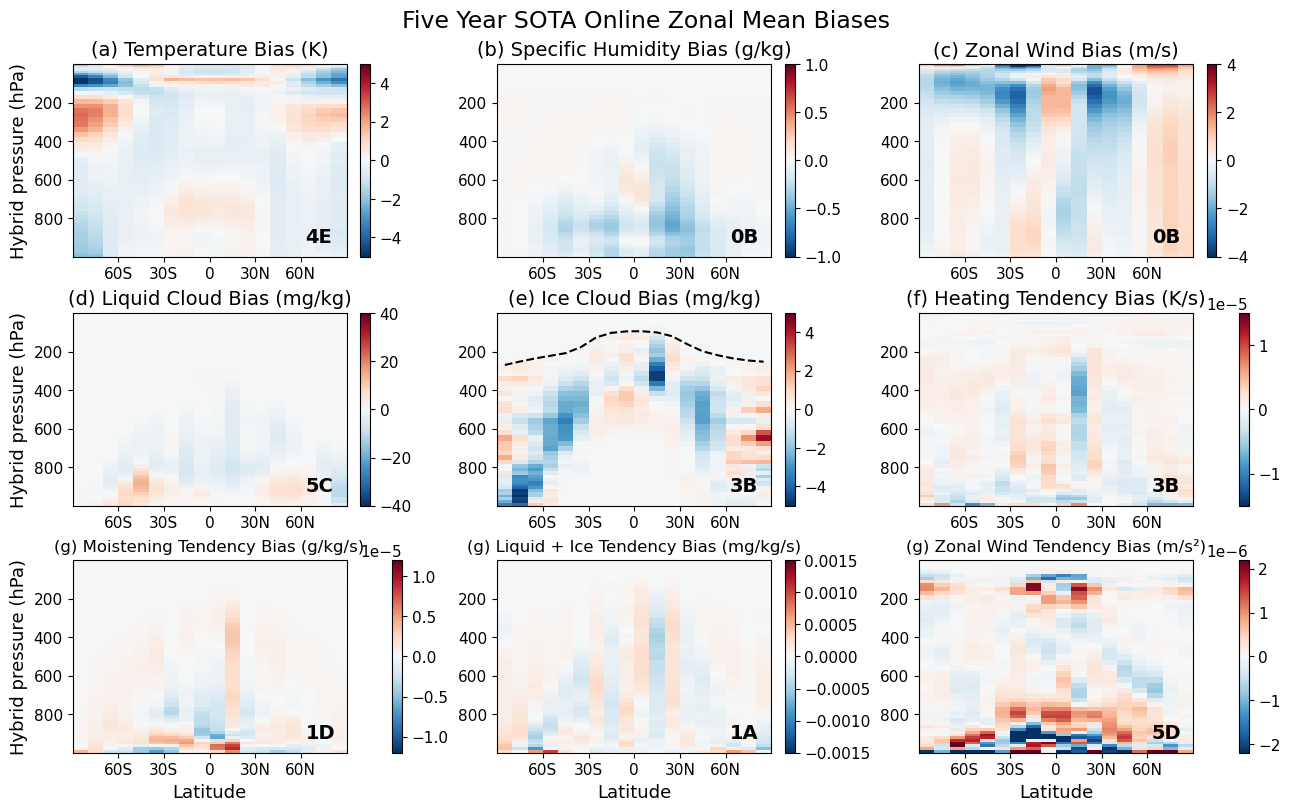}
 \setlength{\belowcaptionskip}{-1em}%

\caption{Online zonal mean bias for the architectures with the lowest global RMSE. In a similar fashion to Figure \ref{fig:five_year_online_sota_results}, the architecture and configuration responsible for each zonal mean bias plot is denoted with a number-letter combination where the numbers (0-6) corresponds to the choice of architecture (i.e. U-Net, Squeezeformer, Pure ResLSTM, Pao Model, ConvNeXt, and Encoder-Decoder LSTM, respectively) while the letters (A-E) corresponds to the choice of configuration (i.e. standard, confidence loss, difference loss, multirepresentation, and expanded variable list, respectively).}

 \label{fig:five_year_sota_online_zonal_mean_biases}
\end{figure}

\begin{figure}
    \centering
    \includegraphics[width=\textwidth]{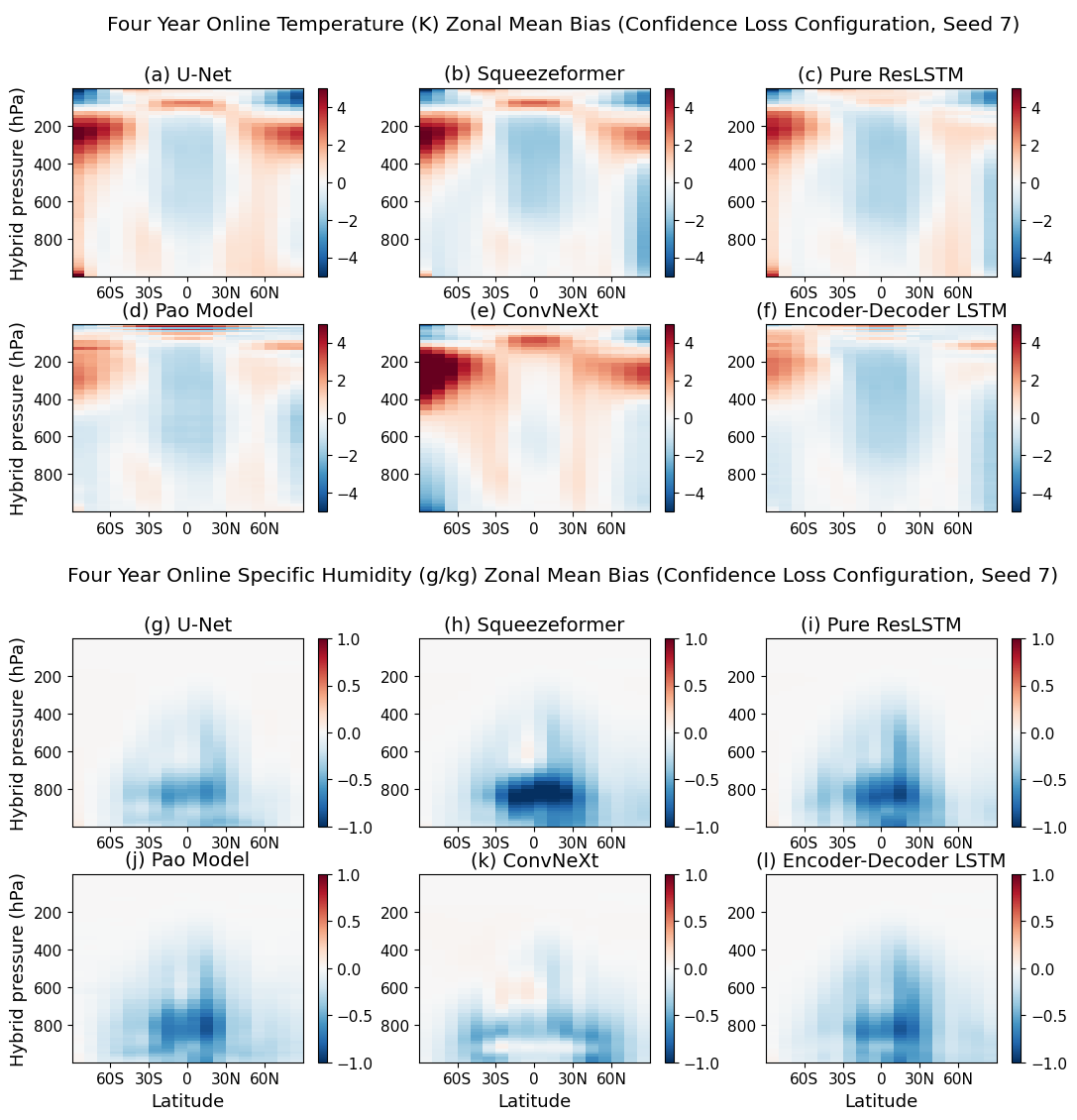}
 \setlength{\belowcaptionskip}{-1em}%

\caption{This figure shows online zonal mean biases for temperature and moisture across architectures in the confidence loss configuration, trained using a common seed.}

 \label{fig:four_year_online_t_and_q_bias_conf_loss}
\end{figure}

In Figure \ref{fig:five_year_sota_online_zonal_mean_biases}, we show the online zonal mean biases corresponding to the hybrid simulations with state-of-the-art online five year global RMSE for temperature, specific humidity, zonal wind, liquid cloud, ice cloud, heating tendencies, moistening tendencies, liquid and ice cloud tendencies, and zonal wind tendencies. While some of these zonal mean biases represent substantial improvements over those seen in \citeA{Hu2025-mf}, the overall structure of these biases persists. Online zonal mean biases with similar structure (e.g. warm bias at higher altitudes over the poles) have also been reported in multiple independent studies (e.g. \citeA{Rasp2018-fk, Wang2022-po, iglesias2024causally, Lin2025-ya, Hu2025-mf}) and are \textit{nearly universal} for all online simulations conducted in this study. As an illustrative example, we show the online zonal mean temperature and moisture biases for different architectures in the confidence loss configuration using a common seed in Figure \ref{fig:four_year_online_t_and_q_bias_conf_loss}. A similar phenomenon exists across all variables, as seen in Figures S9, S10, S11, and S12.

\begin{figure}
    \centering
    \includegraphics[width=\textwidth]{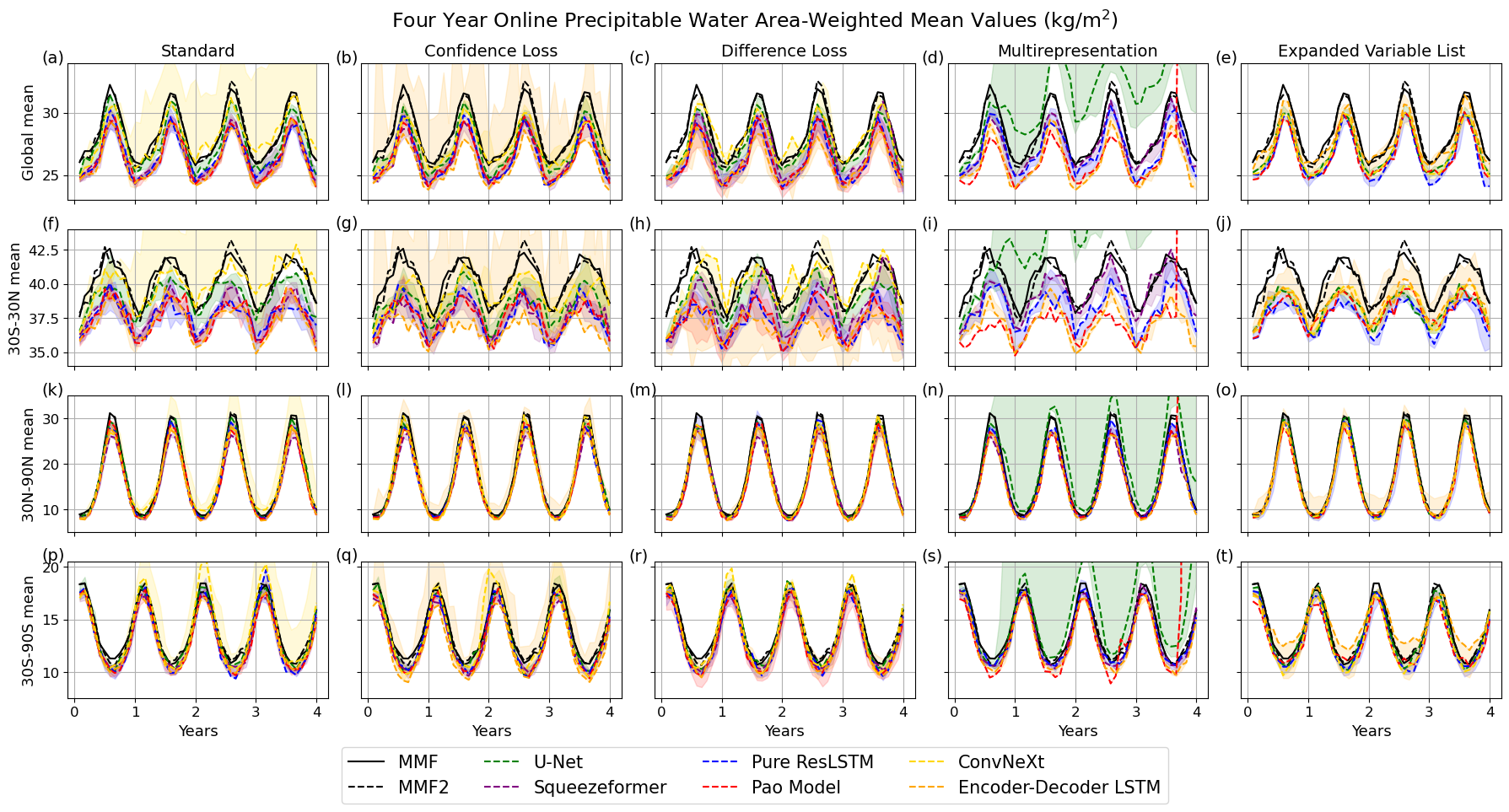}
 \setlength{\belowcaptionskip}{-1em}%

\caption{Four-year area-weighted mean values for precipitable water in four regions (global, 30S–30N, 30N–90N, and 30S–90S) for all architectures across all configurations. Solid and dashed black lines show two independent MMF reference simulations. For each architecture and configuration combination, shading depicts the min-to-max range across seeds, and the dashed colored line shows the median-performing seed by mean absolute difference from MMF.}

 \label{fig:four_year_online_precipitable_water_mean}
\end{figure}

\begin{figure}
    \centering
    \includegraphics[width=\textwidth]{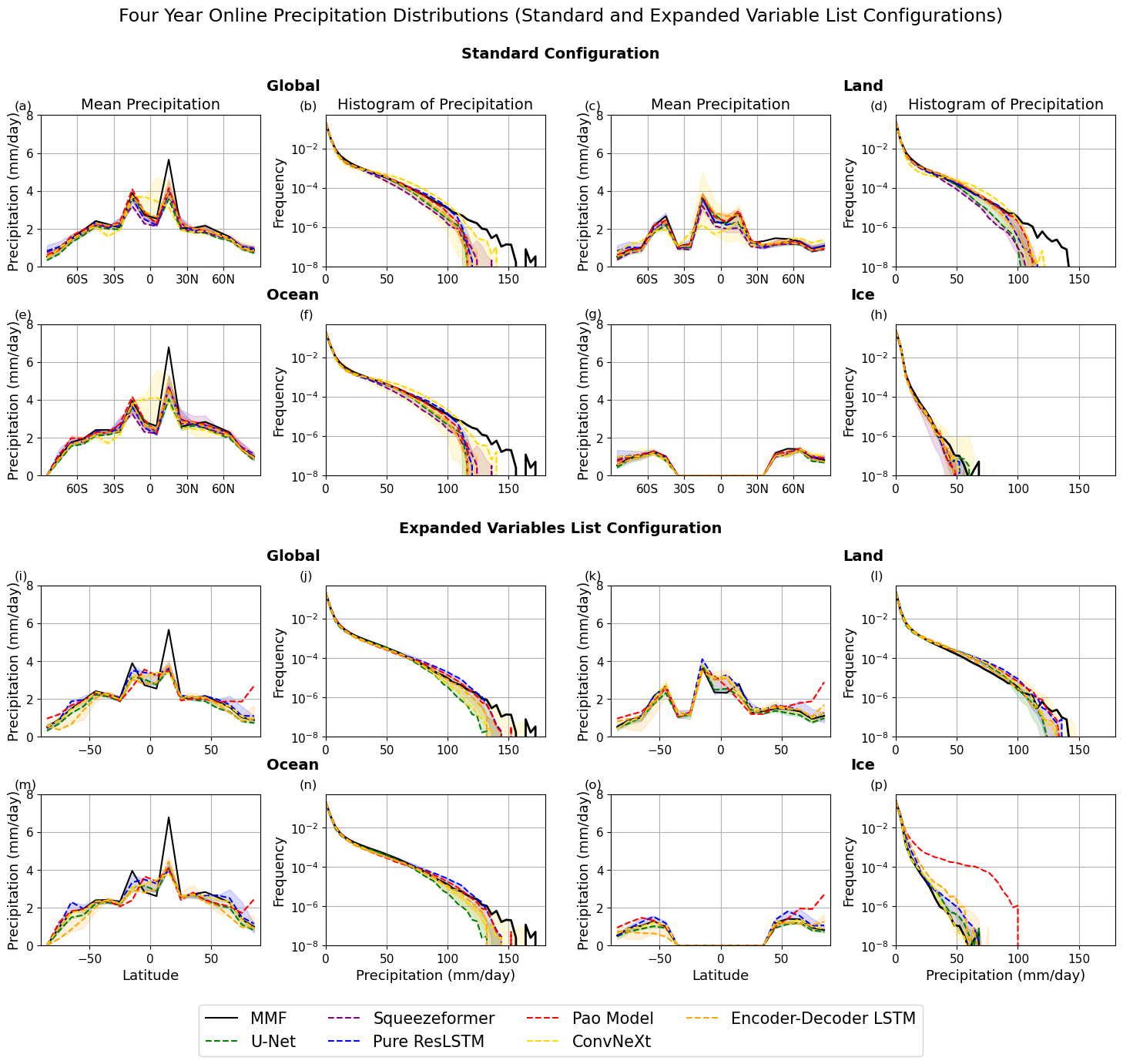}
 \setlength{\belowcaptionskip}{-1em}%

\caption{Four-year online precipitation distributions for the standard (top half) and expanded variable list (bottom half) configurations. Each configuration block shows four surface types (global, land, ocean, and ice), with paired subplots per surface type: zonal mean precipitation (mm/day) as a function of latitude (left) and an area-weighted histogram of hourly precipitation on a log-frequency scale (right). The solid black line shows the MMF reference simulation. For each architecture, shading depicts the min-to-max range across seeds, and the dashed colored line shows the median-performing seed by mean absolute difference from the median.}

 \label{fig:four_year_online_precc_dists}
\end{figure}

In addition to exhibiting similar zonal mean biases, all architectures across all configurations systematically underpredict both the global average and standard deviation of precipitable water online, particularly in the tropics, as seen in Figure \ref{fig:four_year_online_precipitable_water_mean} and Figure S13. Precipitation extremes are also underestimated across architectures in the standard configuration; however, this is partially mitigated in the expanded variable list configuration, as seen in Figure \ref{fig:four_year_online_precc_dists}. Corresponding figures for other configurations are shown in Figures S14, S15, and S16. In line with the findings from \citeA{Hwong2023-hc, Shamekh2023-ed, Beucler2025}, we hypothesize that the improved representation of precipitation extremes when using an expanded variable list may stem from the ability of convective memory to provide more information about underlying subgrid-scale cloud structure and organization \citeA{Colin2019-ks}. If so, future work could improve performance on precipitation extremes by making the ML parameterizations inherently stochastic and more explicitly incorporating CRM state information, which persists between time steps but is currently excluded from ML inputs due to data volume constraints \cite{Leutbecher2017-pu, Christensen2024-dh, Schneider2024-jx}. An alternative explanation could be that the expanded variable list also offers large-scale forcings, which provide additional information about the overall model state prior to running the dynamical core. However, a more detailed ablation study of the expanded variables would be needed to understand the actual cause. 

\subsection{Universal Offline Biases}

\begin{figure}
    \centering
    \includegraphics[width=\textwidth]{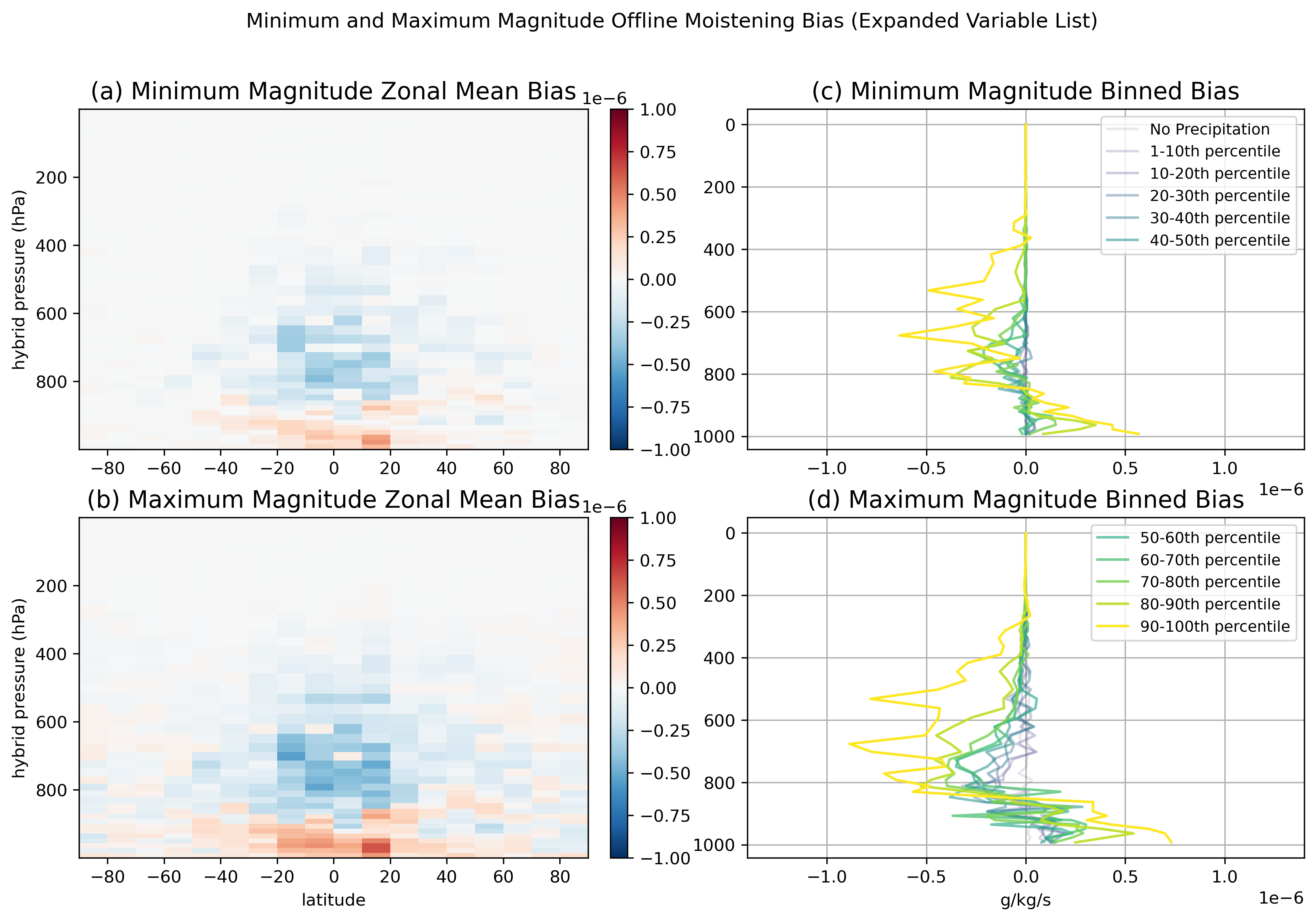}
 \setlength{\belowcaptionskip}{-1em}%

\caption{Figures 9a and 9b show the minimum and maximum offline zonal-mean moistening tendency biases across the expanded variable list configuration. The minima and maxima are calculated over the architecture dimension after averaging across seeds. Figure 9c and 9d show the minima and maxima of the vertical profiles of these biases binned by precipitation percentile (also after averaging across seeds).}

 \label{fig:offline_DQ1PHYS_bias_minmax_v6}
\end{figure}

Pathologies that transcend the choice of architecture are even more pronounced offline. Because offline errors compound across timesteps online, offline biases are a natural suspect for stubborn online biases. Offline zonal mean biases across vertically-resolved variables are remarkably similar across seeds, architectures, and configurations, as seen in Figures 9 and SI Figures S17-S45. For offline moistening, these biases are strongest in the tropics with a wet bias near the surface and dry bias higher in the atmosphere. Figure 9 and the aforementioned SI figures also reveal the state-dependent nature of this bias, whose magnitude increases with convective activity (approximated here using precipitation percentile). This is in line with findings from \citeA{Heuer2025-gu}, who showed that the ML parameterization was most ``uncertain" in moist, unstable conditions, which occur more frequently in lower latitudes. Interestingly enough, the direction of this bias is consistent across seeds, architectures, and configurations, indicating that offline biases are a systemic issue that is unlikely to be resolved by simply using a more sophisticated deterministic architecture. We did not expect to find such self-similar offline biases across the top architectures in the Kaggle leaderboard, which could be viewed as a limitation of the $R^2$-based reward metric used in the competition. This could also be viewed as motivation to explicitly penalize zonal mean bias in the loss function to avoid the accumulation of systematic biases online. This is analogous in aim, though not in type of bias, to the penalization of spectral bias in \citeA{Kochkov2024-uy}.

Multiple other variables like zonal and meridional wind tendencies also show state-dependent universal offline biases, but there are also exceptions. For example, the offline zonal mean heating tendency bias is greatly reduced and loses its distinctive structure (i.e. a cooling bias over the tropics) with an expanded variable list. When it comes to offline liquid and ice cloud tendency biases, those are indeed seemingly universal across all models trained in this study. However, such biases were not seen when evaluating predictions produced directly by the first and second place Kaggle teams (shown in Figures S46 and S47). A plausible cause for this discrepancy is the use of the microphysics constraint developed by \citeA{Hu2025-mf} as it was implemented across all of our models but not used by the Kaggle competitors. However, a more thorough ablation is required to fully confirm this hypothesis.

\subsection{Offline Training and Online Simulation Computational Efficiency}

\begin{figure}
    \centering
    \includegraphics[width=\textwidth]{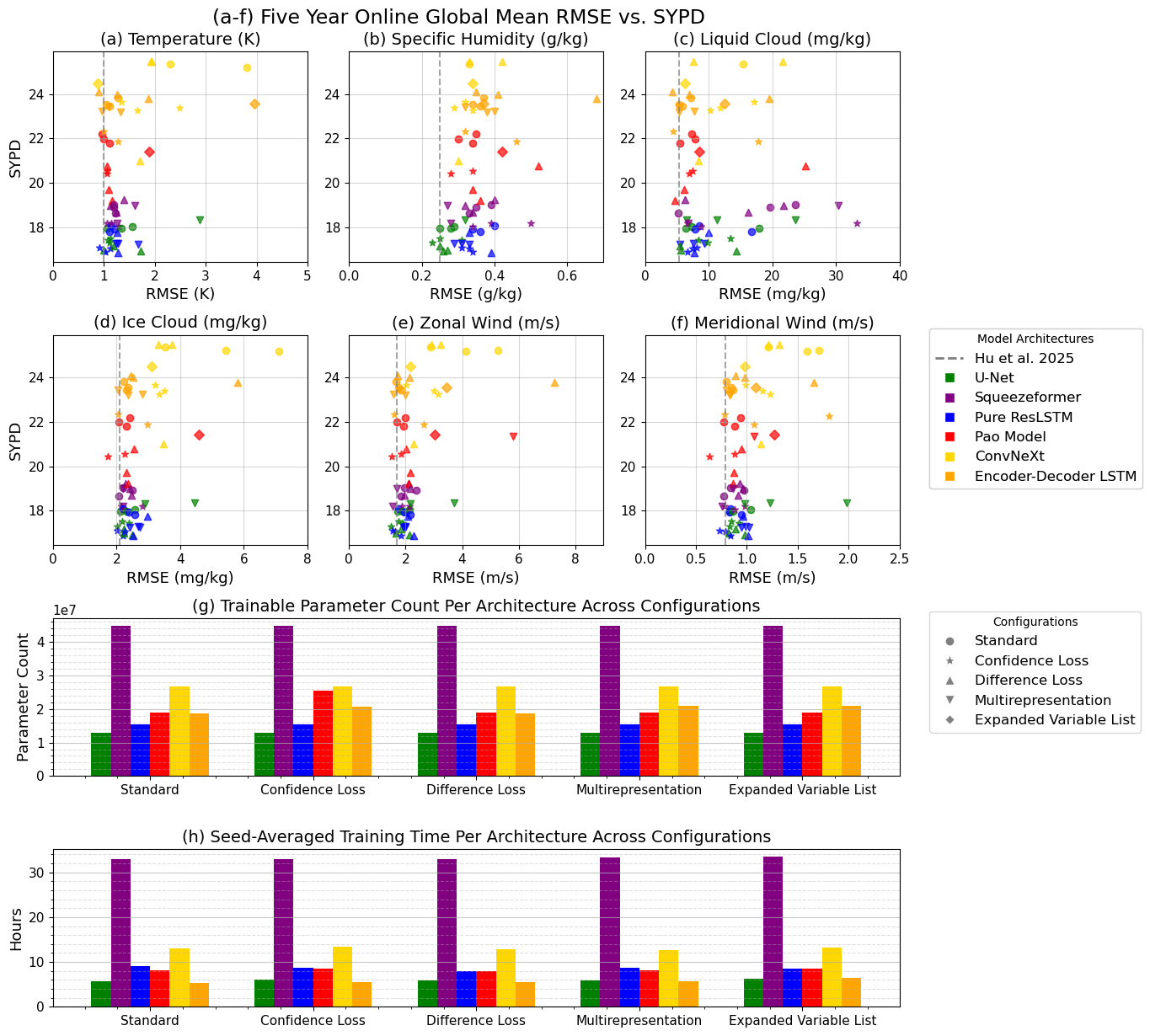}
 \setlength{\belowcaptionskip}{-1em}%

\caption{Figures 10a-f shows Simulation Years Per Day (SYPD) vs. Global Mean RMSE for temperature, specific humidity, liquid cloud, ice cloud, zonal wind, and meridional wind for all hybrid simulations that integrated five years. The best five year Global Mean RMSE for each variable from \citeA{Hu2025-mf} is depicted with a dashed vertical line for reference. Figure 10g shows the number of trainable parameters for each architecture and configuration, and Figure 10h shows the seed-averaged training time for each architecture and configuration combination. Each model was trained using a Distributed Data Parallel (DDP) strategy across 4 NVIDIA A100 GPUs, and each online simulation was conducted with GPU acceleration across 8 NVIDIA A100 GPUs.}

 \label{fig:five_year_rmse_vs_sypd_figure}
\end{figure}

In addition to being able to accurately emulate the coarse-grained effects of subgrid physics in MMF, we would like ML parameterizations to be able to do so efficiently. To measure tradeoffs between these competing priorities, we evaluate five year online global mean RMSE against Simulation Years Per Day (SYPD) in Figures \ref{fig:five_year_rmse_vs_sypd_figure}a-f. In our case, all online simulations are run with GPU acceleration on 8 NVIDIA A100 GPUs using an FTorch binding. The reference MMF simulation is also GPU accelerated, and achieves 9.9 SYPD \cite{Hu2025-mf}. While earlier figures comparing online error across different variables were mostly ambiguous with regards to which architectures were more performant, a clearer hierachy emerges with SYPD. The U-Net and Pure ResLSTM show similar SYPD, but they are surpassed by the Pao Model which is in turn surpassed by the Encoder-Decoder LSTM and finally by the ConvNeXt architecture. It is important to note that this hierarchy is not predictable a priori using the respective trainable parameter count for each architecture, shown in Figure \ref{fig:five_year_rmse_vs_sypd_figure}g. For example, the U-Net and Pure ResLSTM have lower trainable parameter counts and lower SYPD than all other architectures evaluated in this study. By contrast, the ConvNeXt architecture has the second highest trainable parameter count and achieves the highest SYPD across all architectures. When it comes to training time, the number of trainable parameters is critical (as seen in Figures \ref{fig:five_year_rmse_vs_sypd_figure}g-h), but it is still not perfectly predictive of computational performance. For example, the Encoder-Decoder LSTM has $\sim43\%$ more trainable parameters than the U-Net but takes roughly the same amount of time to train. 

When looking at which architectures strike the best balance between online accuracy and computational efficiency, Figure \ref{fig:five_year_rmse_vs_sypd_figure} shows that the Encoder-Decoder LSTM performs favorably across multiple variables. Indeed, multiple hybrid simulations with Encoder-Decoder LSTMs achieve lower online global mean RMSE than that seen in \citeA{Hu2025-mf} for temperature, liquid cloud, ice cloud, and zonal wind. However, in the case of moisture, hybrid simulations using the U-Net architecture consistently demonstrate lower online RMSE, as evidenced by Figures \ref{fig:five_year_online_sota_results}b and \ref{fig:five_year_rmse_vs_sypd_figure}b. Nevertheless, we believe that the high SYPD and competitive RMSE achieved by the Encoder-Decoder LSTM warrant the use of this architecture as a reasonable baseline in future work.

It is important not to overthink this preliminary assessment of the computational trade-offs of these architecture implementations, which we readily admit were approached pragmatically for research purposes and not performance optimization. There is a rich tradition of inefficient research prototyping followed by impactful performance optimizations in the ML literature, and it is logical to expect that the SYPD hierarchy depicted in this paper would change if the implementations of these architectures were optimized for target hardware. That said, the hierarchy shown could prove to be representative and, at the very least, points to tradeoffs that are readily apparent with existing implementations in the software repository attached to this paper. Benchmark competitions that directly reward model simplicity or efficiency could be an interesting path to discovering more Pareto-optimal architectures \cite{Heuer2025-gu}.

\section{Conclusion}

We implemented, trained, and prognostically tested 90 ML parameterizations of cloud resolving physics of turbulence, radiation, and moist convection, inspired by a subset of novel architectures and design decisions pioneered by the winning teams in the 2024 ClimSim Kaggle competition. 

The results reveal previously unprecedented skill across multiple independent online metrics of prognostic error, confirming our working hypothesis that organized benchmarks can lead to measurable progress for hybrid climate simulation even when formulated as offline training strategies. 

Online stability with tolerable drift, which has been difficult to achieve in the limit of MMF CRM emulation, especially when including full microphysical emulation and land surface feedback, now appears to be reproducible with multiple architecture and design decision combinations, at least in the low-resolution real-geography ClimSim setup we have explored, which we consider an important milestone.

Despite the large diversity in architectures sampled, we have also identified unexpectedly systematic secondary biases that could speak to deeper issues of the MMF emulation problem design. These stubborn symptoms include time mean online temperature bias patterns and systematic online underprediction of precipitable water in the tropics and precipitation extremes in general. Targeting the shared offline and online biases that transcend the choice of seed, architecture, and configuration may be the next frontier for hybrid physics-ML climate simulation in the low-resolution real-geography setting. However, certain aspects of online performance, like variation stemming from inter-seed variability, are still best described as emergent properties, and prioritizing the treatment of the more deterministic offline biases may be the more efficient avenue for continued progress. It is possible that the shared failure modes described in this paper are escaped when deviating from the factors we controlled for to facilitate easier inter-comparisons. However, a more straightforward means of directly addressing systematic biases may be to explicitly include a bias penalty in the loss function \cite{Kochkov2024-uy}. \citeA{Kochkov2024-uy} used a multi-component loss function that penalized spectral bias, but this was done in the context of end-to-end online training to prevent accumulation of errors across timesteps. Finally, we expect including information about subgrid-scale cloud structure and organization in the input to also reduce systematic errors \cite{Colin2019-ks, Hwong2023-hc, Shamekh2023-ed, Beucler2025}.

When comparing architectures, drawing empirically robust conclusions is complicated by the fact that only three seeds were trained for each architecture and configuration pair. Nevertheless, our results in Figures \ref{fig:five_year_online_monthly_rmse_t_and_q} and \ref{fig:five_year_online_sota_results} suggest emerging patterns that we believe warrant further investigation in future work. Rather than presenting these observations as definitive findings, we highlight them here to generate questions for future research, focusing on aspects of the machine learning part of the problem that can be investigated without enhancements to the underlying ClimSim benchmark data:

\begin{obsquestion}{Observation \#1}

\item[\textbf{Observation:}] The U-Net achieves the lowest online moisture error across all architectures, and none of the competing architectures surpass the online moisture error from the best U-Net in \citeA{Hu2025-mf}.

\item[\textbf{Motivated Question:}] Are there multi-scale mechanisms utilized by the U-Net that make it particularly well-suited for reduced online moisture biases? To what extent are these mechanisms portable to other architectures?
\end{obsquestion}

\begin{obsquestion}{Observation \#2}

\item[\textbf{Observation:}] Different architectures respond differently online to different design decisions. For example, results shown in Figure \ref{fig:five_year_online_monthly_rmse_t_and_q} hint that input variable expansion may favor architectures with purely convolutional elements while multirepresentation may favor architectures with purely recurrent NN elements.

\item[\textbf{Motivated Question:}] What makes various architectures respond differently to design decisions online? Is the online response to a given design decision entirely emergent or can it be predicted ahead of time?
\end{obsquestion}

\begin{obsquestion}{Observation \#3}

\item[\textbf{Observation:}] Some architectures seem to have greater inter-seed online variability than others. For example, while some hybrid simulations using the Encoder-Decoder LSTM architecture strike a favorable balance between SYPD and global mean RMSE relative to those from other architectures, this is not universally the case across seeds.

\item[\textbf{Motivated Question:}] Are there ways to reduce inter-seed online variability and improve the reliability of hybrid physics-ML climate simulations? Which architectural components are most sensitive to weight initialization?

\end{obsquestion}

\begin{obsquestion}{Observation \#4}

\item[\textbf{Observation:}] Architectures that used transformer-encoder elements, like the Squeezeformer and Pao Model, experienced numerical instability across almost all seeds in the expanded variable list configuration.

\item[\textbf{Motivated Question:}] Does learning potentially unphysical non-vertically-local patterns that fail to generalize online make transformer-encoder elements ill-suited for ML parameterization (compared to more traditional convolutional or recurrent NN layers)? 

\end{obsquestion}

Democratizing the problem of hybrid physics-ML climate simulation in the context of MMF to the machine learning and data science community on Kaggle has yielded new insights that narrow the still-formidable gap between proof-of-concept and operational capability. Being such a large-scale competition, the Kaggle competition offered arguably the largest search across ML architectures that has been carried out for Earth system parameterizations. It is interesting that the winning models were primarily composed of neural network based architectures, suggesting that today’s modern neural networks offer the best flexibility for modeling the highly complex phenemomena within a grid-cell. This contrasts with other Kaggle competitions (particularly ones centered around tabular datasets) in which lower complexity models such as random forests and gradient boosted trees were able to outcompete higher complexity neural networks, which can be biased to overly smooth solutions \cite{Grinsztajn2022-pi, Januschowski2022-mz}. This affirms the idea that ML parameterization development is a complex and nonlinear enough problem to deserve complex neural network based architectures going forward. Unfortunately, the emergent discrepancy between offline and online skill limits the degree to which crowdsourced solutions tailored to a singular offline benchmark can be relied upon for breakthrough progress online.

Although we achieve new SOTA values across multiple online metrics in this study, no single hybrid simulation conducted here presents a pareto-improvement over online results from \citeA{Hu2025-mf}. Instead, our main contributions are demonstrating that stable and accurate online simulations are reproducibly achievable across diverse architectures going forward and identifying universal failure points and emerging patterns worthy of exploration in future research. With the democratization of the online aspect of this problem via ClimSim-Online and the identification of universal offline and online failure modes, we expect follow-up work to result in rapid, continued progress \cite{Yu2025-cl}. 

Looking ahead, such progress may eventually result in reproducibly stable, high-resolution hybrid simulations with minimal bias or drift within the ClimSim framework. However, actual downstream impact (in the form of a hybrid physics-ML climate model usable for operational climate simulation) will inevitably require moving beyond the confines of the current ClimSim benchmark dataset, and potentially MMF in general. Although ClimSim has already facilitated remarkable progress in hybrid physics-ML climate modeling and will likely continue to do so going forward, the MMF it is based around also possesses significant limitations. Chief among them is the lack of support for aerosol-cloud interaction (ACI) in the GPU-enabled version of E3SM-MMF---a deficiency that will only be addressed with additional funding. Provided this is fixed, there are other limitations that are likely best addressed with a new dataset, benchmarks, containerized climate model, and online competition. In \citeA{Heuer2025-gu}, the fact that radiative and convective tendencies were not separated in ClimSim meant that the authors had to subtract radiative tendencies by approximating them using the ``RTE+RRTMGP" scheme \cite{Pincus2019-ek, Pincus_RTE_RRTMGP_2023}. In \citeA{Beucler2021-qc}, the authors developed analytically constrained NNs that enforced conservation laws to within machine precision. However, this is not currently possible with the current variable list, which is missing top of atmosphere (TOA) radiative energy fluxes. TOA irradiance would also allow for incorporating the influence of cloud radiative effects, which have been shown to be important for accurately representing extreme rainfall \cite{Medeiros2021-sg}. Prior work has also shown the utility of multiclimate data for training, validation, testing, and developing or learning climate-invariant feature transformations \cite{Clark2022-sr, Bhouri2023-pe, Lin2025-ya, Liu2025-yf, Han2025-js}. Unfortunately, ClimSim only contains data from a single climate. Additionally, detecting differences in emergent online behavior with sufficient statistical power can require large ($\mathrm{O(100)}$) ensembles, motivating the inclusion of even simpler climate models that can facilitate online testing with large sample sizes without inordinate computational expense \cite{Mansfield2023-am, Lin2025-ya}.

Nevertheless, the long-term ambition of hybrid physics-ML climate modeling is to potentially \textit{exceed}, and not just match, the accuracy of the expensive physics-based simulations on which such models are trained. Encouragingly, data-driven emulators for weather and PDE systems have in some cases even demonstrated the ability to outperform their own training data \cite{Koehler2025-js, Christensen2026-sc}, and hybrid physics-ML climate models designed from the ground up to be end-to-end differentiable (e.g., \citeA{Kochkov2024-uy, Davenport2026-ro}) enable direct training on observational data in ways that are extremely impractical or impossible for conventional Fortran-based physics models \cite{Yuval2026-pk}. Moreover, there is some emerging evidence suggesting that NN parameterizations may be capable of generalizing across climates and even across climate models \cite{Han2025-js, Heuer2025-gu}. In \citeA{Han2025-js}, the authors showed for the first time that a NN parameterization trained exclusively on simulation data using present-day SSTs could produce a stable, decadal hybrid GCM simulation under a +4K SST warm climate with real geography, accurately reproducing climate responses across thermodynamic states, circulations, and extreme precipitation. In \citeA{Heuer2025-gu}, the authors successfully coupled a ML convective parameterization---trained on high-resolution ClimSim data with a Kaggle-inspired architecture---to ICON, achieving stable 20-year AMIP simulations. Such cross-model transferability is particularly promising for future work, as mastering hybrid simulation with MMF may be a necessary stepping stone toward achieving success in the harder problem of hybrid simulations using coarse-grained GCRMs. Unlike MMF, GCRMs eliminate not only the convenient scale-separation used for ML parameterizations but also the unnatural artifacts caused by the inability of the large-scale GCM flow to advect small-scale CRM fluctuations \cite{Pritchard2011-fa, Brenowitz2019-by, Brenowitz2020-bv, Hannah2020-ez, Jansson2022-vv, Hannah2022-im, Heuer2024-vt, Schneider2024-jx, Heuer2025-gu}. The progress reported here---spanning crowdsourced architecture discovery, reproducibly stable online integration, and systematic failure-mode analysis---is yet another important step towards this grander ambition.

\section*{Open Research Section}
The GitHub repository associated with this work is available via Apache License 2.0 and developed openly at \url{https://github.com/leap-stc/climsim-kaggle-edition}. The version of E3SM-MMF made compatible with FTorch is available at \newline \mbox{\url{https://github.com/leap-stc/E3SM_nvlab}}. All experiments were run on NVIDIA A100 GPUs. Approximately 4,477 GPU-hours were used for training, 3,833 GPU-hours were used for online simulation, and less than 10 GPU-hours were used for offline inference. In total, we estimate 8,320 GPU-hours were used to conduct this study. All models, checkpoints, and normalization files are uploaded to Hugging Face and are available at \newline \mbox{\url{https://hf.co/collections/jlin404/climsim-kaggle-models}}. Software repositories, online simulation output, and various online simulation metadata are preserved at \mbox{\url{https://zenodo.org/records/18883121}}.

\section*{Conflict of Interest Disclosure Statement}
The authors declare that they have no conflicts of interest related to this manuscript. All authors have fully disclosed any potential conflicts of interest, and all contributing authors have reviewed and approve of this conflict of interest disclosure statement.

\acknowledgments
This work is primarily funded by National Science Foundation (NSF) Science and Technology Center (STC) Learning the Earth with Artificial Intelligence and Physics (LEAP), Award \# 2019625-STC. High-performance computing was conducted on NERSC Perlmutter. T. Beucler acknowledges funding from the Swiss State Secretariat for Education, Research and Innovation (SERI) for the Horizon Europe project AI4PEX (Grant agreement ID: 101137682 and SERI no 23.00546). H. Christensen acknowledges funding through a Leverhulme Trust Research Leadership Award, and from the UK Research and Innovation (UKRI) under the UK government’s Horizon Europe funding guarantee (grant number 10049639) to support the Horizon Europe EERIE project (Grant Agreement No 101081383) funded by the European Union. W. Hannah acknowledges support for their contribution to this work from the U.S. Department of Energy by Lawrence Livermore National Laboratory under Contract DE-AC52-07NA27344. H. Heuer received funding for this study from the European Research Council (ERC) Synergy Grant ``Understanding and Modelling the Earth System with Machine Learning (USMILE)'' under the Horizon 2020 research and innovation programme (Grant agreement No. 855187). Views and opinions expressed are however those of the author(s) only and do not necessarily reflect those of the European Union or the European Climate Infrastructure and Environment Executive Agency (CINEA). Neither the European Union nor the granting authority can be held responsible for them. We are thankful to the Kaggle staff who made this competition possible, in particular Ashley Chow, Walter Reade, and Maggie Demkin, and to the anonymous reviewers at JAMES for for their constructive comments and careful attention to detail. We also extend our gratitude to the hundreds of Kaggle participants who contributed their time and brilliant ideas to pushing science forward.

%
\bibliography{references}

%


%
%
%
%
%
\clearpage

\title{Supporting Information for ``Crowdsourcing the Frontier: Advancing Hybrid Physics-ML Climate Simulation via a \$50,000 Kaggle Competition"}

\authors{Jerry Lin$^{1,2}$, Zeyuan Hu$^{3}$, Tom Beucler$^{4,5}$, Katherine Frields$^{1}$, Hannah Christensen$^{6}$, Walter Hannah$^{7}$, Helge Heuer$^{8}$, Peter Ukkonen$^{6}$, Laura A. Mansfield$^{6}$, Tian Zheng$^{9}$, Liran Peng$^{1}$, Ritwik Gupta$^{10,11}$, Pierre Gentine$^{12}$, Yusef Al-Naher, Mingjiang Duan$^{13}$, Kyo Hattori$^{14}$, Weiliang Ji$^{13}$, Chunhan Li$^{13}$, Kippei Matsuda$^{15}$, Naoki Murakami$^{16}$, Shlomo Ron, Marec Serlin$^{17}$, Hongjian Song$^{13}$, Yuma Tanabe, Daisuke Yamamoto, Jianyao Zhou, Mike Pritchard$^{1,3}$}

\affiliation{1}{Department of Earth System Sciences, University of California at Irvine, Irvine, CA, USA}
\affiliation{2}{Department of Computing \& Data Sciences, Boston University, Boston, MA, USA}
\affiliation{3}{NVIDIA Research}
\affiliation{4}{Faculty of Geosciences and Environment, University of Lausanne, Lausanne, VD, Switzerland}
\affiliation{5}{Expertise Center for Climate Extremes, University of Lausanne, Lausanne, VD, Switzerland}
\affiliation{6}{Department of Physics, University of Oxford, Oxford, United Kingdom}
\affiliation{7}{Lawrence Livermore National Laboratory}
\affiliation{8}{Deutsches Zentrum f{\"u}r Luft- und Raumfahrt, Institut f{\"u}r Physik der Atmosph{\"a}re, Oberpfaffenhofen, Germany}
\affiliation{9}{Department of Statistics, Columbia University, New York, NY, USA}
\affiliation{10}{University of California, Berkeley, Berkeley, CA, USA}
\affiliation{11}{University of Maryland, College Park, MD, USA}
\affiliation{12}{LEAP Science and Technology Center, School of Engineering and Applied Sciences, Climate School, Columbia University}
\affiliation{13}{Z Lab, China}
\affiliation{14}{ABEJA Inc., Japan}
\affiliation{15}{Kawasaki Heavy Industries, Ltd., Japan}
\affiliation{16}{DeNA Co., Ltd}
\affiliation{17}{Uber Technologies, Inc.}

\renewcommand{\thefigure}{S\arabic{figure}}
\renewcommand{\thetable}{S\arabic{table}}
\setcounter{figure}{0}
\setcounter{table}{0}

\noindent\textbf{Contents of this file}

\begin{enumerate}

\item Text S1 to S3

\item Figures S1 to S47

\end{enumerate}

\noindent\textbf{Introduction}

Text S1 explains how offline $R^2$ was calculated. Text S2 explains how online monthly RMSE, area-weighted means, and global RMSE were calculated. Text S3 explains the method used to zonally average on E3SM-MMF's unstructured, cubed-sphere grid.

Figures S1 through S3 show additional offline $R^2$ values across architectures in the confidence loss, difference loss, and multirepresentation configurations. Figure S4 shows how offline $R^2$ from the expanded variables list configuration compares to offline $R^2$ from \citeA{Hu2025-mf}. Figures S5 through S8 show online monthly RMSE for zonal and meridional wind, liquid and ice cloud, heating and moistening tendencies, and liquid and ice cloud tendencies. Figures S9 through S12 show online zonal mean biases across architectures with a fixed seed in the confidence loss configuration for zonal wind, meridional wind, liquid cloud, ice cloud, heating tendency, moistening tendency, liquid cloud tendency, and ice cloud tendency. Figure S13 shows the online monthly area-weighted standard deviation for precipitable water across all architectures and configurations. Figures S14 through S16 show the latitudinal distribution of mean precipitation and histogram of precipitation values for confidence loss, difference loss, and multirepresentation configurations. Figures S17 through S45 show offline zonal mean biases across variables and configurations as well as the corresponding vertical profiles of these biases binned by precipitation percentile.

\noindent\textbf{Text S1.}

To calculate offline $R^2$ we simply average across all samples in the test set separately for each variable and level. Concretely, it is calculated via the formula below:

\begin{equation}
R^2 = 1 - \frac{\sum_{i=1}^{n} (y_i - \hat{y}_i)^2}{\sum_{i=1}^{n} (y_i - \bar{y})^2}
\end{equation}

where $n$ is the number of samples in the test set and $y$, without loss of generality, refers to a given variable at a given level.

\noindent\textbf{Text S2.}

Online monthly RMSE is a mass weighted RMSE using both pressure and area. For any given month $m$ it can be formalized as:

\begin{equation}
\text{RMSE}_{month}(m) = \sqrt{\frac{\sum_{i=1}^{384} \sum_{j=1}^{60} a_i * p_j (\hat{y}_{i,j,m} - y_{i,j,m})^2}{\sum_{i=1}^{384} \sum_{j=1}^{60} a_i * p_j}}
\end{equation}

where:

\begin{itemize}
\item $i$ is the index for a given grid cell 
\item $j$ refers to a given vertical level
\item $a_i$ refers to the area weight for grid cell $i$
\item $p_j$ refers to the pressure weight for vertical level $j$
\item $y_{i,j,m}$ refers to the monthly average of variable $y$ at grid cell $i$ and vertical level $j$
\end{itemize}

Global online RMSE is only calculated for online runs that completed 5 simulation years without memory or instability issues, and it is defined as:

\begin{equation}
\text{RMSE}_{complete} = \sqrt{\frac{\sum_{i=1}^{384} \sum_{j=1}^{60} a_i * p_j (\sum_{m=1}^{60}\hat{y}_{i,j,m}/60 - \sum_{m=1}^{60}y_{i,j,m}/60)^2}{\sum_{i=1}^{384} \sum_{j=1}^{60} a_i * p_j}}
\end{equation}

Area weighting for non-vertically-resolved variables like precipitation or precipitable water is conducted in a similar manner, except without pressure levels.

\noindent\textbf{Text S3.}

To zonally-average on E3SM-MMF's unstructured cubed-sphere grid, we separate latitudes into 18 ten degree latitudinal bins, assign grid cells to each latitudinal bin, and normalize values based on the sum of the areas of the grid cells in each latitudinal bin.

Mathematically, it can be formalized as:

\begin{equation}
\overline{F}_j = \frac{\sum_{i \in B_j} A_i \, F_i}{\sum_{i \in B_j} A_i}
\end{equation}

where:

\begin{itemize}
    \item $j$ is the $j$-th latitudinal bin.
    \item $i$ is the $i$-th grid cell assigned to the $j$-th latitudinal bin.
    \item $B_j$ is the set of all grid cell indices that lie within the $j$-th latitudinal bin.
    \item $A_i$ is the area of the $i$-th grid cell.
    \item $F_i$ is the value of the field (e.g., temperature) in the $i$-th grid cell.
\end{itemize}

\begin{figure}[!htbp]
 \centering
 \includegraphics[width=\textwidth]{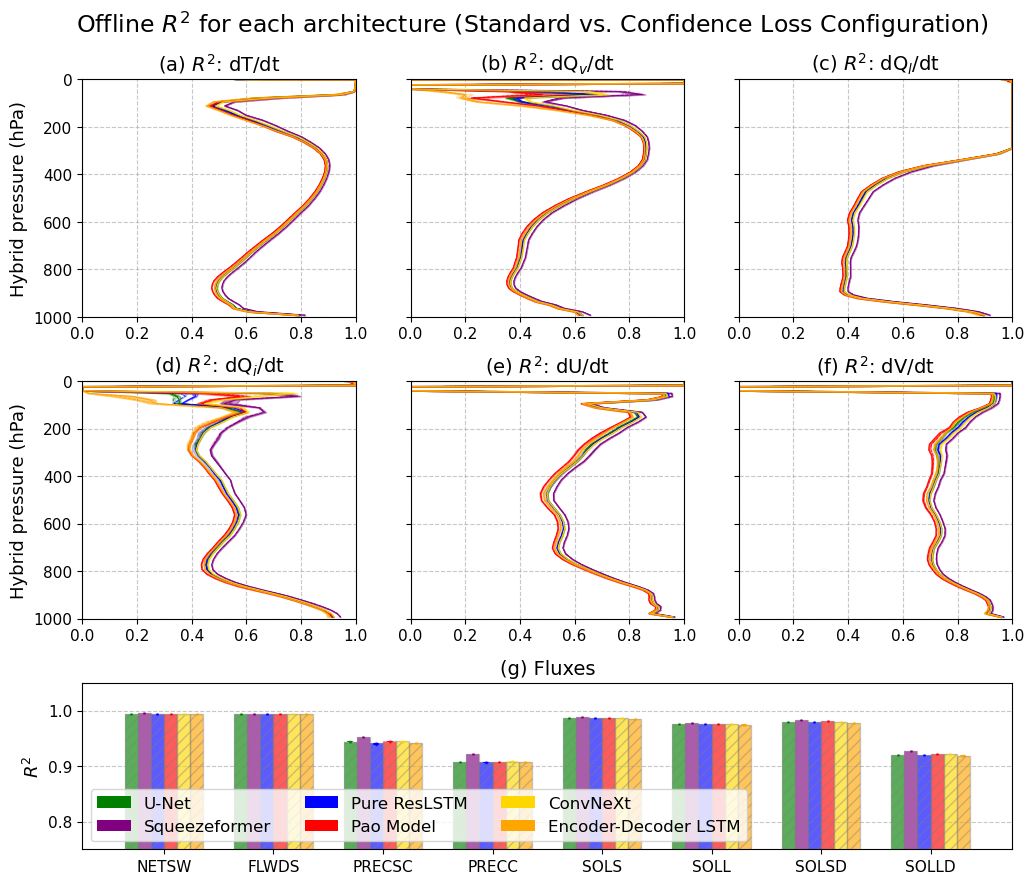}

\caption{This figure shows offline $R^2$ values for each variable across architectures for the standard configuration (depicted using dashed lines and hatched bar charts) and the confidence loss configuration. For vertically-resolved variables, the colored lines depict the median $R^2$ while the shading shows the min and max across seeds for each architecture. For scalar variables, the bars show the medians while the vertical lines at the top of each bar show the min-max range.}
 \label{fig:offline_r2_lines_standard_vs_conf_loss}
\end{figure}

\begin{figure}[!htbp]
 \centering
 \includegraphics[width=\textwidth]{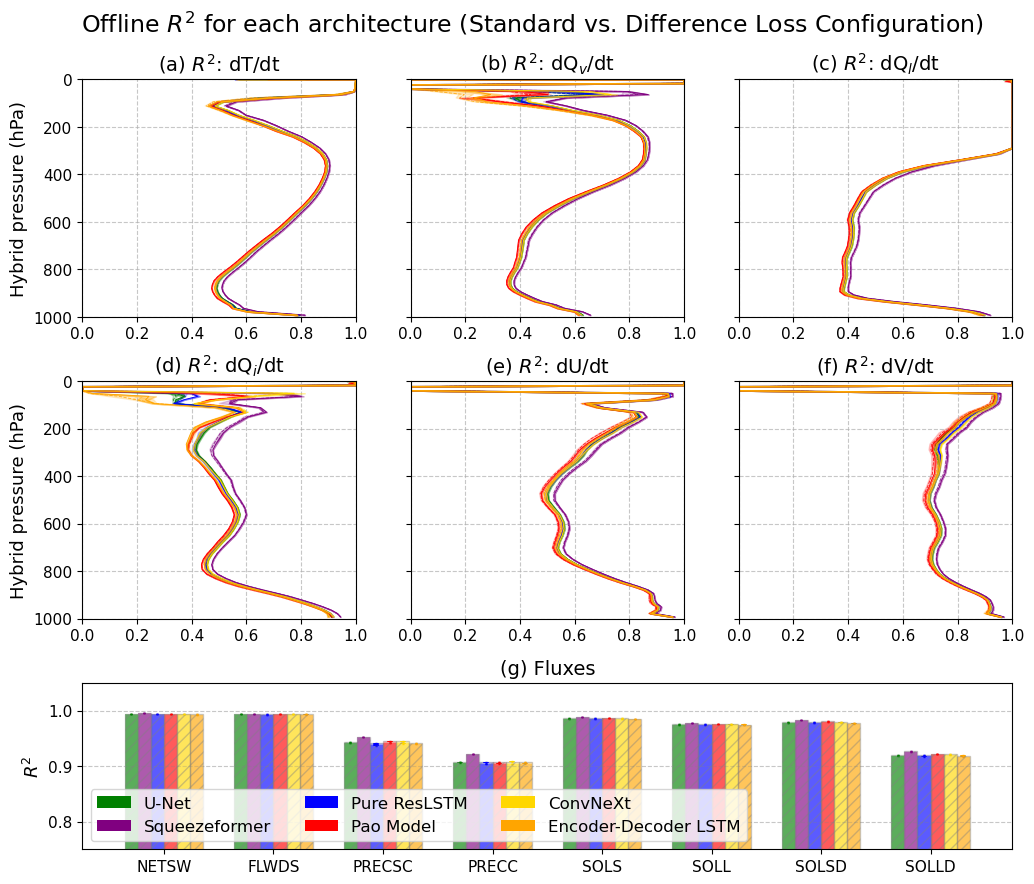}

\caption{This figure shows offline $R^2$ values for each variable across architectures for the standard configuration (depicted using dashed lines and hatched bar charts) and the difference loss configuration. For vertically-resolved variables, the colored lines depict the median $R^2$ while the shading shows the min and max across seeds for each architecture. For scalar variables, the bars show the medians while the vertical lines at the top of each bar show the min-max range.}
 \label{fig:offline_r2_lines_standard_vs_diff_loss}
\end{figure}

\begin{figure}[!htbp]
 \centering
 \includegraphics[width=\textwidth]{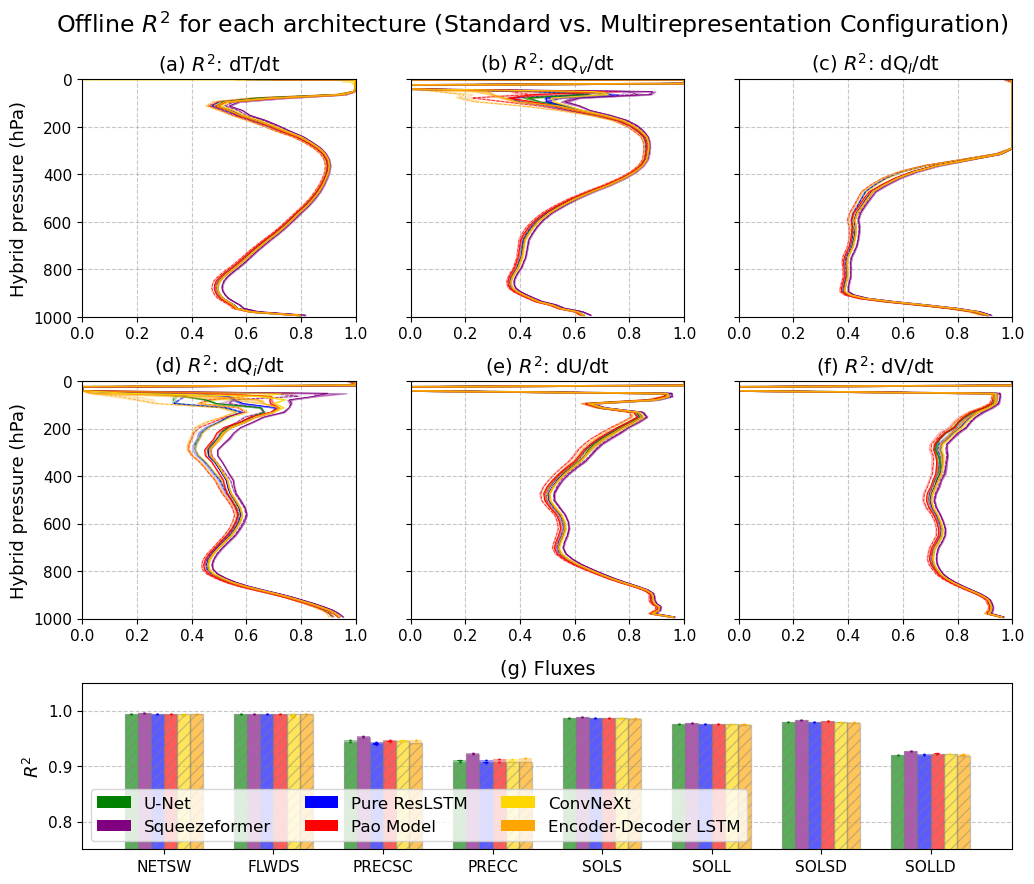}

\caption{This figure shows offline $R^2$ values for each variable across architectures for the standard configuration (depicted using dashed lines and hatched bar charts) and the multirepresentation configuration. For vertically-resolved variables, the colored lines depict the median $R^2$ while the shading shows the min and max across seeds for each architecture. For scalar variables, the bars show the medians while the vertical lines at the top of each bar show the min-max range.}
 \label{fig:offline_r2_lines_standard_vs_multirep}
\end{figure}

\begin{figure}[!htbp]
 \centering
 \includegraphics[width=\textwidth]{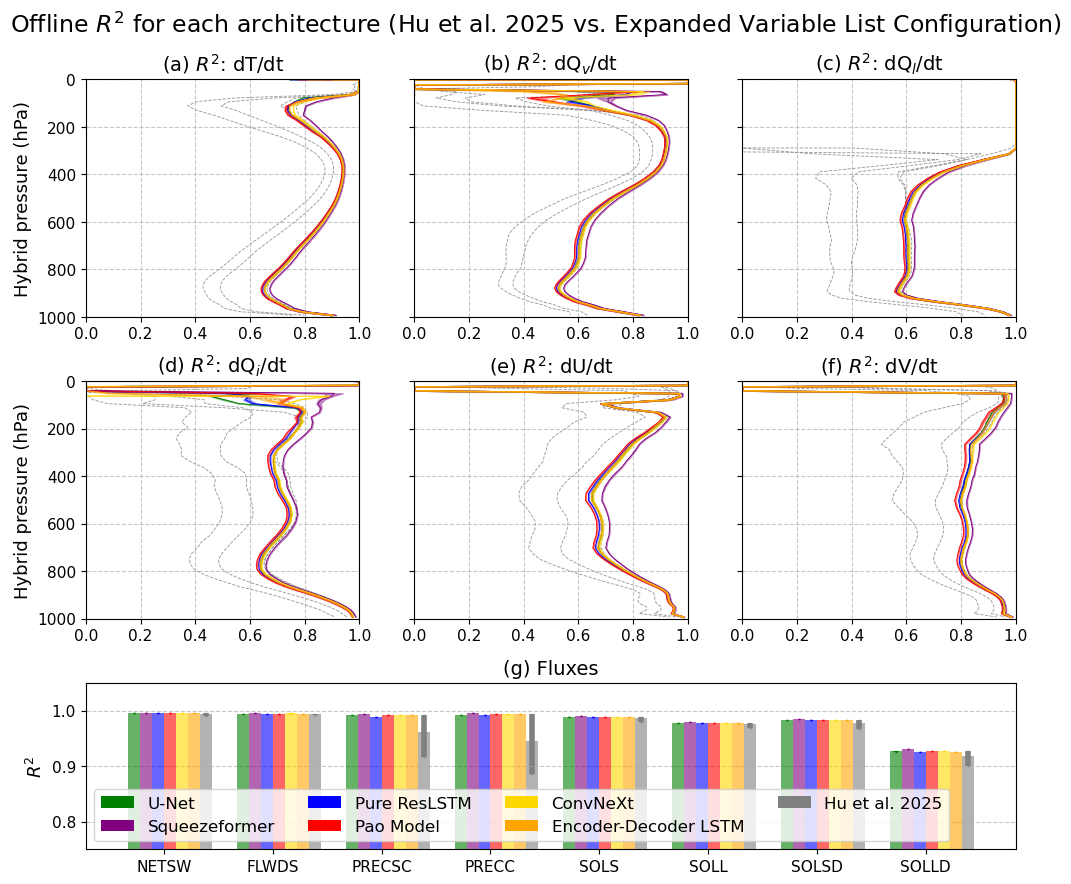}

\caption{This figure shows offline $R^2$ values for each variable across architectures in the expanded variables list configuration as well as those from models in \citeA{Hu2025-mf}. For vertically-resolved variables, the lines depict the median $R^2$ while the shading shows the min and max across seeds for each architecture. For scalar variables, the bars show the medians while the vertical lines at the top of each bar show the min-max range.}
 \label{fig:offline_r2_lines_huetal2025_vs_v6}
\end{figure}

\begin{figure}[!htbp]
 \centering
 \includegraphics[width=\textwidth]{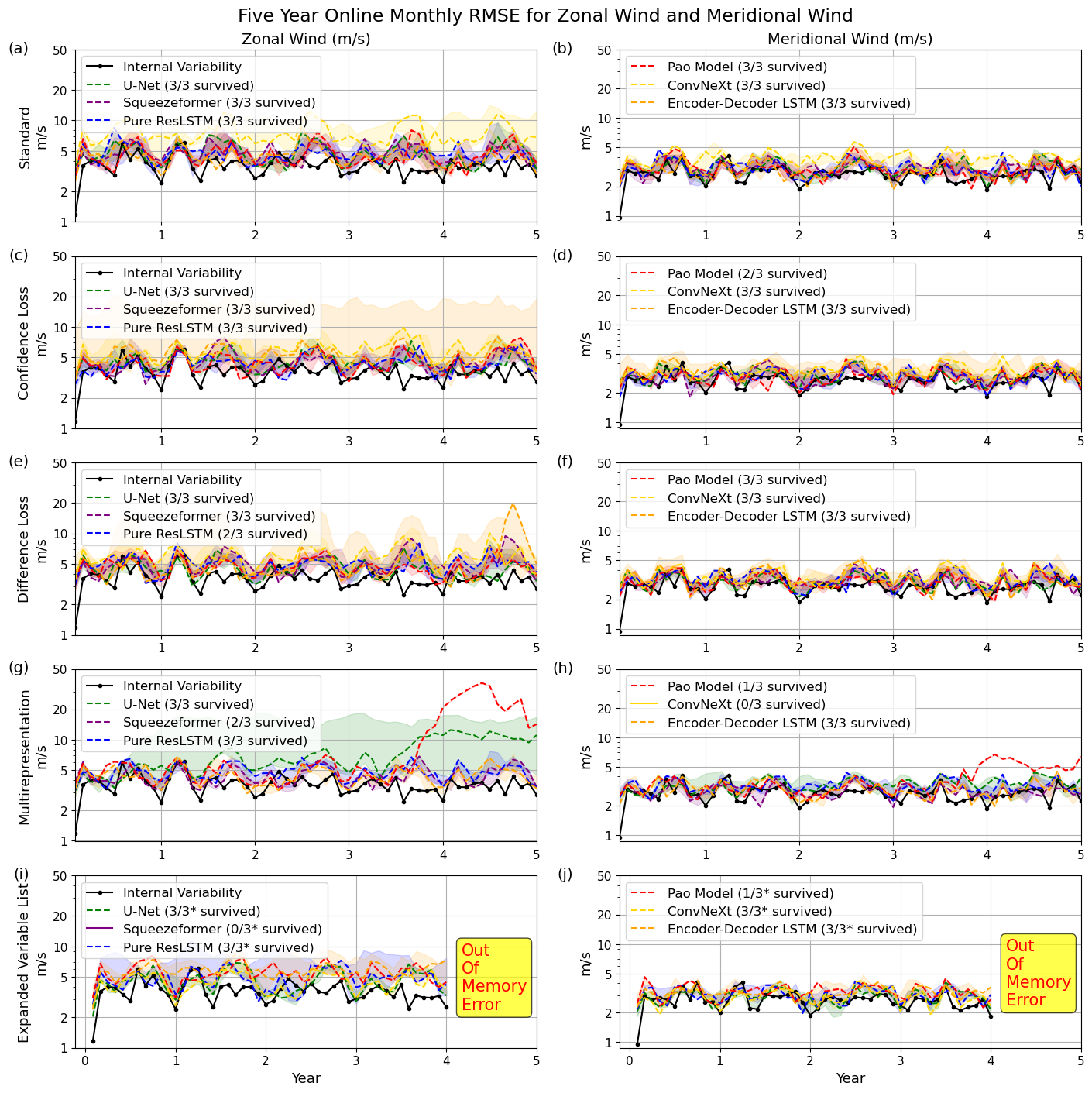}

\caption{This figure shows online monthly RMSE for zonal and meridional wind for all architectures in each configuration. Shading indicates the inter-seed range (min to max RMSE across three seeds per month). Dashed lines show RMSE from the seed whose monthly mean absolute deviation from MMF RMSE is closest to the median absolute deviation across seeds. Subplots i and j only show data up to four years because of out-of-memory issues that caused many hybrid simulations to terminate in the fifth year. Asterisk (*) indicates that survival is assessed via integrating for four, and not five, simulation years. For visual clarity, RMSE for hybrid simulations that crash due to numerical instability are not shown.}
 \label{fig:five_year_online_monthly_rmse_u_and_v}
\end{figure}

\begin{figure}[!htbp]
 \centering
 \includegraphics[width=\textwidth]{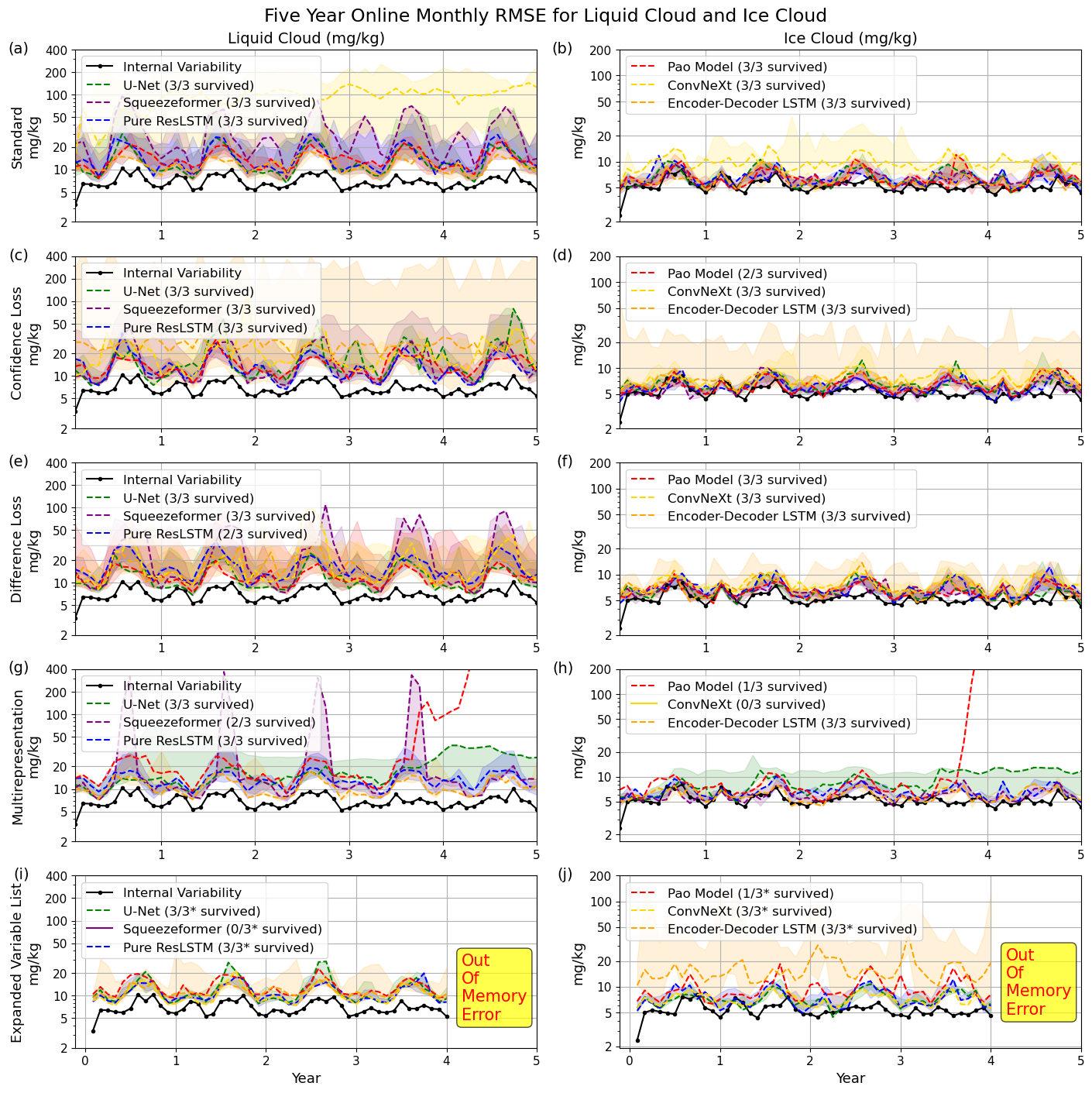}

\caption{This figure shows online monthly RMSE for liquid and ice cloud for all architectures in each configuration. Shading indicates the inter-seed range (min to max RMSE across three seeds per month). Dashed lines show RMSE from the seed whose monthly mean absolute deviation from MMF RMSE is closest to the median absolute deviation across seeds. Subplots i and j only show data up to four years because of out-of-memory issues that caused many hybrid simulations to terminate in the fifth year. Asterisk (*) indicates that survival is assessed via integrating for four, and not five, simulation years. For visual clarity, RMSE for hybrid simulations that crash due to numerical instability are not shown.}
 \label{fig:five_year_online_monthly_rmse_cldliq_and_cldice}
\end{figure}

\begin{figure}[!htbp]
 \centering
 \includegraphics[width=\textwidth]{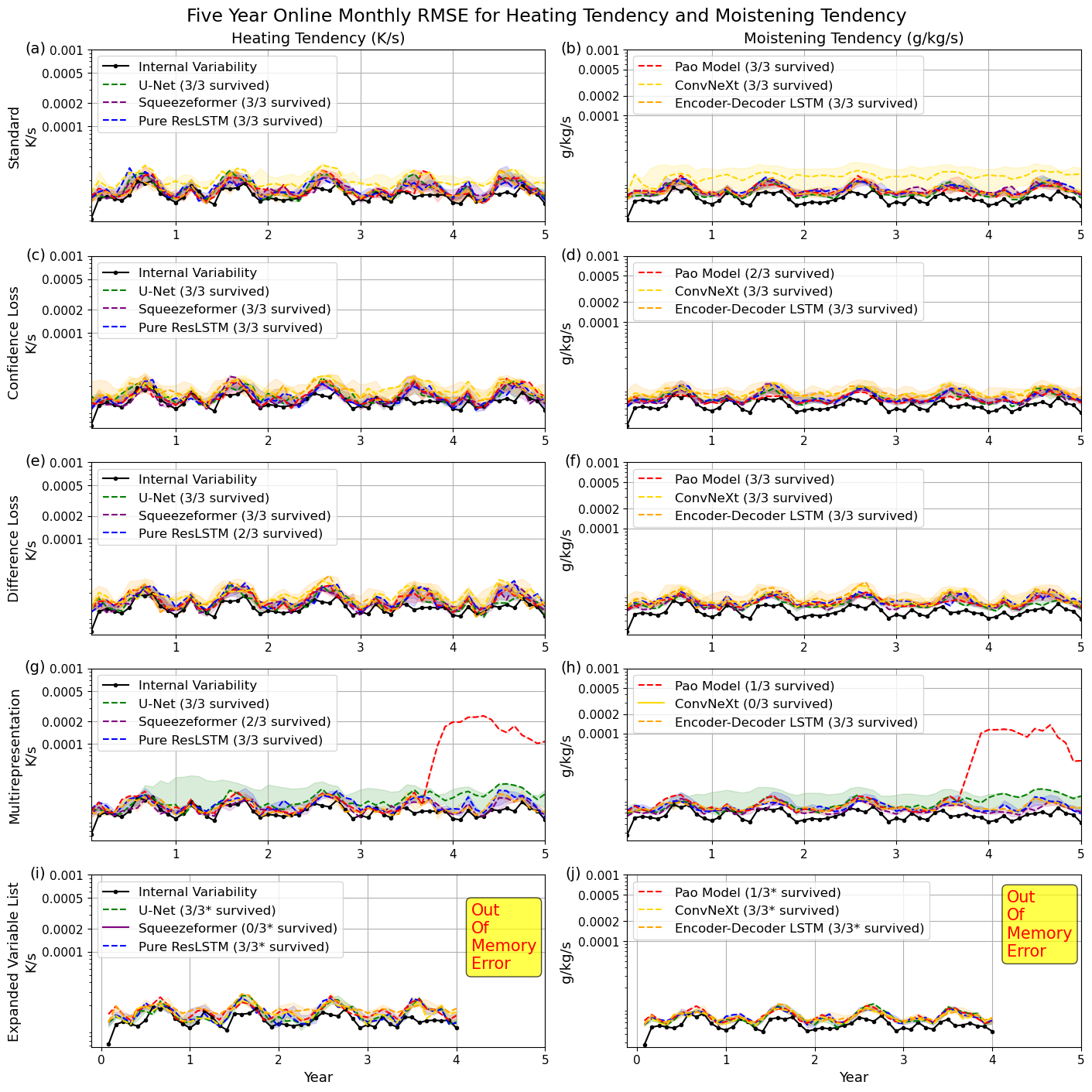}

\caption{This figure shows online monthly RMSE for heating and moistening tendencies for all architectures in each configuration. Shading indicates the inter-seed range (min to max RMSE across three seeds per month). Dashed lines show RMSE from the seed whose monthly mean absolute deviation from MMF RMSE is closest to the median absolute deviation across seeds. Subplots i and j only show data up to four years because of out-of-memory issues that caused many hybrid simulations to terminate in the fifth year. Asterisk (*) indicates that survival is assessed via integrating for four, and not five, simulation years. For visual clarity, RMSE for hybrid simulations that crash due to numerical instability are not shown.}
 \label{fig:five_year_online_monthly_rmse_dtphys_and_dq1phys}
\end{figure}

\begin{figure}[!htbp]
 \centering
 \includegraphics[width=\textwidth]{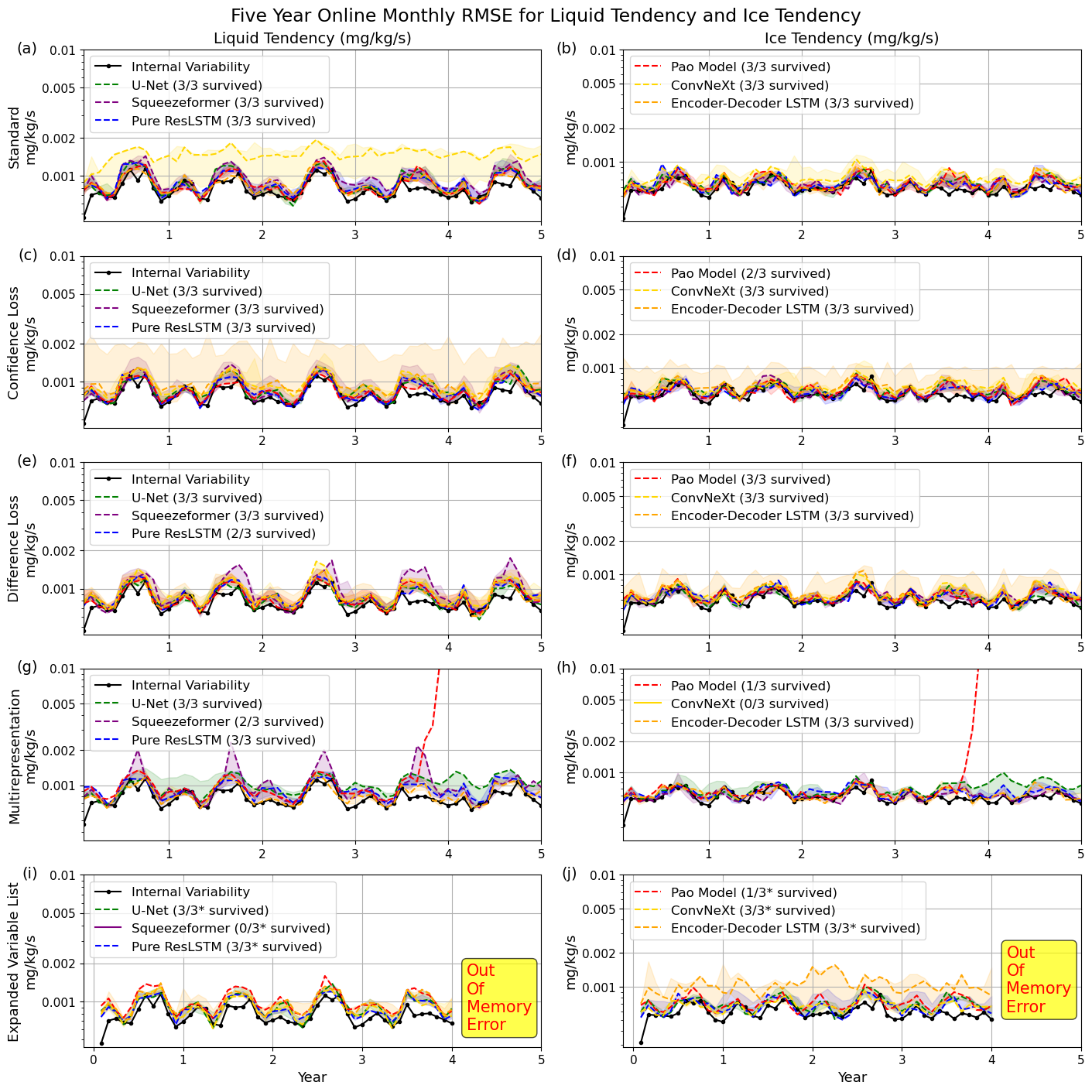}

\caption{This figure shows online monthly RMSE for liquid and ice cloud tendencies for all architectures in each configuration. Shading indicates the inter-seed range (min to max RMSE across three seeds per month). Dashed lines show RMSE from the seed whose monthly mean absolute deviation from MMF RMSE is closest to the median absolute deviation across seeds. Subplots i and j only show data up to four years because of out-of-memory issues that caused many hybrid simulations to terminate in the fifth year. Asterisk (*) indicates that survival is assessed via integrating for four, and not five, simulation years. For visual clarity, RMSE for hybrid simulations that crash due to numerical instability are not shown.}
 \label{fig:five_year_online_monthly_rmse_dq2phys_and_dq3phys}
\end{figure}

\begin{figure}[!htbp]
 \centering
 \includegraphics[width=\textwidth]{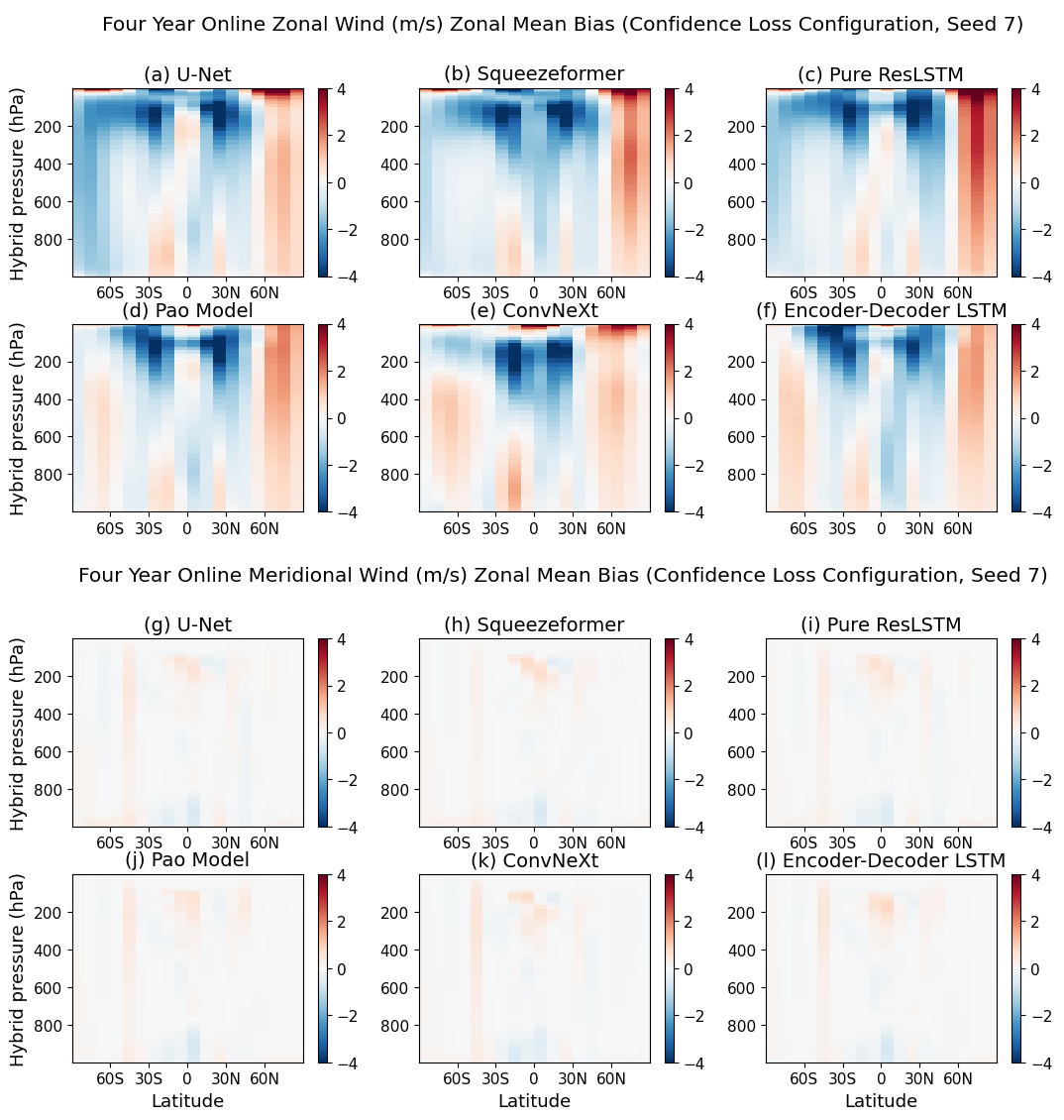}

\caption{This figure shows online zonal mean biases for zonal and meridional wind across architectures in the confidence loss configuration, trained using a common seed.}
 \label{fig:four_year_online_u_and_v_bias_conf_loss}
\end{figure}

\begin{figure}[!htbp]
 \centering
 \includegraphics[width=\textwidth]{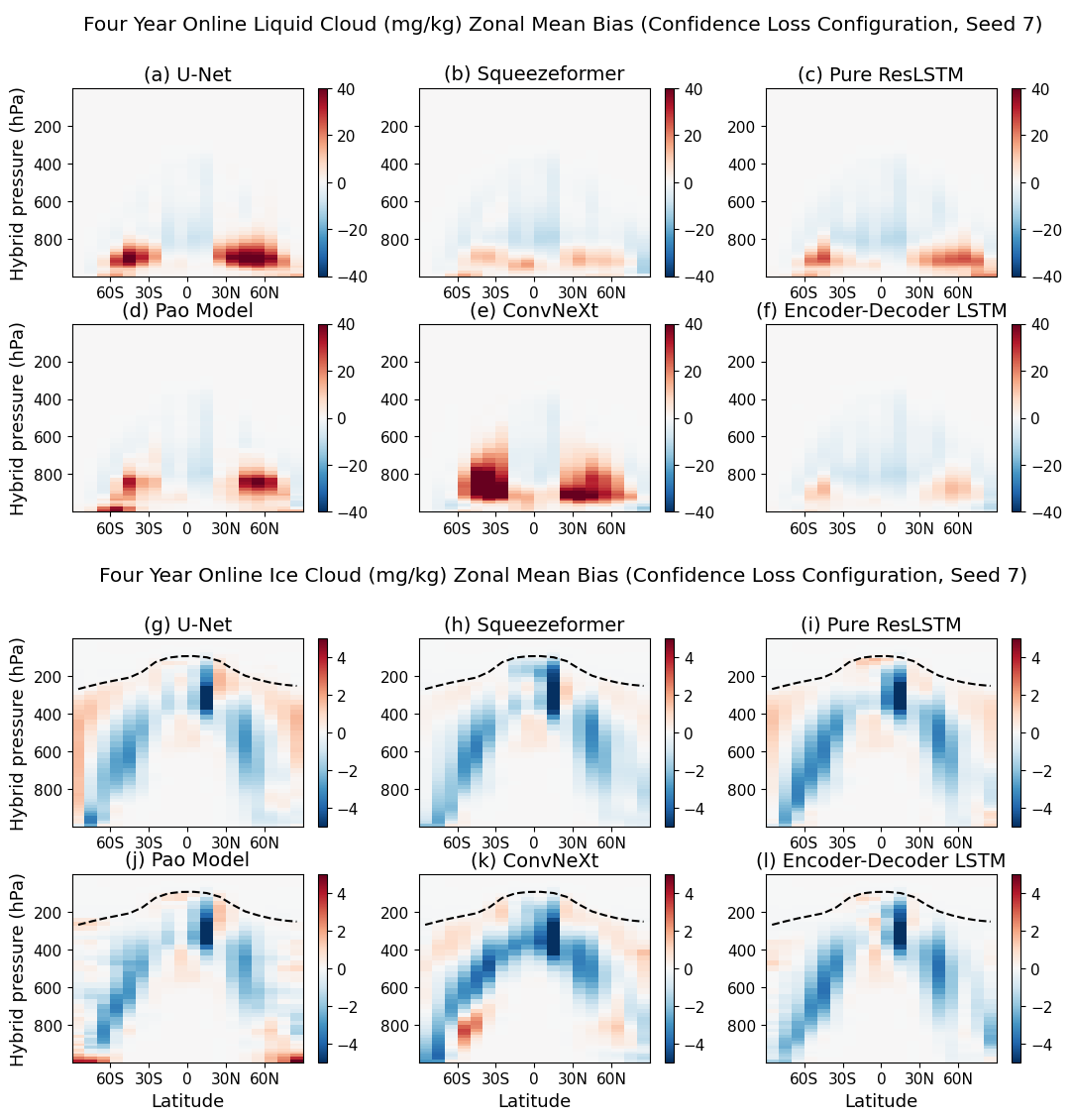}

\caption{This figure shows online zonal mean biases for liquid and ice cloud across architectures in the confidence loss configuration, trained using a common seed.}
 \label{fig:four_year_online_cldliq_and_cldice_bias_conf_loss}
\end{figure}

\begin{figure}[!htbp]
 \centering
 \includegraphics[width=\textwidth]{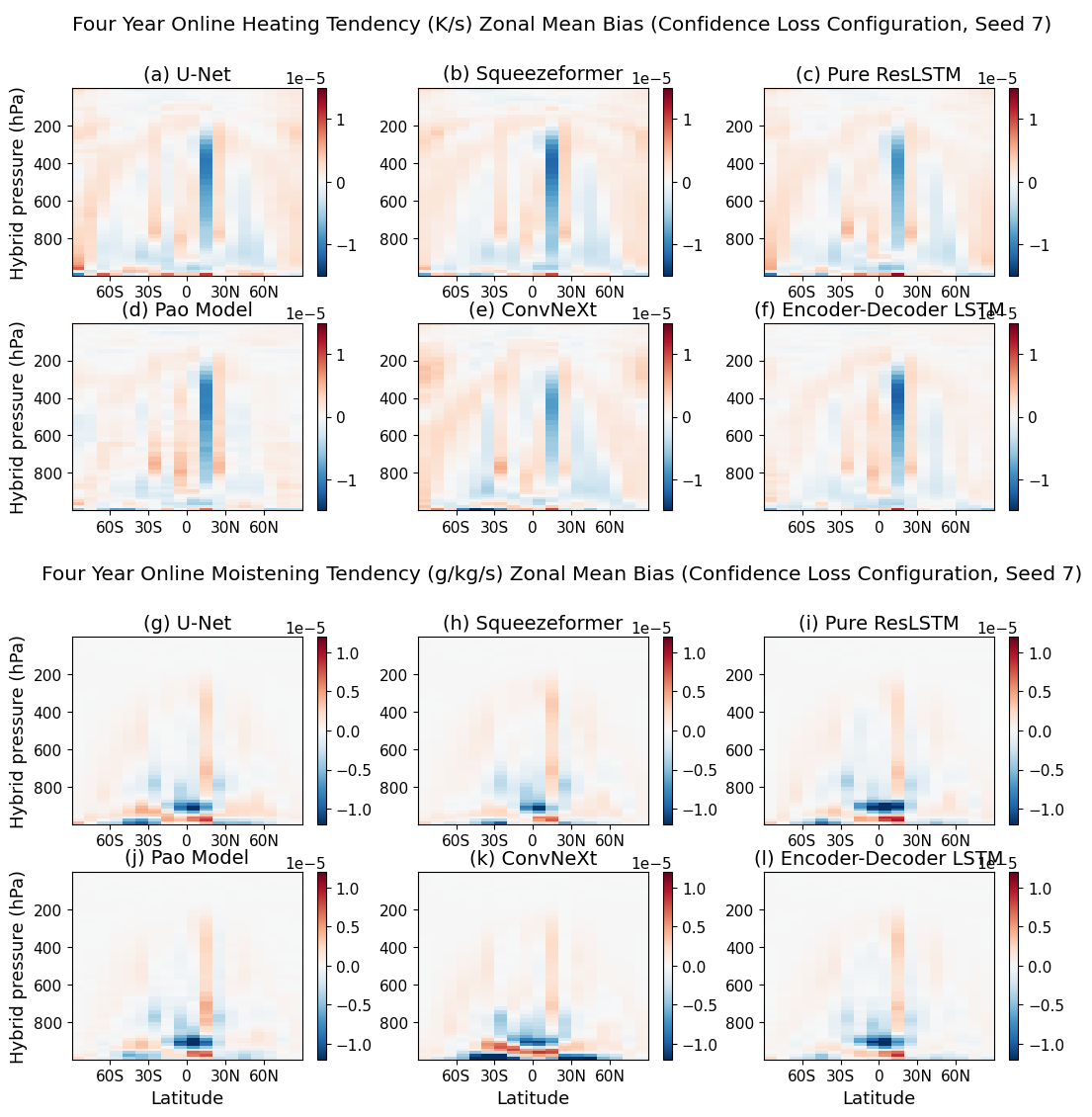}

\caption{This figure shows online zonal mean biases for heating and moistening tendencies across architectures in the confidence loss configuration, trained using a common seed.}
 \label{fig:four_year_online_dtphys_and_dq1phys_bias_conf_loss}
\end{figure}

\begin{figure}[!htbp]
 \centering
 \includegraphics[width=\textwidth]{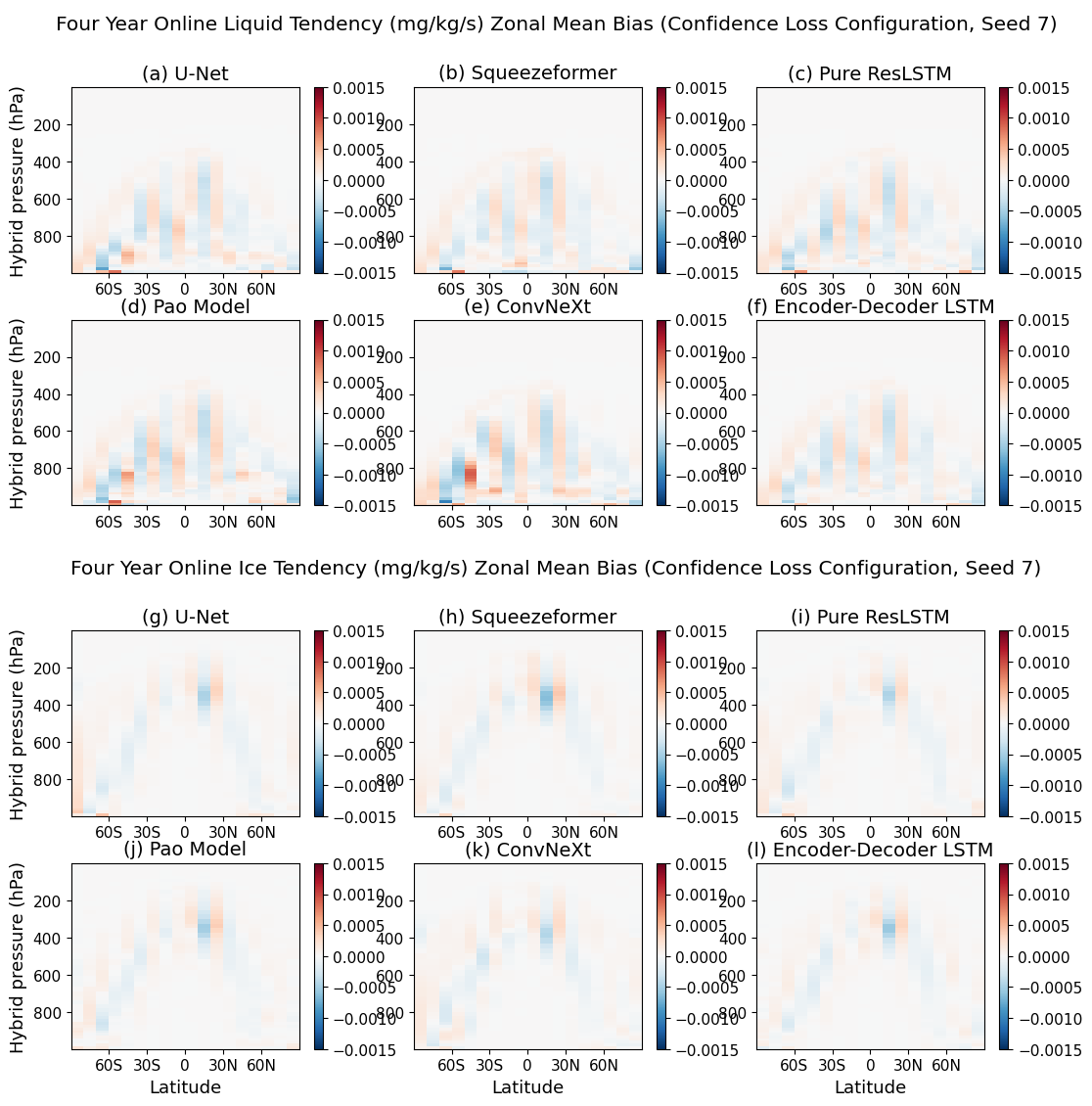}

\caption{This figure shows online zonal mean biases for liquid and ice cloud tendencies across architectures in the confidence loss configuration, trained using a common seed.}
 \label{fig:four_year_online_dq2phys_and_dq3phys_bias_conf_loss}
\end{figure}

\begin{figure}[!htbp]
 \centering
 \includegraphics[width=\textwidth]{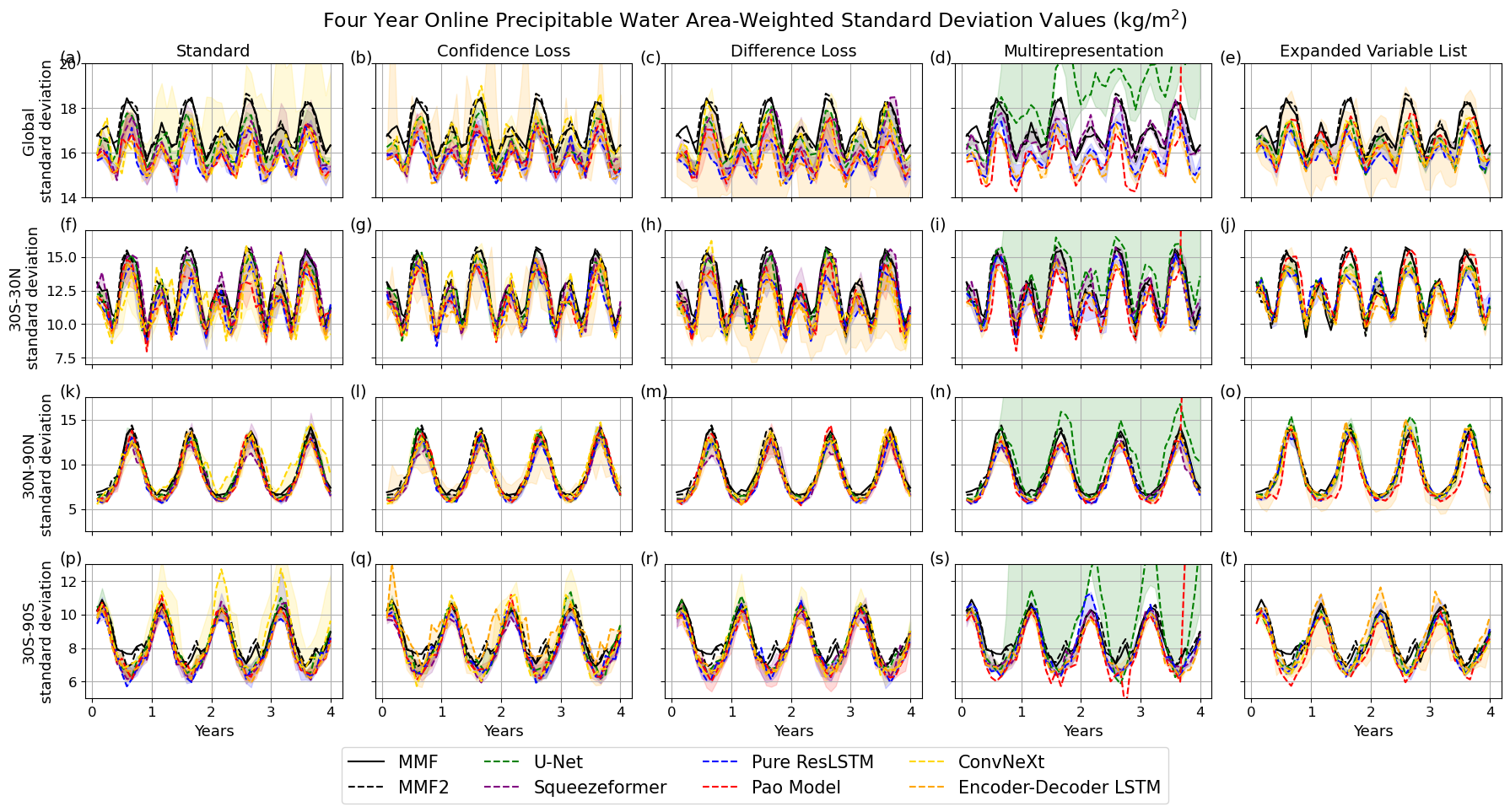}

\caption{This figure shows four year area-weighted standard deviation values for precipitable water in four regions (global, 30S-30N, 30N-90N, and 30S-90S) for all architectures across all configurations.}
 \label{fig:four_year_online_precipitable_water_std}
\end{figure}

\begin{figure}[!htbp]
 \centering
 \includegraphics[width=\textwidth]{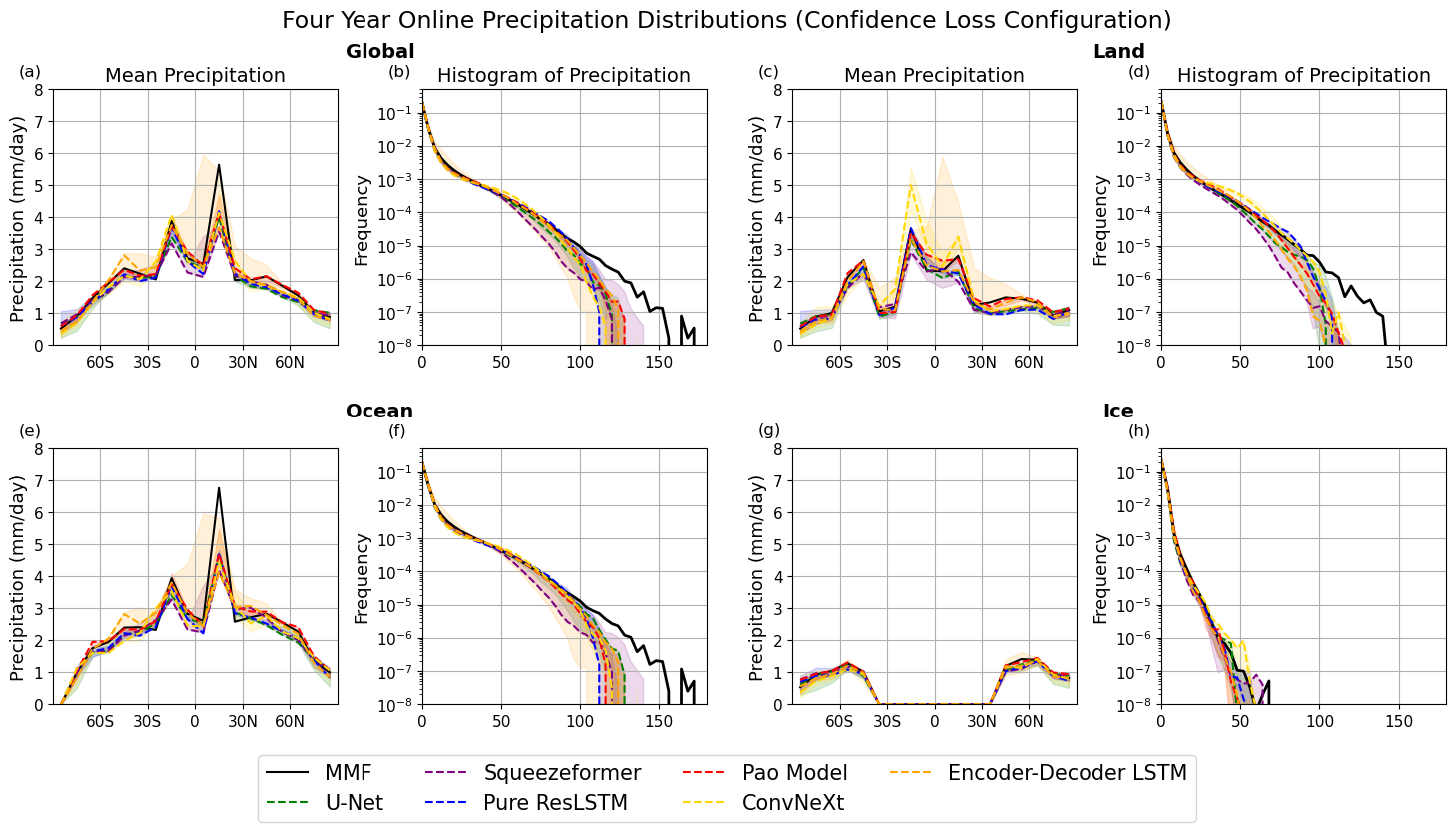}

\caption{This figure shows the latitudinal distribution of mean precipitation and histogram of precipitation values compared to those from MMF for all architectures in the confidence loss configuration.}
 \label{fig:four_year_online_precc_dists_conf_loss}
\end{figure}

\begin{figure}[!htbp]
 \centering
 \includegraphics[width=\textwidth]{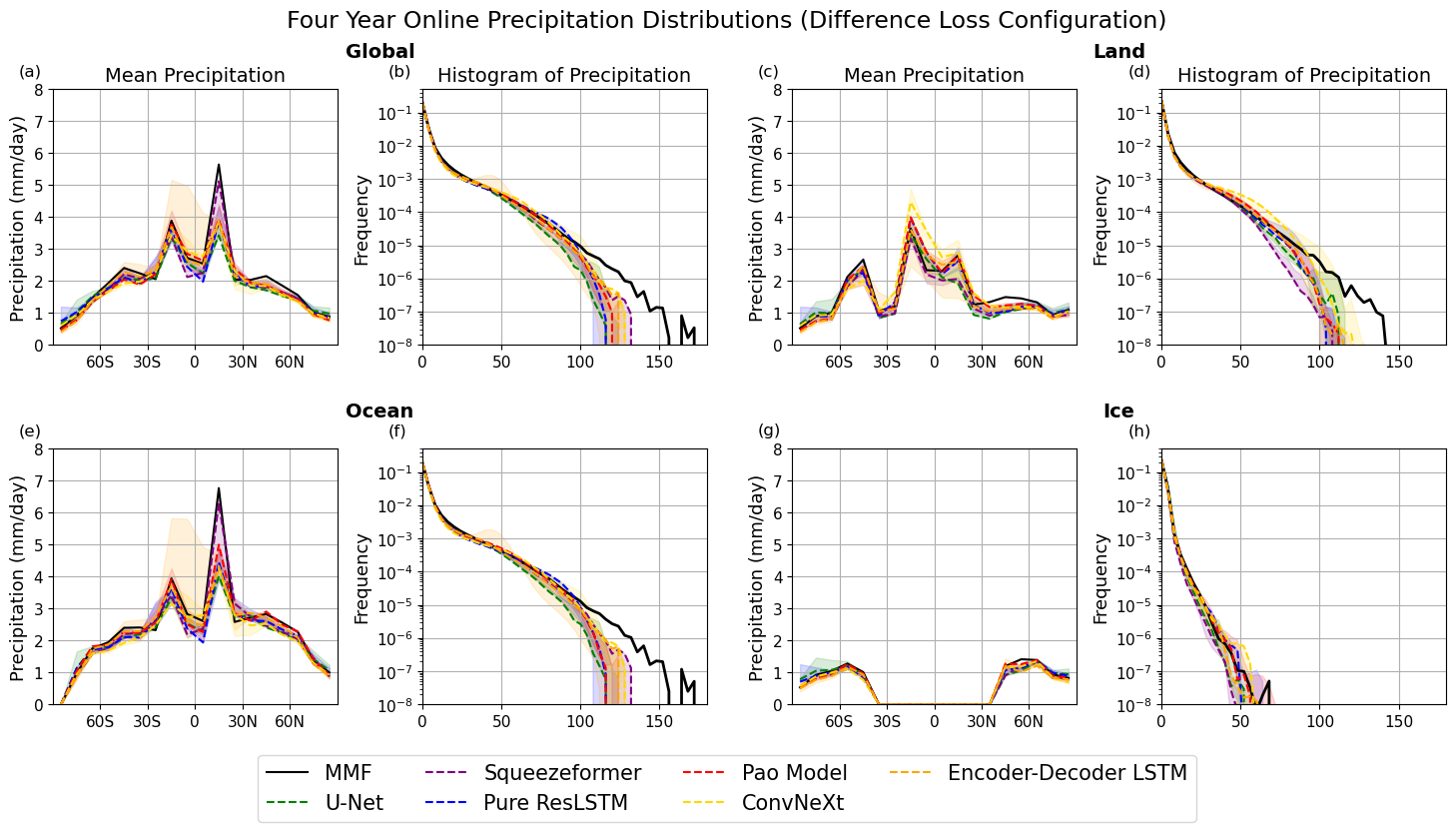}

\caption{This figure shows the latitudinal distribution of mean precipitation and histogram of precipitation values compared to those from MMF for all architectures in the difference loss configuration.}
 \label{fig:four_year_online_precc_dists_diff_loss}
\end{figure}

\begin{figure}[!htbp]
 \centering
 \includegraphics[width=\textwidth]{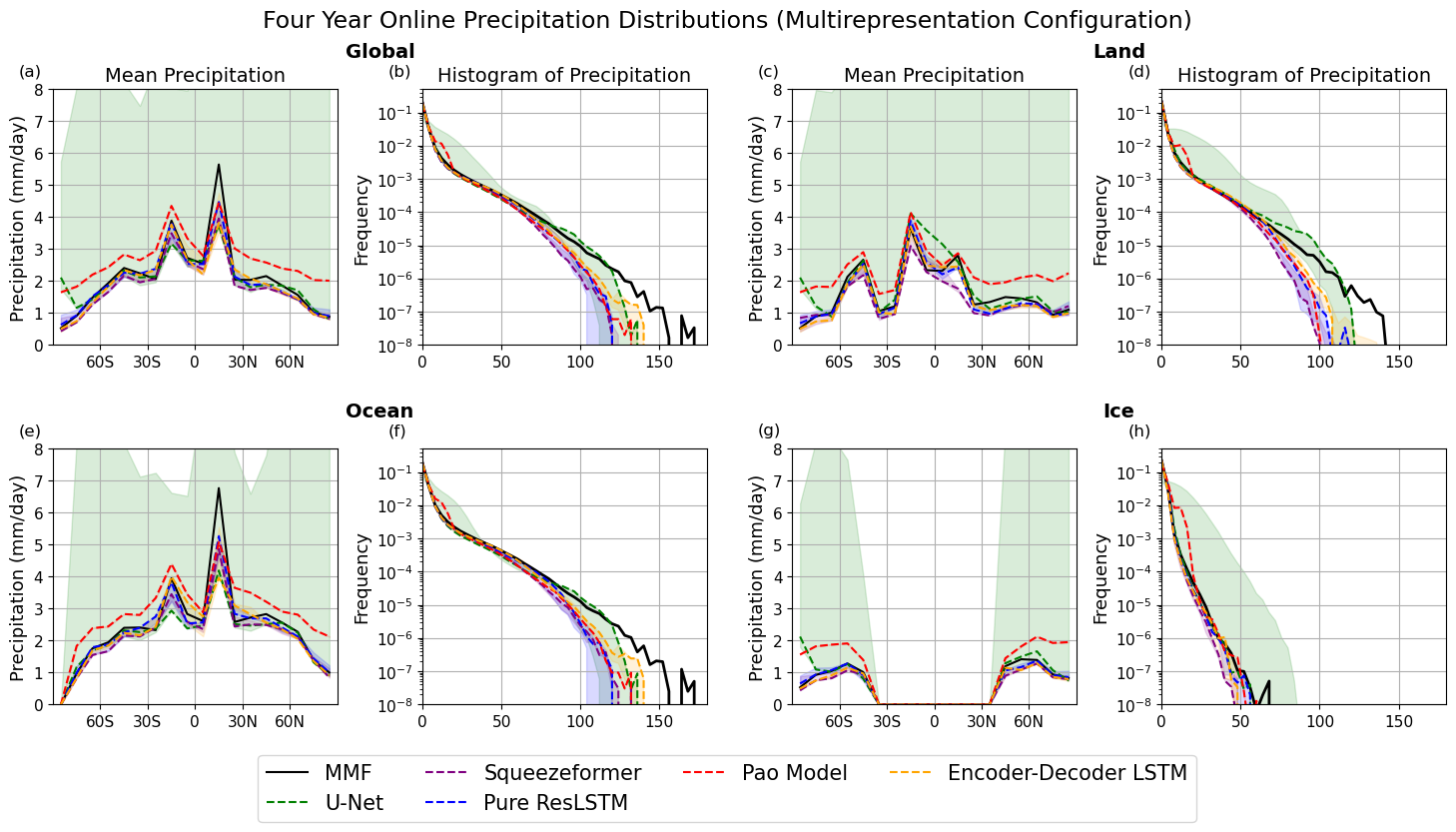}

\caption{This figure shows the latitudinal distribution of mean precipitation and histogram of precipitation values compared to those from MMF for all architectures in the multirepresentation configuration.}
 \label{fig:four_year_online_precc_dists_multirep}
\end{figure}

\begin{figure}[!htbp]
 \centering
 \includegraphics[width=\textwidth]{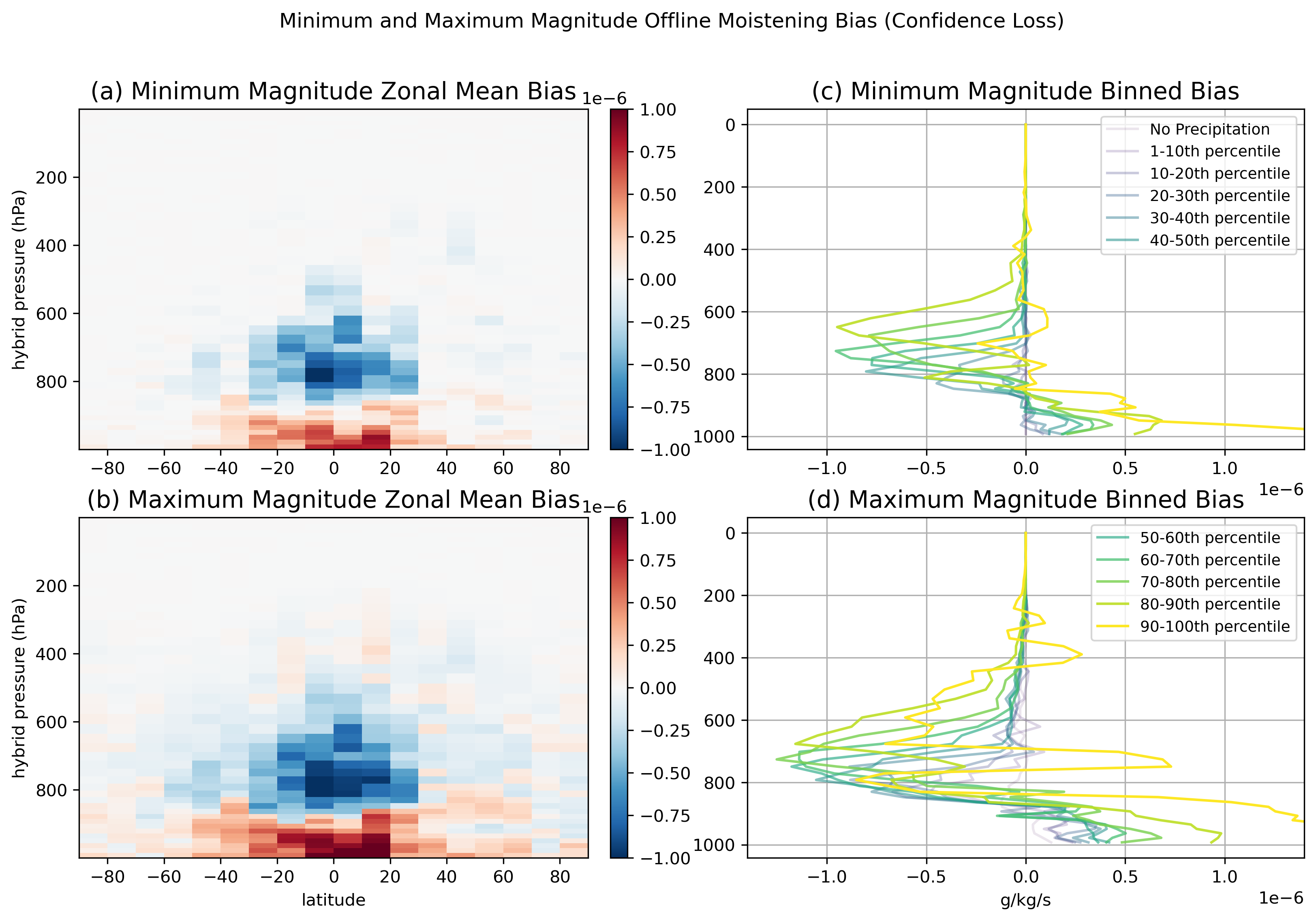}

\caption{Figures S17a and S17b shows minimum and maximum offline zonal mean moistening tendency biases across architectures in the confidence loss configuration after averaging across seeds. Figures S17c and S17d show the vertical profiles of minimum and maximum offline moistening tendency biases across architectures (again after averaging across seeds) when binned by precipitation percentile.}
 \label{fig:offline_DQ1PHYS_bias_minmax_conf_loss}
\end{figure}

\begin{figure}[!htbp]
 \centering
 \includegraphics[width=\textwidth]{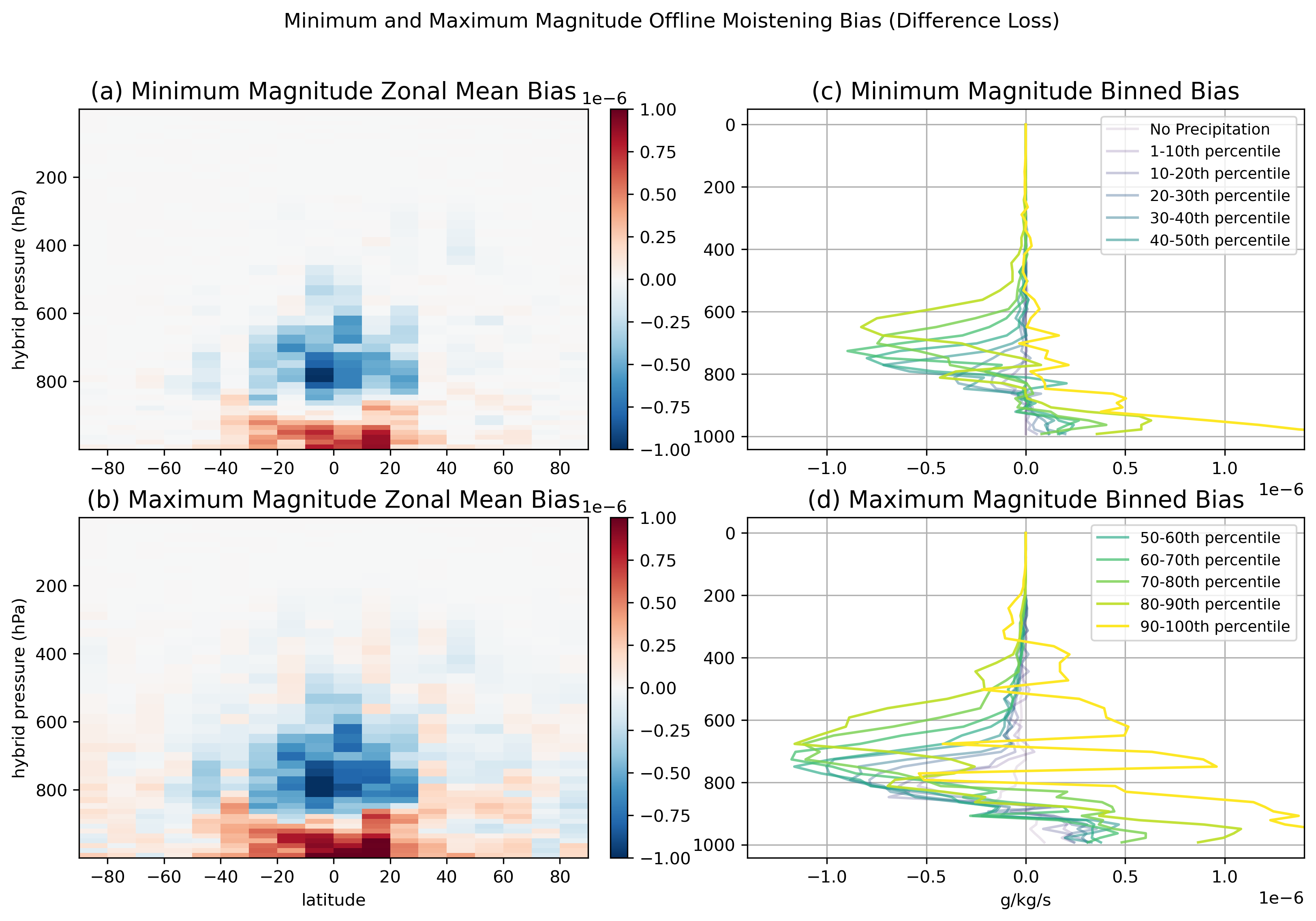}

\caption{Figures S18a and S18b shows minimum and maximum offline zonal mean moistening tendency biases across architectures in the difference loss configuration after averaging across seeds. Figures S18c and S18d show the vertical profiles of minimum and maximum offline moistening tendency biases across architectures (again after averaging across seeds) when binned by precipitation percentile.}
 \label{fig:offline_DQ1PHYS_bias_minmax_diff_loss}
\end{figure}

\begin{figure}[!htbp]
 \centering
 \includegraphics[width=\textwidth]{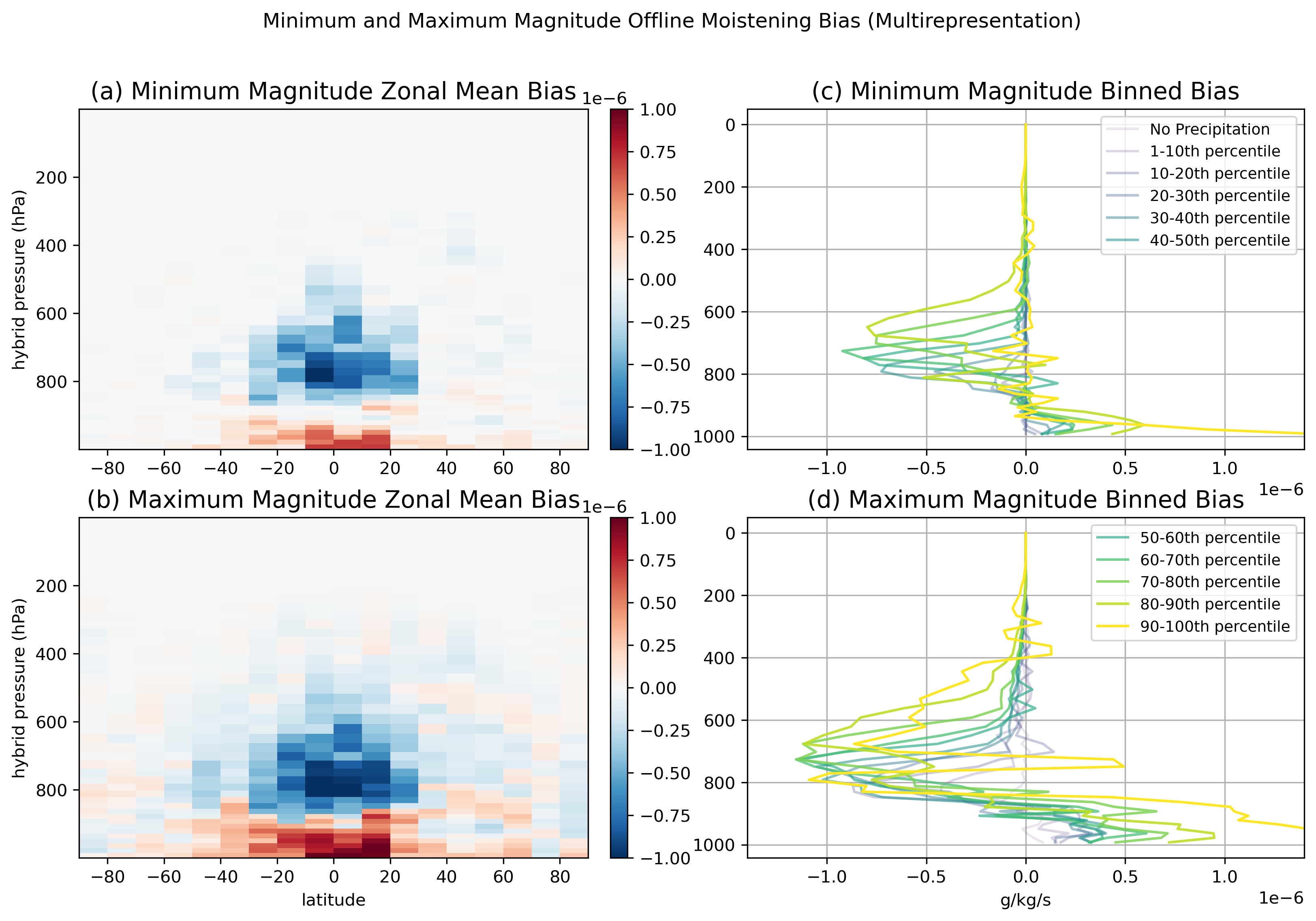}

\caption{Figures S19a and S19b shows minimum and maximum offline zonal mean moistening tendency biases across architectures in the multirepresentation configuration after averaging across seeds. Figures S19c and S19d show the vertical profiles of minimum and maximum offline moistening tendency biases across architectures (again after averaging across seeds) when binned by precipitation percentile.}
 \label{fig:offline_DQ1PHYS_bias_minmax_multirep}
\end{figure}

\begin{figure}[!htbp]
 \centering
 \includegraphics[width=\textwidth]{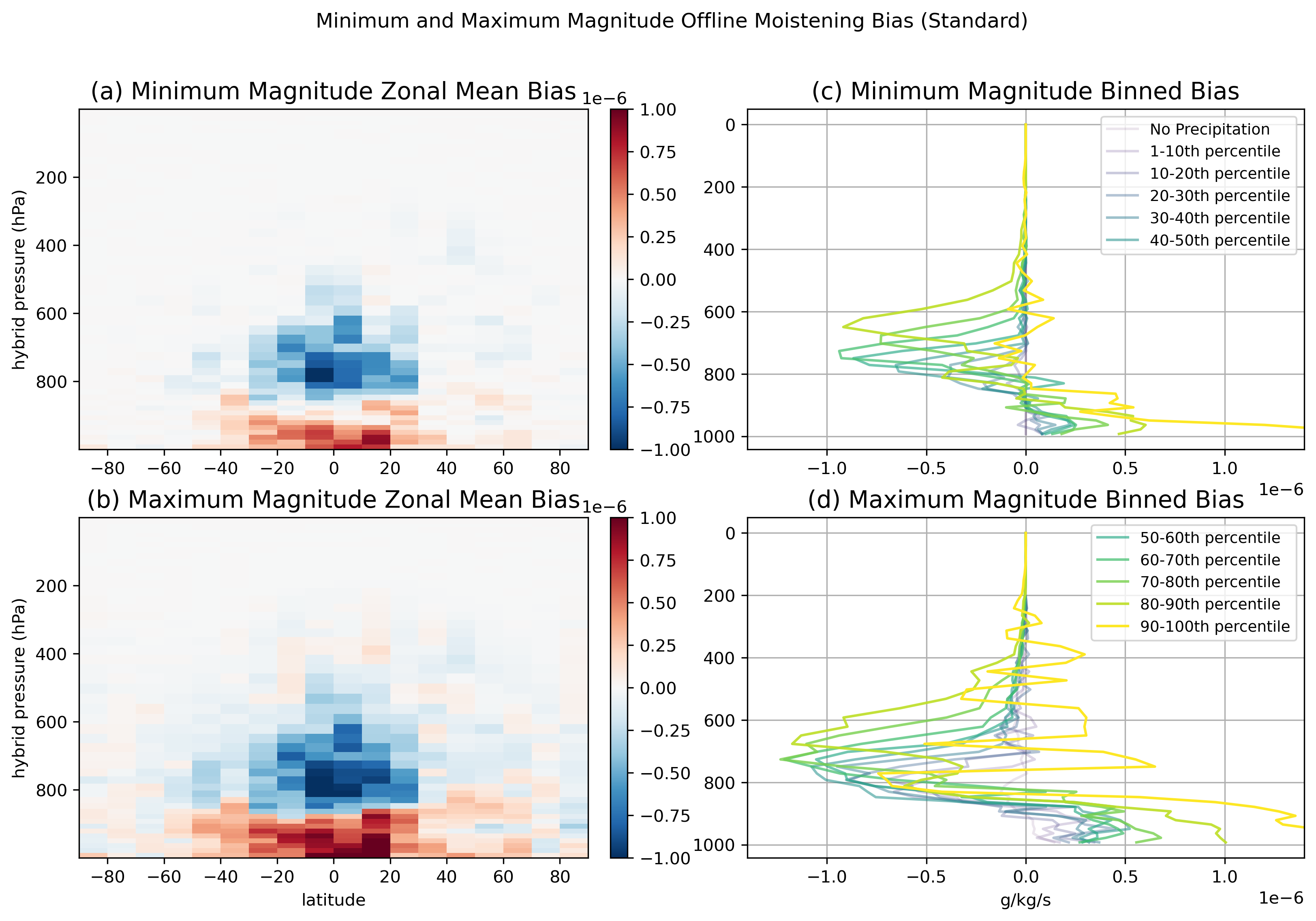}

\caption{Figures S20a and S20b shows minimum and maximum offline zonal mean moistening tendency biases across architectures in the standard configuration after averaging across seeds. Figures S20c and S20d show the vertical profiles of minimum and maximum offline moistening tendency biases across architectures (again after averaging across seeds) when binned by precipitation percentile.}
 \label{fig:offline_DQ1PHYS_bias_minmax_standard}
\end{figure}

\begin{figure}[!htbp]
 \centering
 \includegraphics[width=\textwidth]{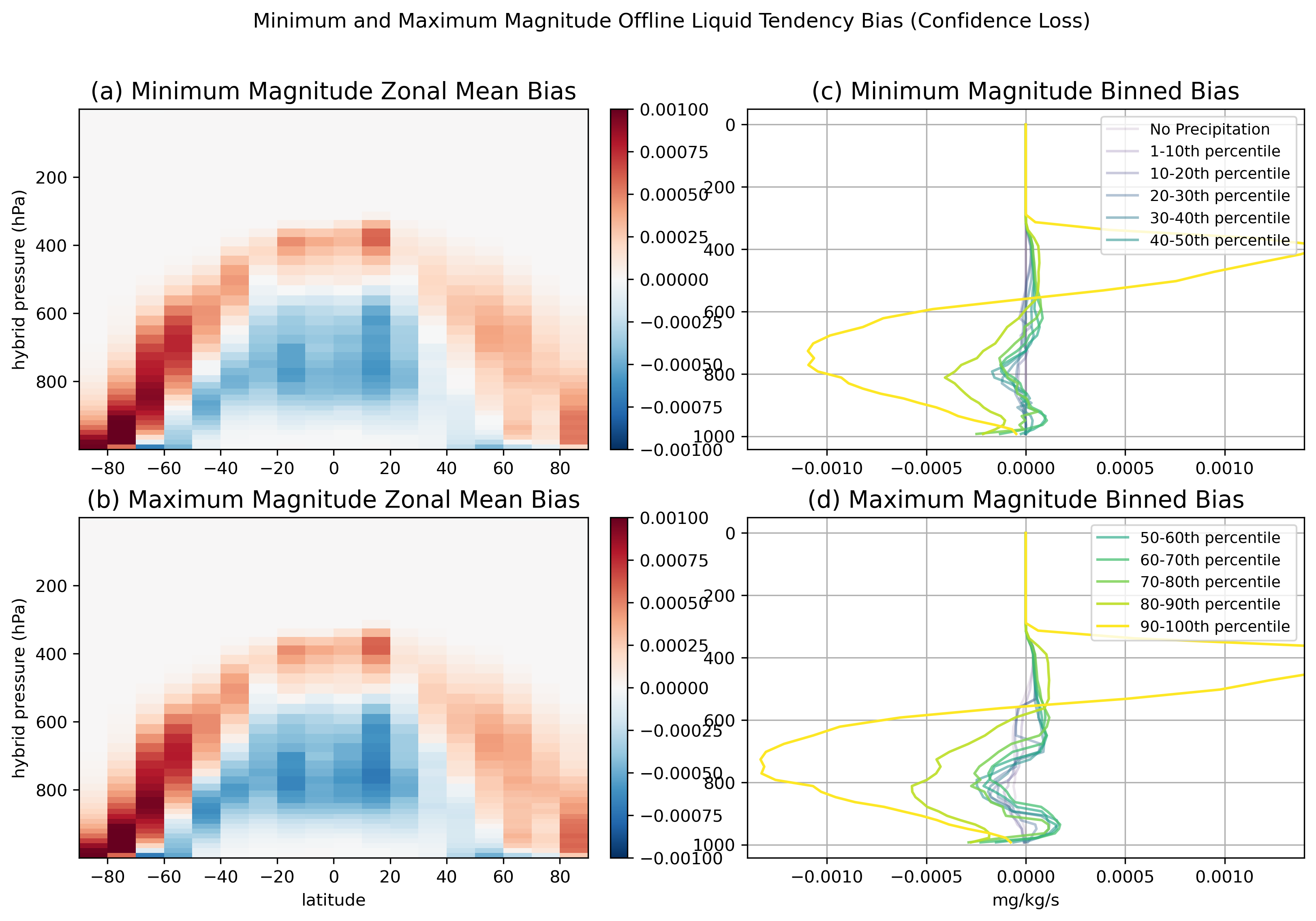}

\caption{Figures S21a and S21b shows minimum and maximum offline zonal mean liquid cloud tendency biases across architectures in the confidence loss configuration after averaging across seeds. Figures S21c and S21d show the vertical profiles of minimum and maximum offline liquid cloud tendency biases across architectures (again after averaging across seeds) when binned by precipitation percentile.}
 \label{fig:offline_DQ2PHYS_bias_minmax_conf_loss}
\end{figure}

\begin{figure}[!htbp]
 \centering
 \includegraphics[width=\textwidth]{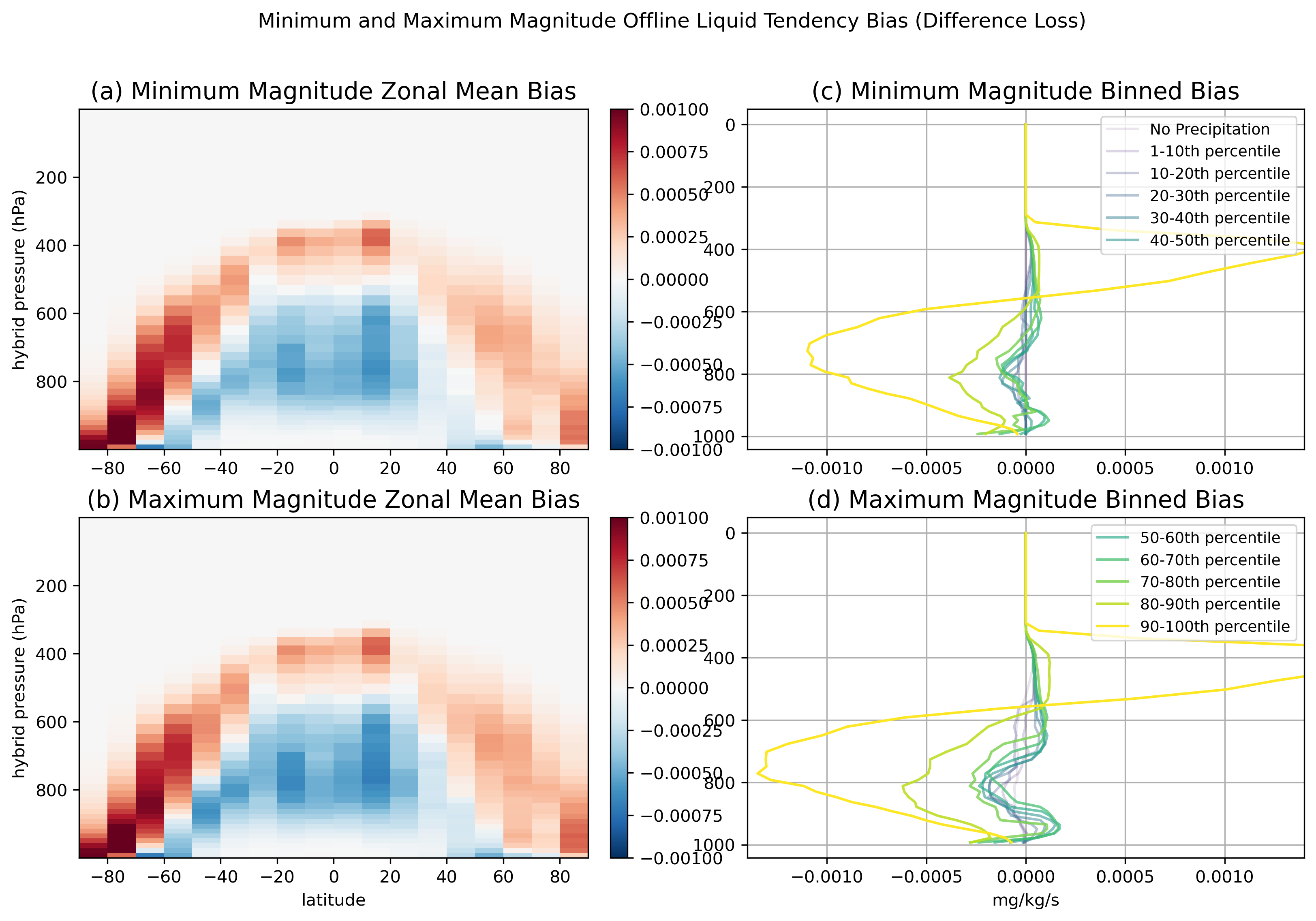}

\caption{Figures S22a and S22b shows minimum and maximum offline zonal mean liquid cloud tendency biases across architectures in the difference loss configuration after averaging across seeds. Figures S22c and S22d show the vertical profiles of minimum and maximum offline liquid cloud tendency biases across architectures (again after averaging across seeds) when binned by precipitation percentile.}
 \label{fig:offline_DQ2PHYS_bias_minmax_diff_loss}
\end{figure}

\begin{figure}[!htbp]
 \centering
 \includegraphics[width=\textwidth]{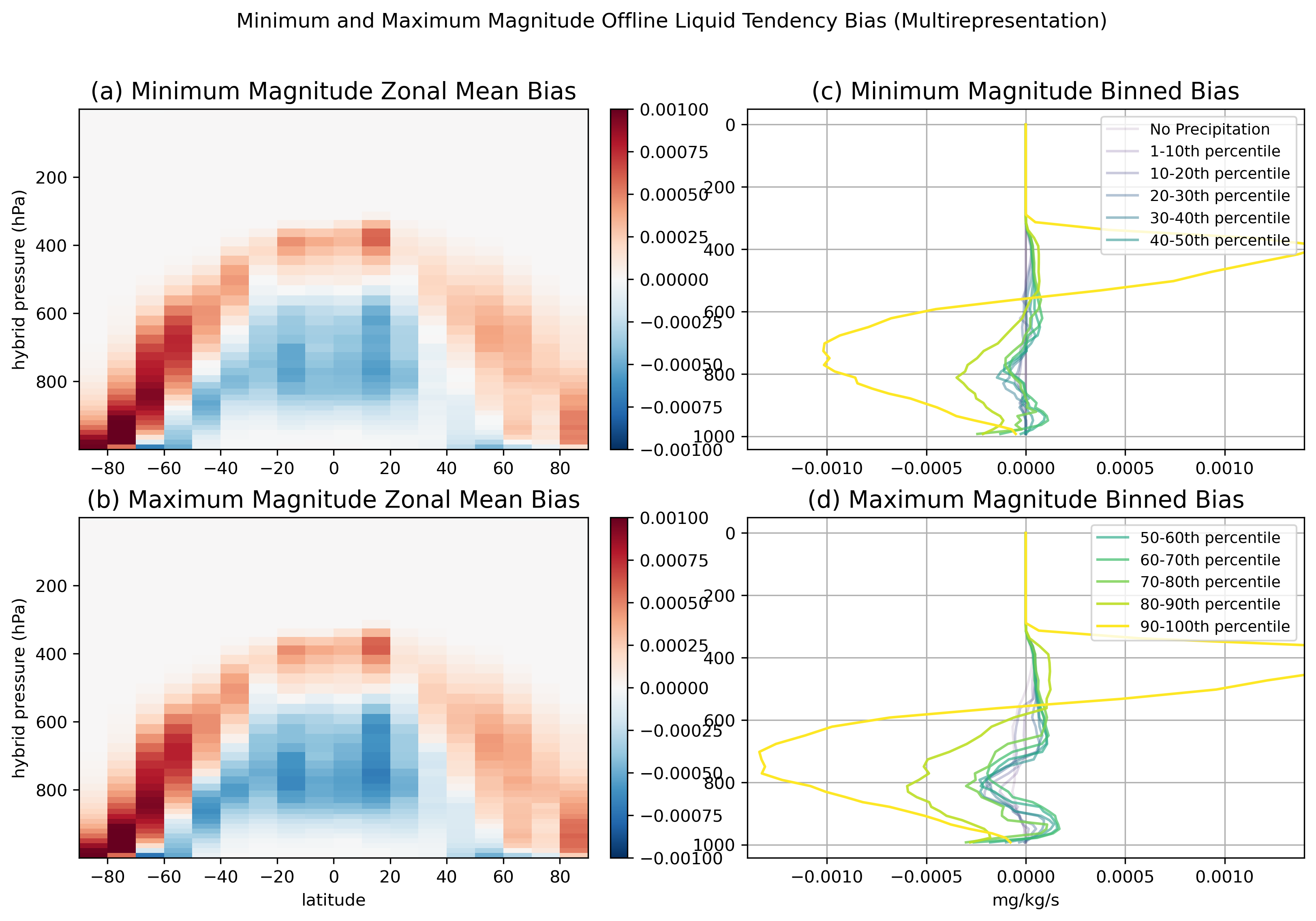}

\caption{Figures S23a and S23b shows minimum and maximum offline zonal mean liquid cloud tendency biases across architectures in the multirepresentation configuration after averaging across seeds. Figures S23c and S23d show the vertical profiles of minimum and maximum offline liquid cloud tendency biases across architectures (again after averaging across seeds) when binned by precipitation percentile.}
 \label{fig:offline_DQ2PHYS_bias_minmax_multirep}
\end{figure}

\begin{figure}[!htbp]
 \centering
 \includegraphics[width=\textwidth]{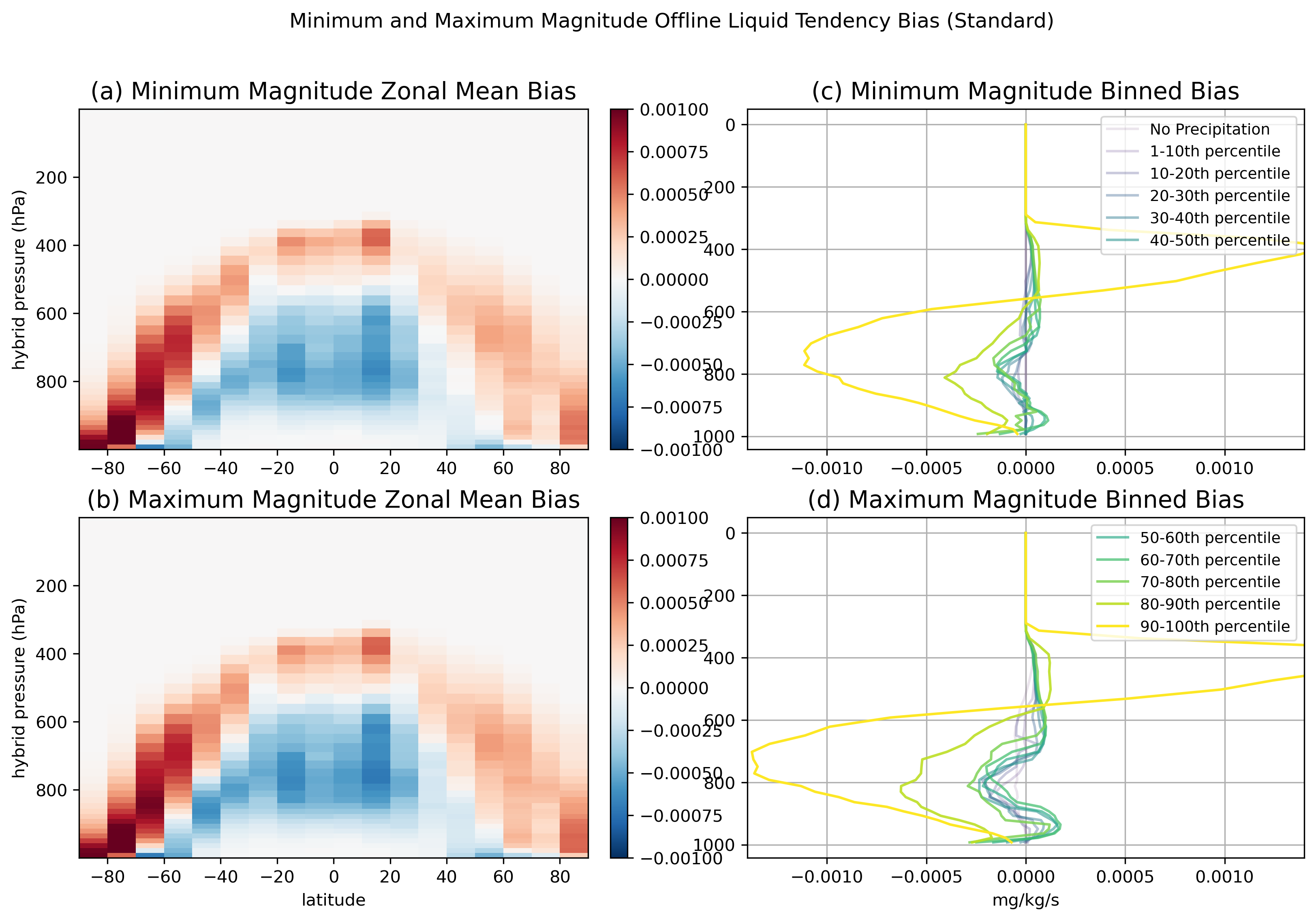}

\caption{Figures S24a and S24b shows minimum and maximum offline zonal mean liquid cloud tendency biases across architectures in the standard configuration after averaging across seeds. Figures S24c and S24d show the vertical profiles of minimum and maximum offline liquid cloud tendency biases across architectures (again after averaging across seeds) when binned by precipitation percentile.}
 \label{fig:offline_DQ2PHYS_bias_minmax_standard}
\end{figure}

\begin{figure}[!htbp]
 \centering
 \includegraphics[width=\textwidth]{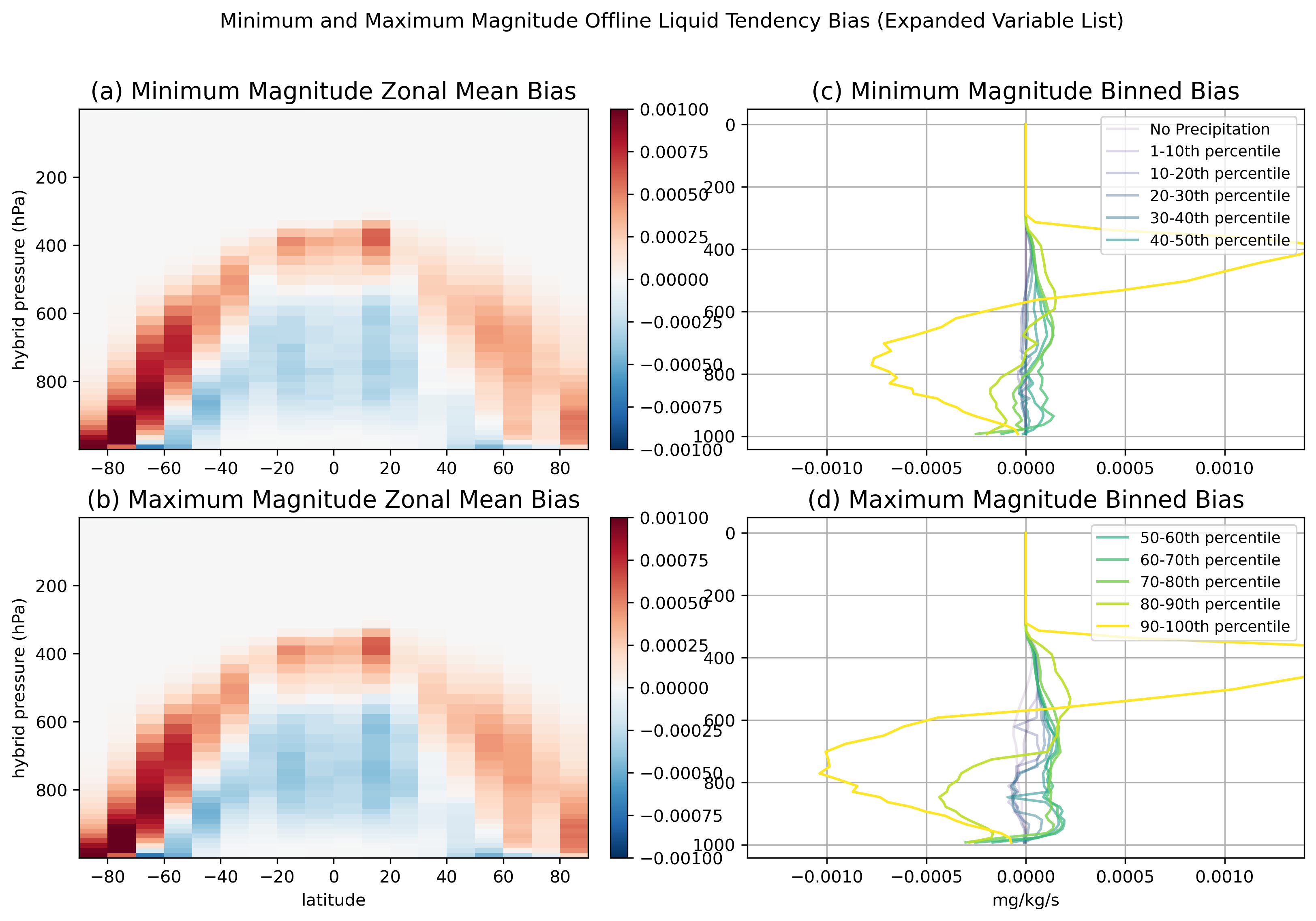}

\caption{Figures S25a and S25b shows minimum and maximum offline zonal mean liquid cloud tendency biases across architectures in the expanded variables list configuration after averaging across seeds. Figures S25c and S25d show the vertical profiles of minimum and maximum offline liquid cloud tendency biases across architectures (again after averaging across seeds) when binned by precipitation percentile.}
 \label{fig:offline_DQ2PHYS_bias_minmax_v6}
\end{figure}

\begin{figure}[!htbp]
 \centering
 \includegraphics[width=\textwidth]{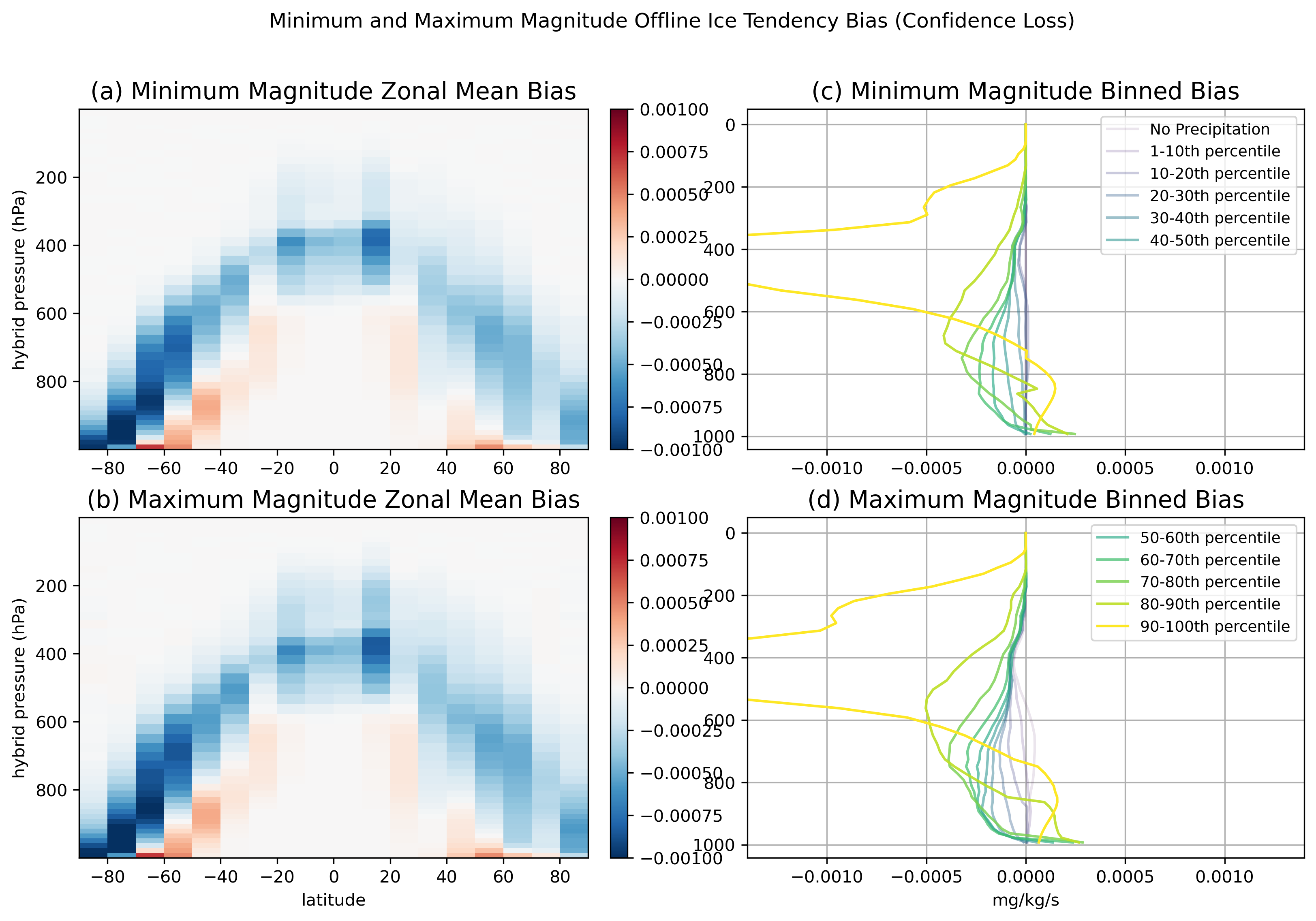}

\caption{Figures S26a and S26b shows minimum and maximum offline zonal mean ice cloud tendency biases across architectures in the confidence loss configuration after averaging across seeds. Figures S26c and S26d show the vertical profiles of minimum and maximum offline ice cloud tendency biases across architectures (again after averaging across seeds) when binned by precipitation percentile.}
 \label{fig:offline_DQ3PHYS_bias_minmax_conf_loss}
\end{figure}

\begin{figure}[!htbp]
 \centering
 \includegraphics[width=\textwidth]{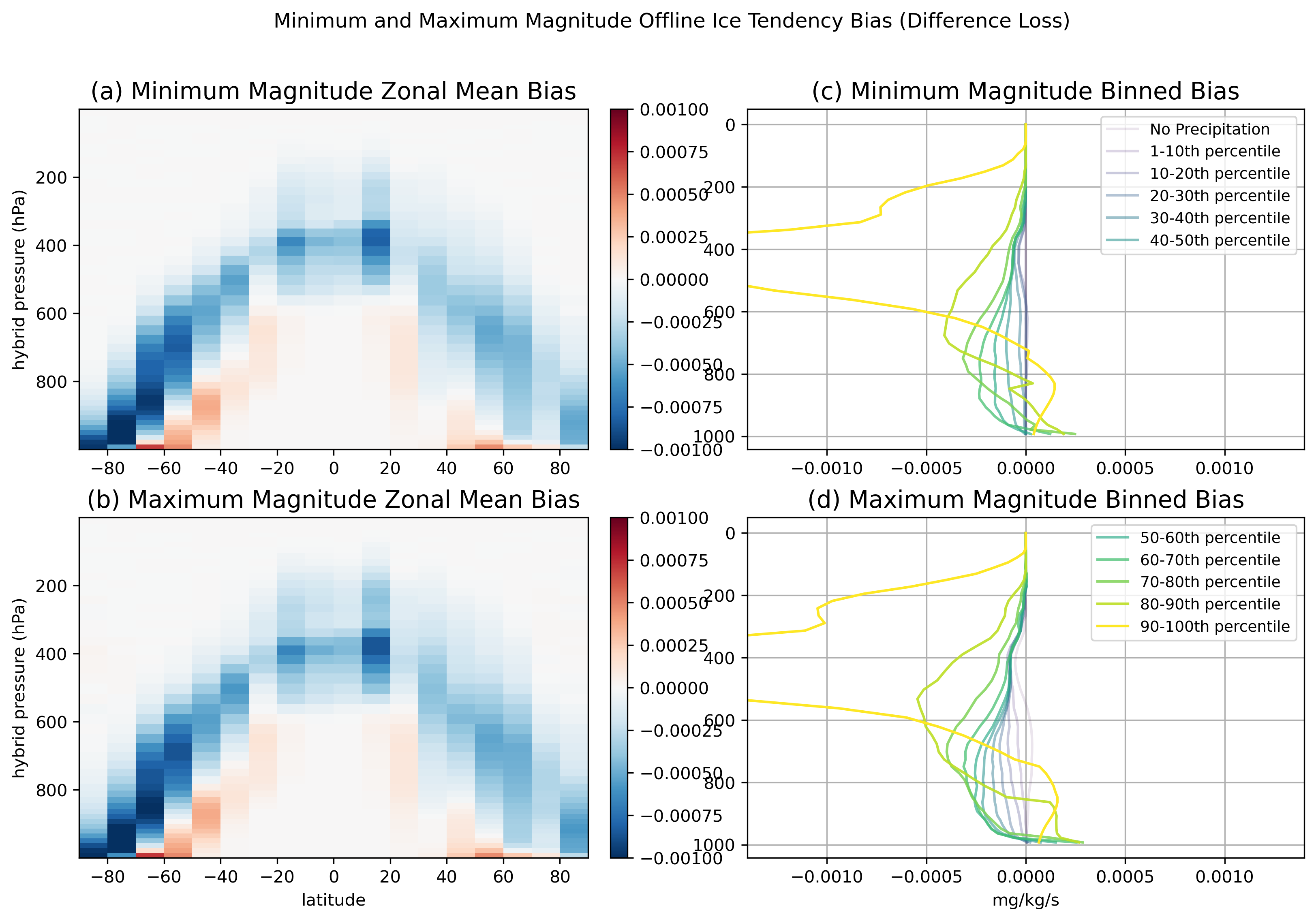}

\caption{Figures S27a and S27b shows minimum and maximum offline zonal mean ice cloud tendency biases across architectures in the difference loss configuration after averaging across seeds. Figures S27c and S27d show the vertical profiles of minimum and maximum offline ice cloud tendency biases across architectures (again after averaging across seeds) when binned by precipitation percentile.}
 \label{fig:offline_DQ3PHYS_bias_minmax_diff_loss}
\end{figure}

\begin{figure}[!htbp]
 \centering
 \includegraphics[width=\textwidth]{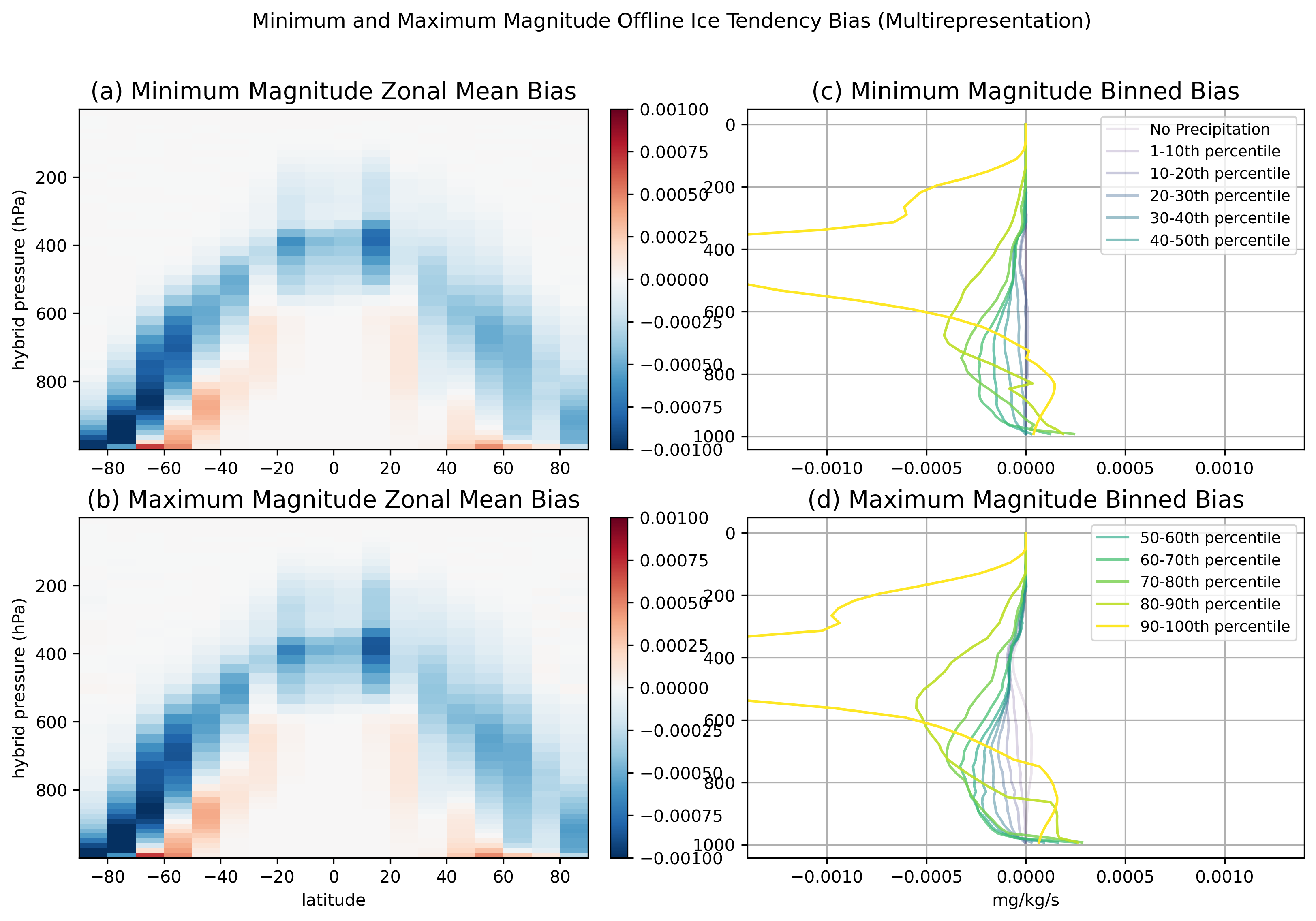}

\caption{Figures S28a and S28b shows minimum and maximum offline zonal mean ice cloud tendency biases across architectures in the multirepresentation configuration after averaging across seeds. Figures S28c and S28d show the vertical profiles of minimum and maximum offline ice cloud tendency biases across architectures (again after averaging across seeds) when binned by precipitation percentile.}
 \label{fig:offline_DQ3PHYS_bias_minmax_multirep}
\end{figure}

\begin{figure}[!htbp]
 \centering
 \includegraphics[width=\textwidth]{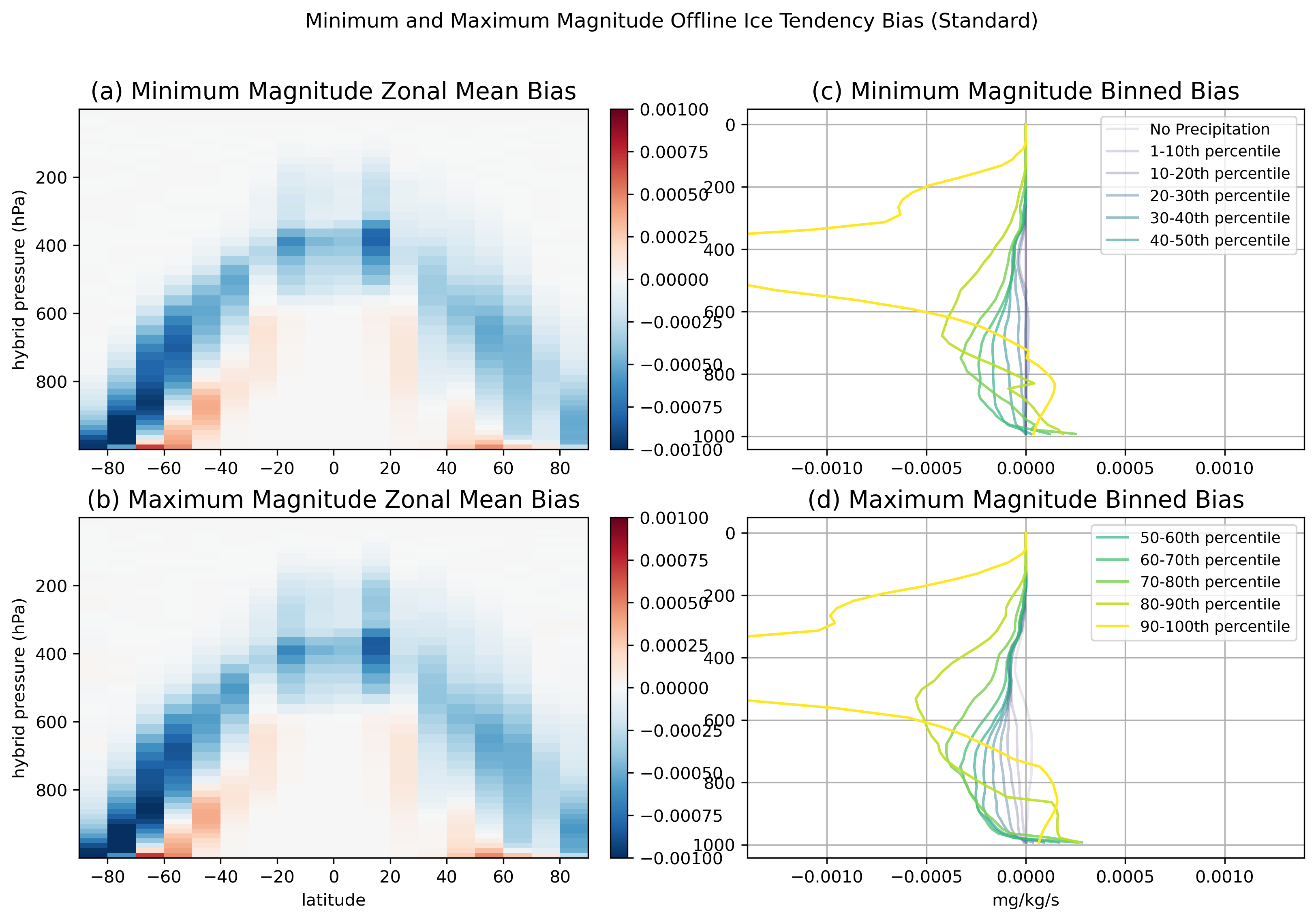}

\caption{Figures S29a and S29b shows minimum and maximum offline zonal mean ice cloud tendency biases across architectures in the standard configuration after averaging across seeds. Figures S29c and S29d show the vertical profiles of minimum and maximum offline ice cloud tendency biases across architectures (again after averaging across seeds) when binned by precipitation percentile.}
 \label{fig:offline_DQ3PHYS_bias_minmax_standard}
\end{figure}

\begin{figure}[!htbp]
 \centering
 \includegraphics[width=\textwidth]{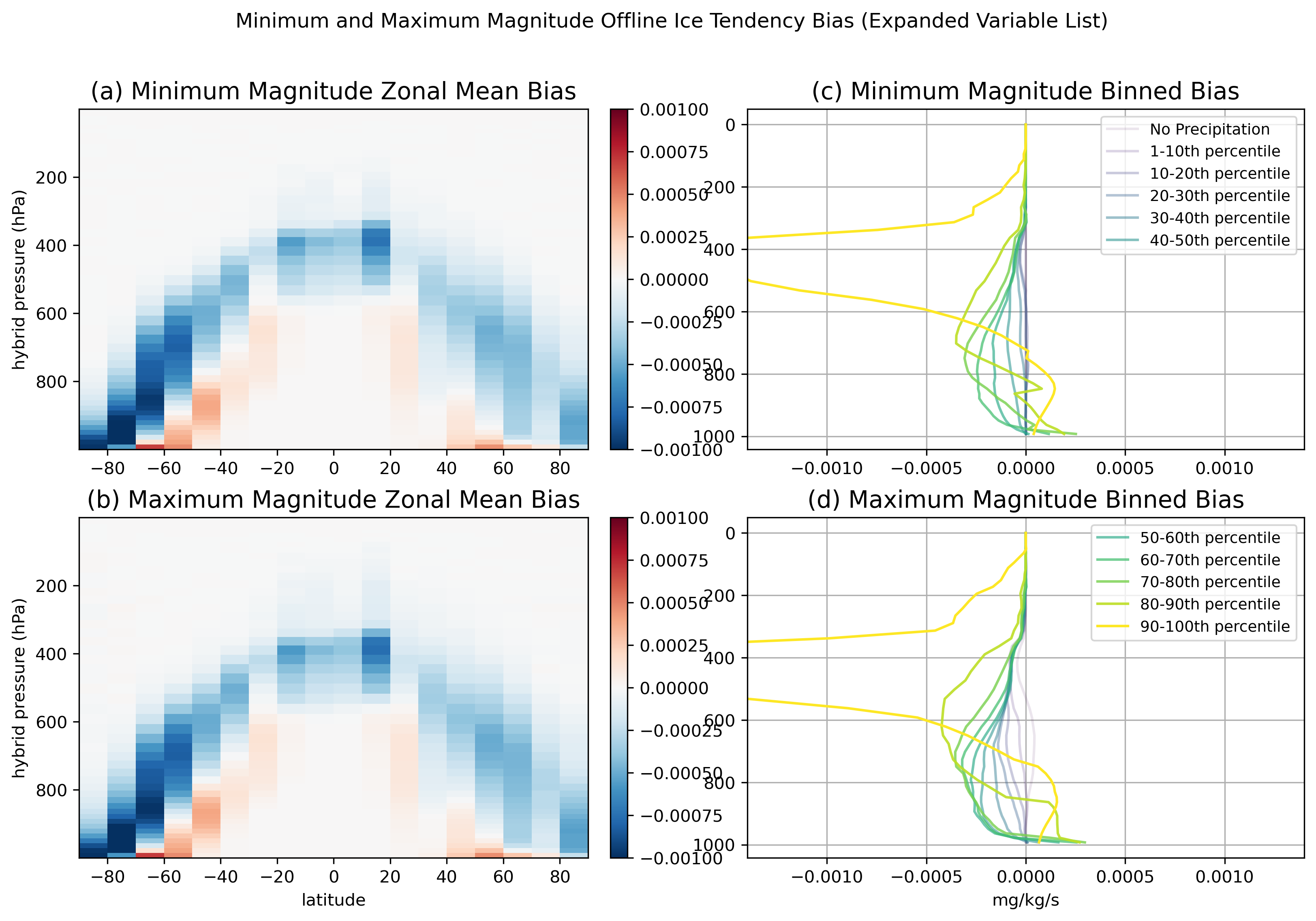}

\caption{Figures S30a and S30b shows minimum and maximum offline zonal mean ice cloud tendency biases across architectures in the expanded variables list configuration after averaging across seeds. Figures S30c and S30d show the vertical profiles of minimum and maximum offline ice cloud tendency biases across architectures (again after averaging across seeds) when binned by precipitation percentile.}
 \label{fig:offline_DQ3PHYS_bias_minmax_v6}
\end{figure}

\begin{figure}[!htbp]
 \centering
 \includegraphics[width=\textwidth]{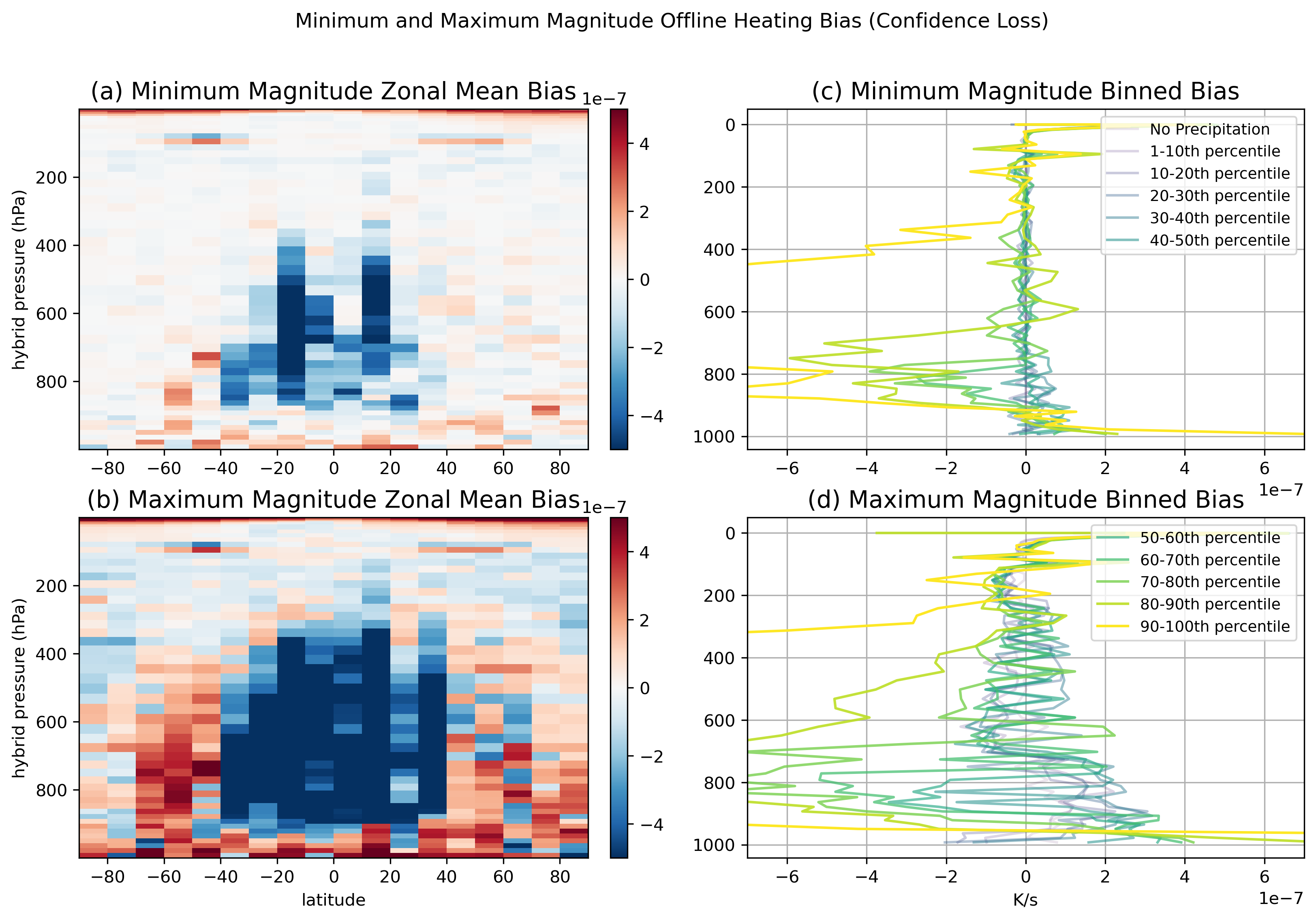}

\caption{Figures S31a and S31b shows minimum and maximum offline zonal mean heating tendency biases across architectures in the confidence loss configuration after averaging across seeds. Figures S31c and S31d show the vertical profiles of minimum and maximum offline heating tendency biases across architectures (again after averaging across seeds) when binned by precipitation percentile.}
 \label{fig:offline_DTPHYS_bias_minmax_conf_loss}
\end{figure}

\begin{figure}[!htbp]
 \centering
 \includegraphics[width=\textwidth]{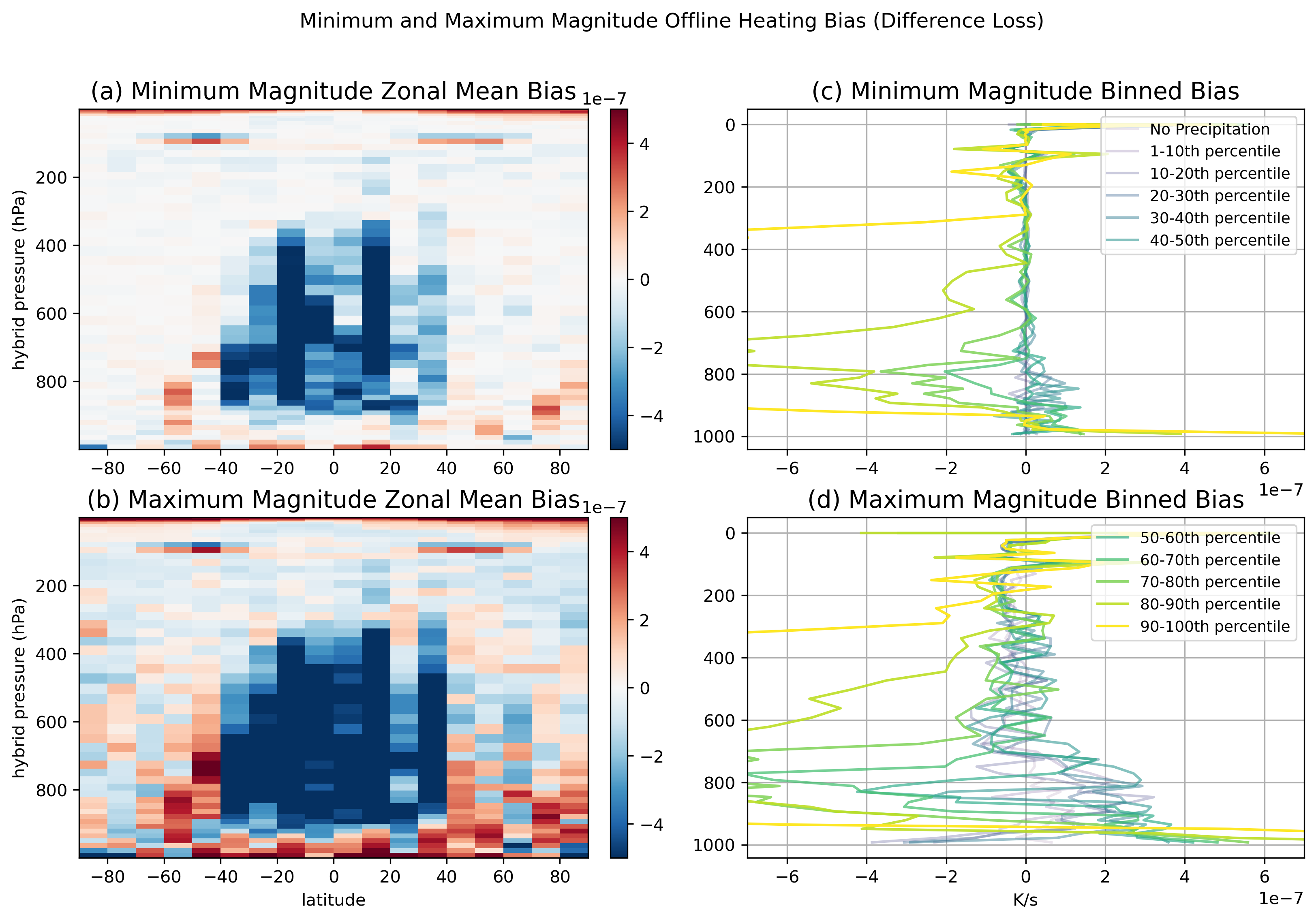}

\caption{Figures S32a and S32b shows minimum and maximum offline zonal mean heating tendency biases across architectures in the difference loss configuration after averaging across seeds. Figures S32c and S32d show the vertical profiles of minimum and maximum offline heating tendency biases across architectures (again after averaging across seeds) when binned by precipitation percentile.}
 \label{fig:offline_DTPHYS_bias_minmax_diff_loss}
\end{figure}

\begin{figure}[!htbp]
 \centering
 \includegraphics[width=\textwidth]{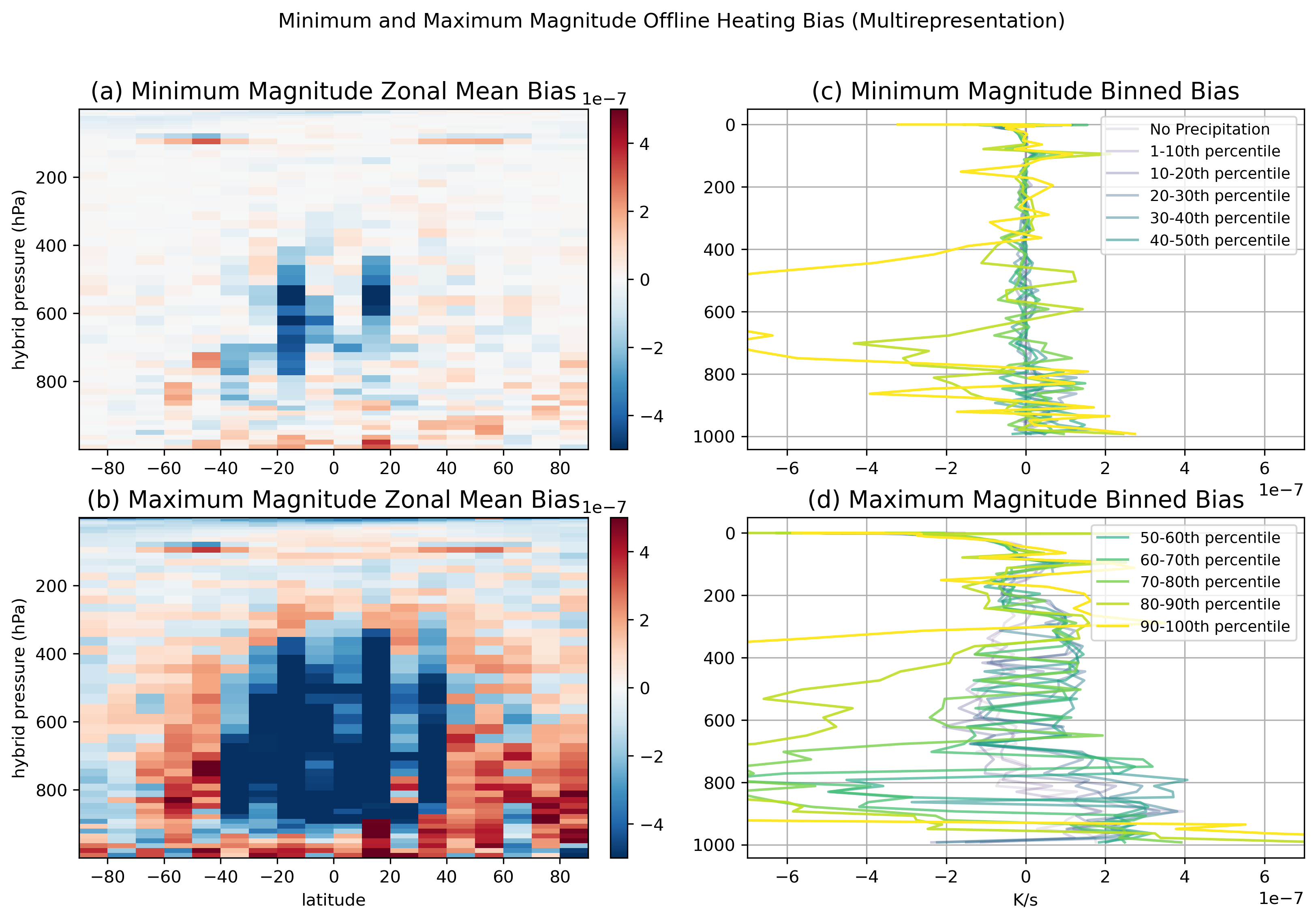}

\caption{Figures S33a and S33b shows minimum and maximum offline zonal mean heating tendency biases across architectures in the multirepresentation configuration after averaging across seeds. Figures S33c and S33d show the vertical profiles of minimum and maximum offline heating tendency biases across architectures (again after averaging across seeds) when binned by precipitation percentile.}
 \label{fig:offline_DTPHYS_bias_minmax_multirep}
\end{figure}

\begin{figure}[!htbp]
 \centering
 \includegraphics[width=\textwidth]{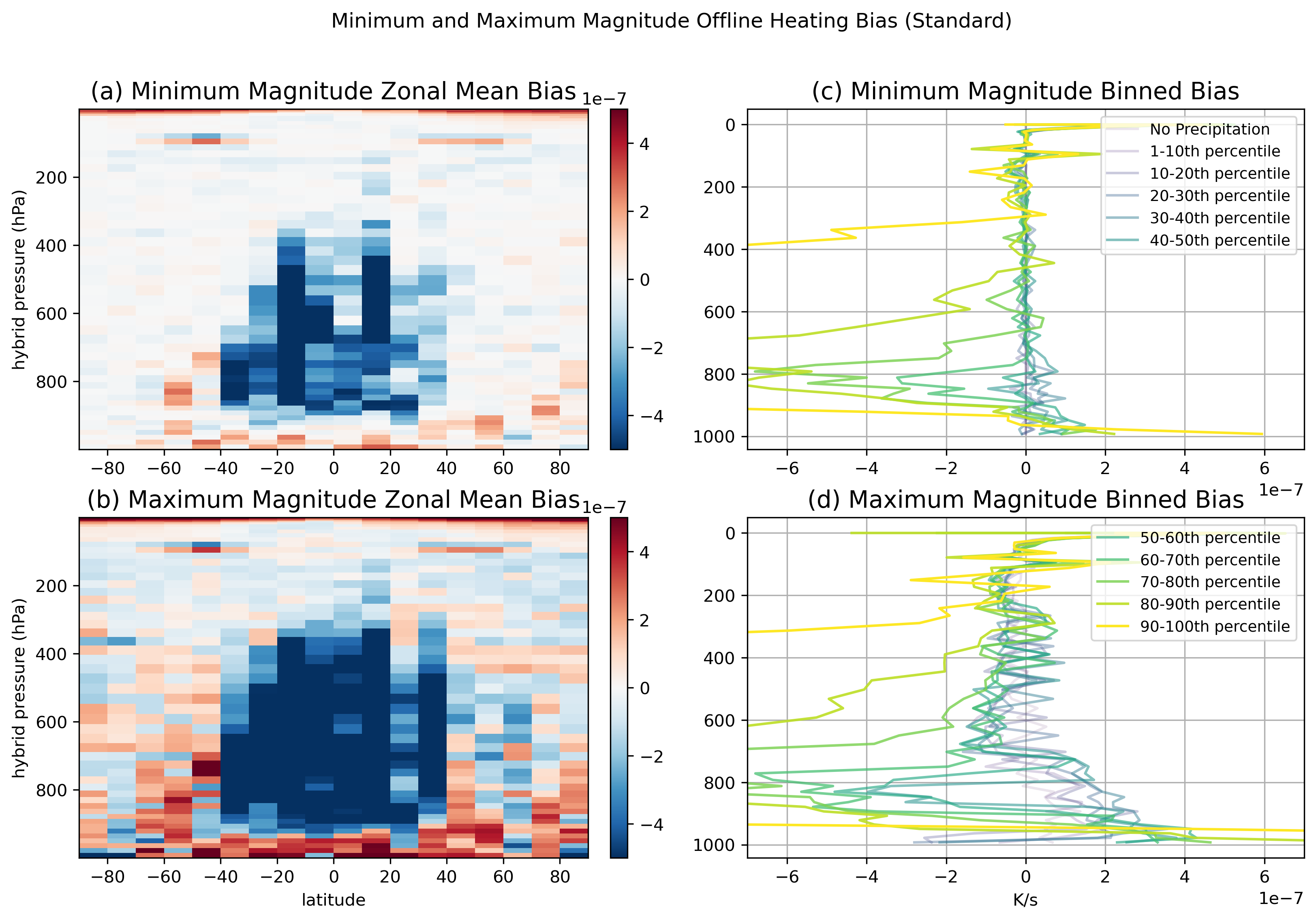}

\caption{Figures S34a and S34b shows minimum and maximum offline zonal mean heating tendency biases across architectures in the standard configuration after averaging across seeds. Figures S34c and S34d show the vertical profiles of minimum and maximum offline heating tendency biases across architectures (again after averaging across seeds) when binned by precipitation percentile.}
 \label{fig:offline_DTPHYS_bias_minmax_standard}
\end{figure}

\begin{figure}[!htbp]
 \centering
 \includegraphics[width=\textwidth]{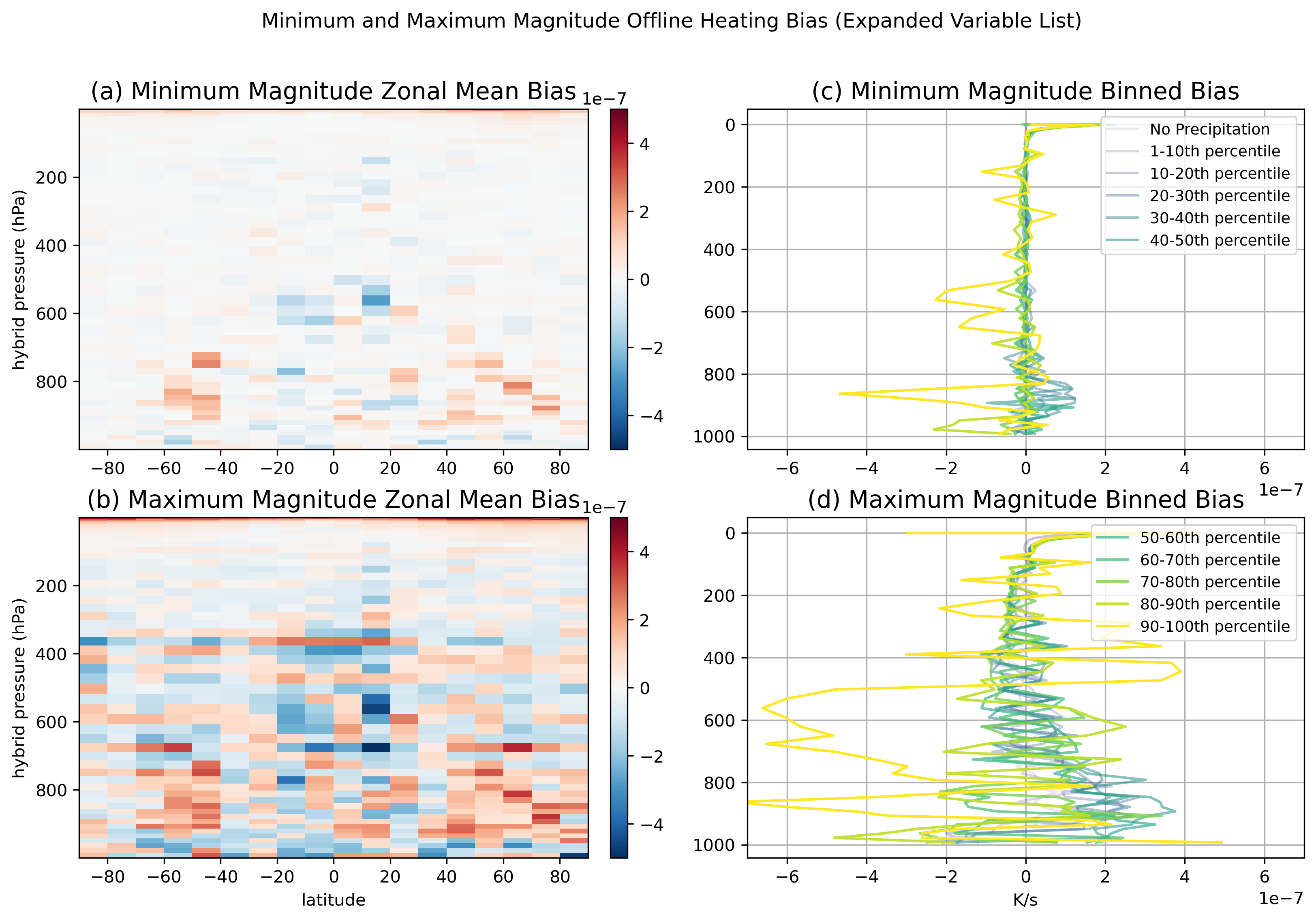}

\caption{Figures S35a and S35b shows minimum and maximum offline zonal mean heating tendency biases across architectures in the expanded variables list configuration after averaging across seeds. Figures S35c and S35d show the vertical profiles of minimum and maximum offline heating tendency biases across architectures (again after averaging across seeds) when binned by precipitation percentile.}
 \label{fig:offline_DTPHYS_bias_minmax_v6}
\end{figure}

\begin{figure}[!htbp]
 \centering
 \includegraphics[width=\textwidth]{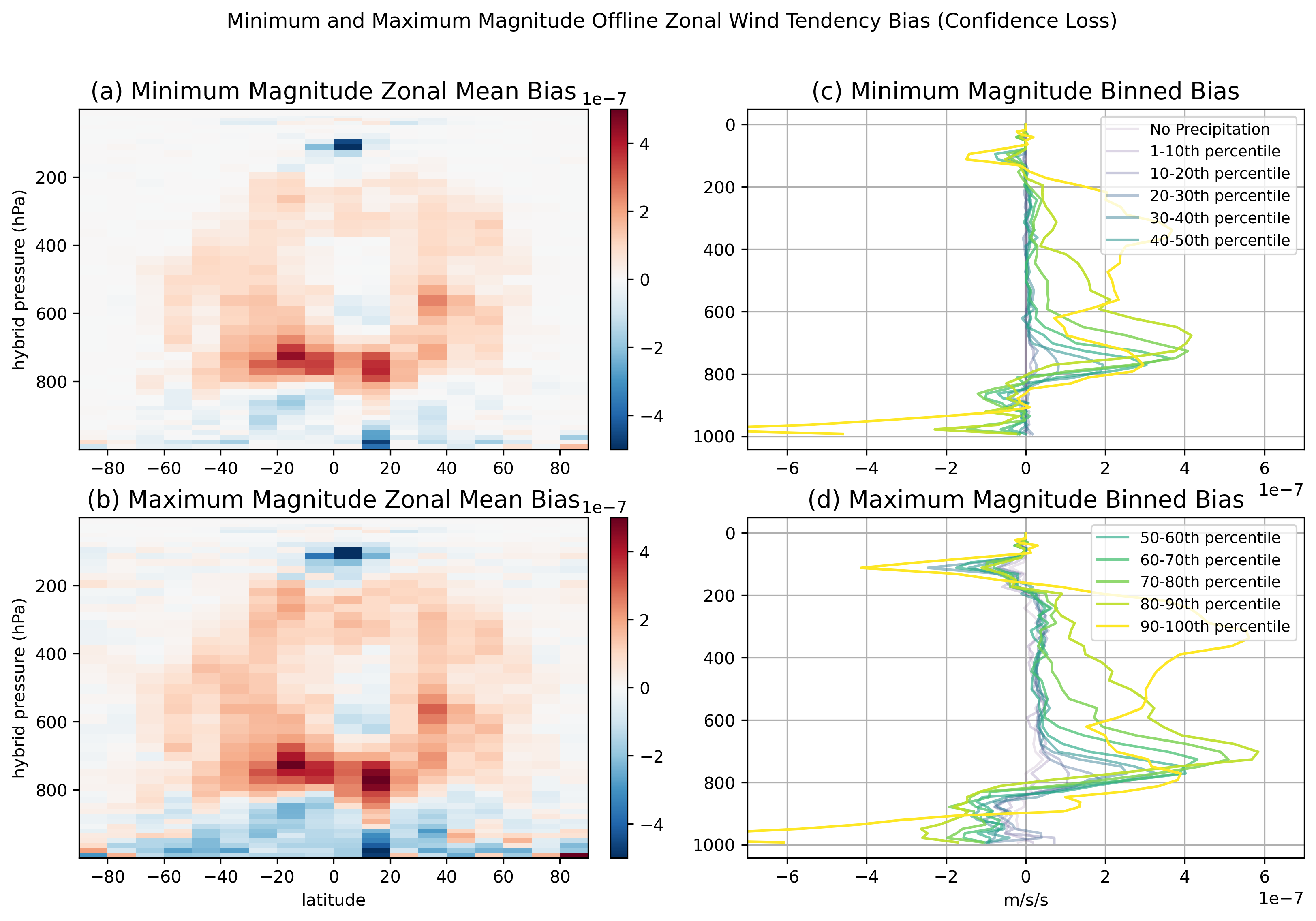}

\caption{Figures S36a and S36b shows minimum and maximum offline zonal mean zonal wind tendency biases across architectures in the confidence loss configuration after averaging across seeds. Figures S36c and S36d show the vertical profiles of minimum and maximum offline zonal wind tendency biases across architectures (again after averaging across seeds) when binned by precipitation percentile.}
 \label{fig:offline_DUPHYS_bias_minmax_conf_loss}
\end{figure}

\begin{figure}[!htbp]
 \centering
 \includegraphics[width=\textwidth]{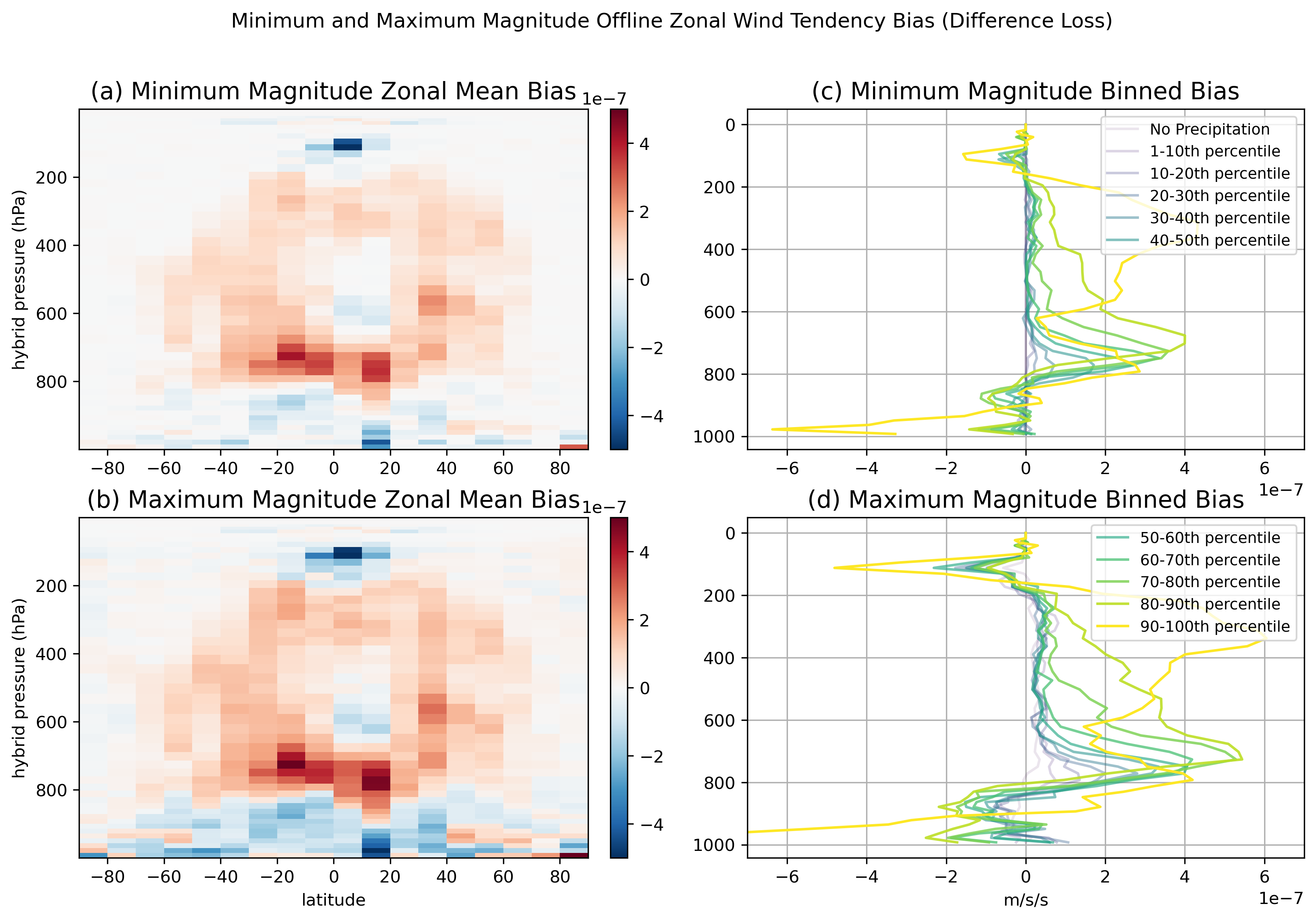}

\caption{Figures S37a and S37b shows minimum and maximum offline zonal mean zonal wind tendency biases across architectures in the difference loss configuration after averaging across seeds. Figures S37c and S37d show the vertical profiles of minimum and maximum offline zonal wind tendency biases across architectures (again after averaging across seeds) when binned by precipitation percentile.}
 \label{fig:offline_DUPHYS_bias_minmax_diff_loss}
\end{figure}

\begin{figure}[!htbp]
 \centering
 \includegraphics[width=\textwidth]{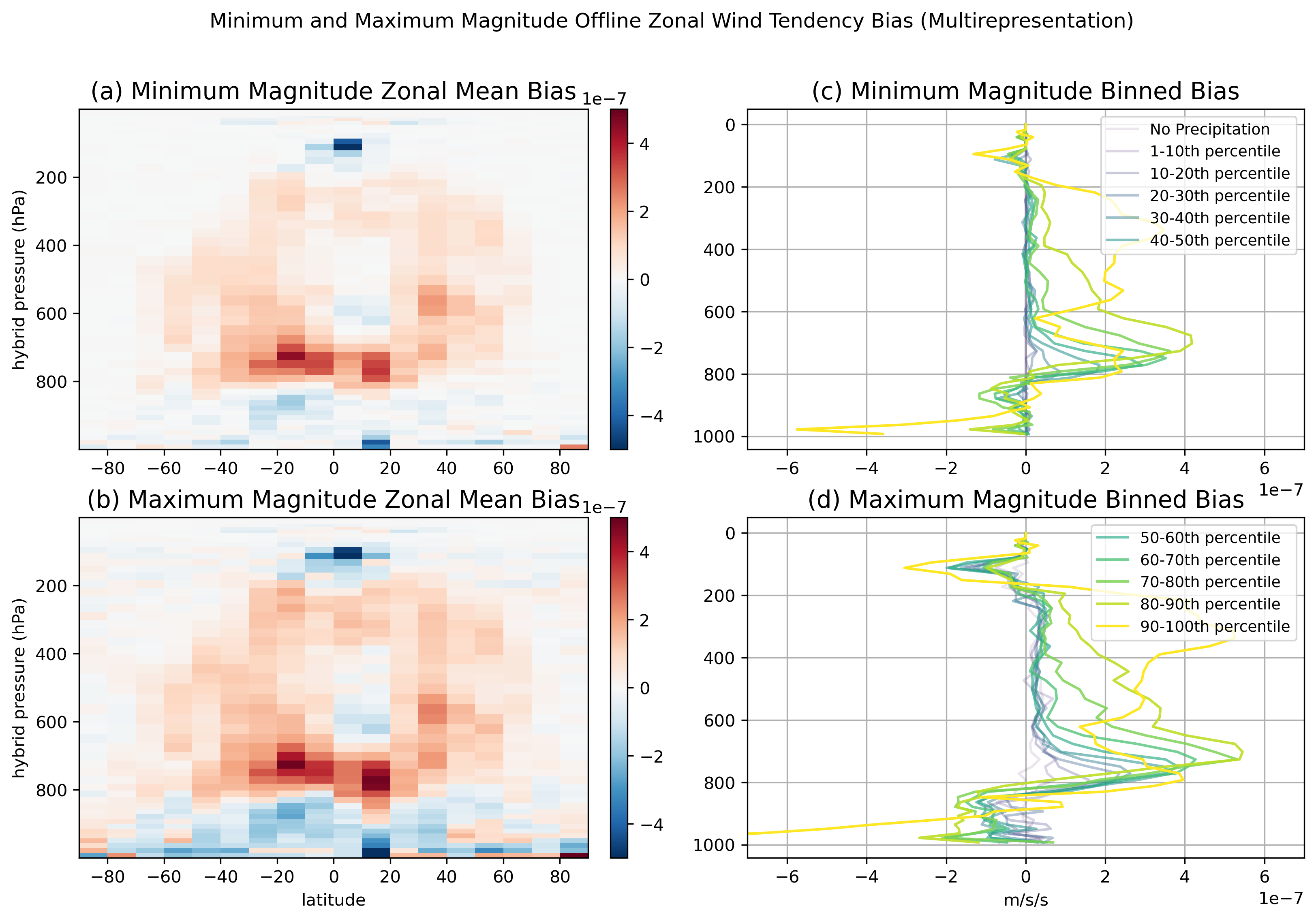}

\caption{Figures S38a and S38b shows minimum and maximum offline zonal mean zonal wind tendency biases across architectures in the multirepresentation configuration after averaging across seeds. Figures S38c and S38d show the vertical profiles of minimum and maximum offline zonal wind tendency biases across architectures (again after averaging across seeds) when binned by precipitation percentile.}
 \label{fig:offline_DUPHYS_bias_minmax_multirep}
\end{figure}

\begin{figure}[!htbp]
 \centering
 \includegraphics[width=\textwidth]{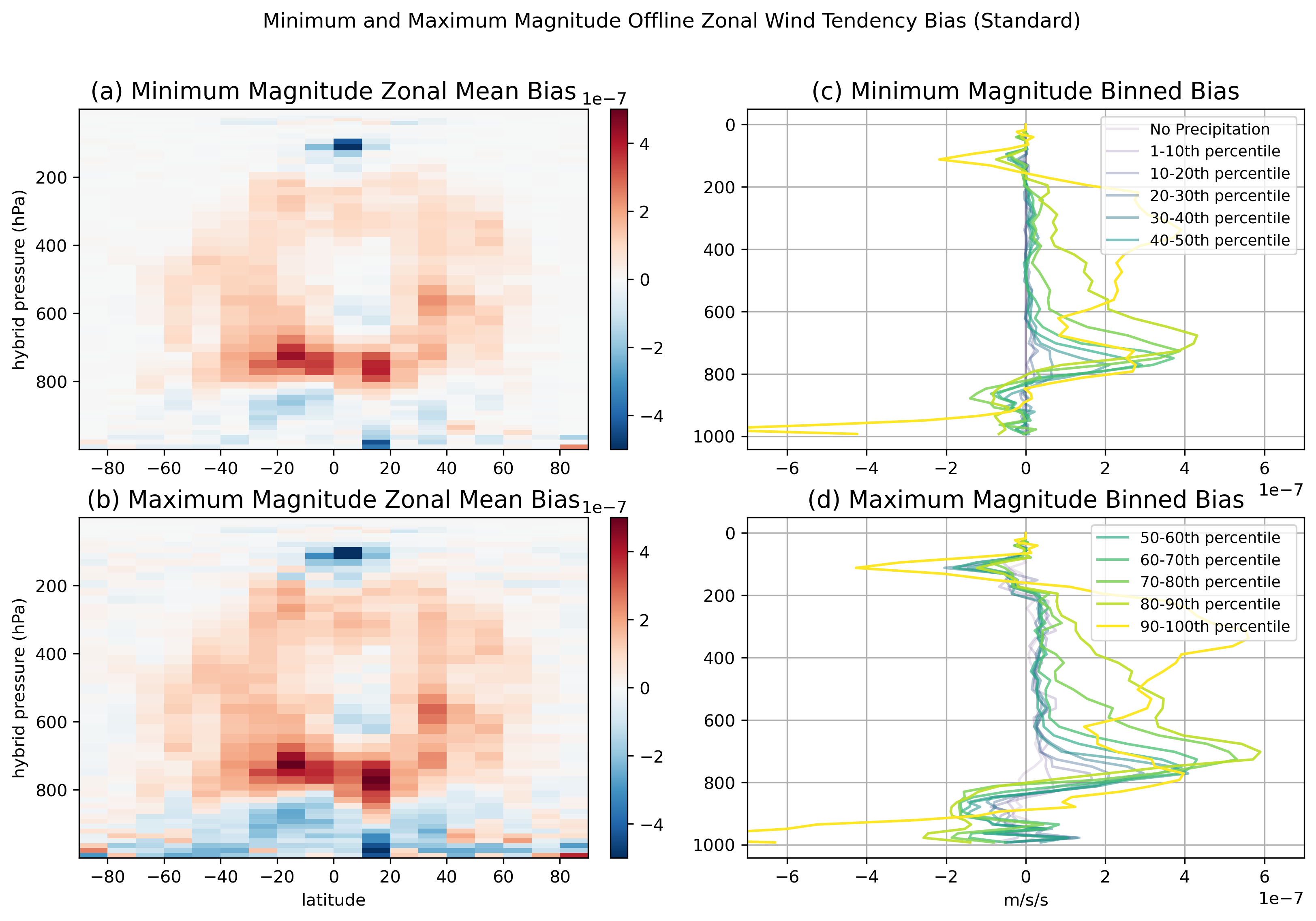}

\caption{Figures S39a and S39b shows minimum and maximum offline zonal mean zonal wind tendency biases across architectures in the standard configuration after averaging across seeds. Figures S39c and S39d show the vertical profiles of minimum and maximum offline zonal wind tendency biases across architectures (again after averaging across seeds) when binned by precipitation percentile.}
 \label{fig:offline_DUPHYS_bias_minmax_standard}
\end{figure}

\begin{figure}[!htbp]
 \centering
 \includegraphics[width=\textwidth]{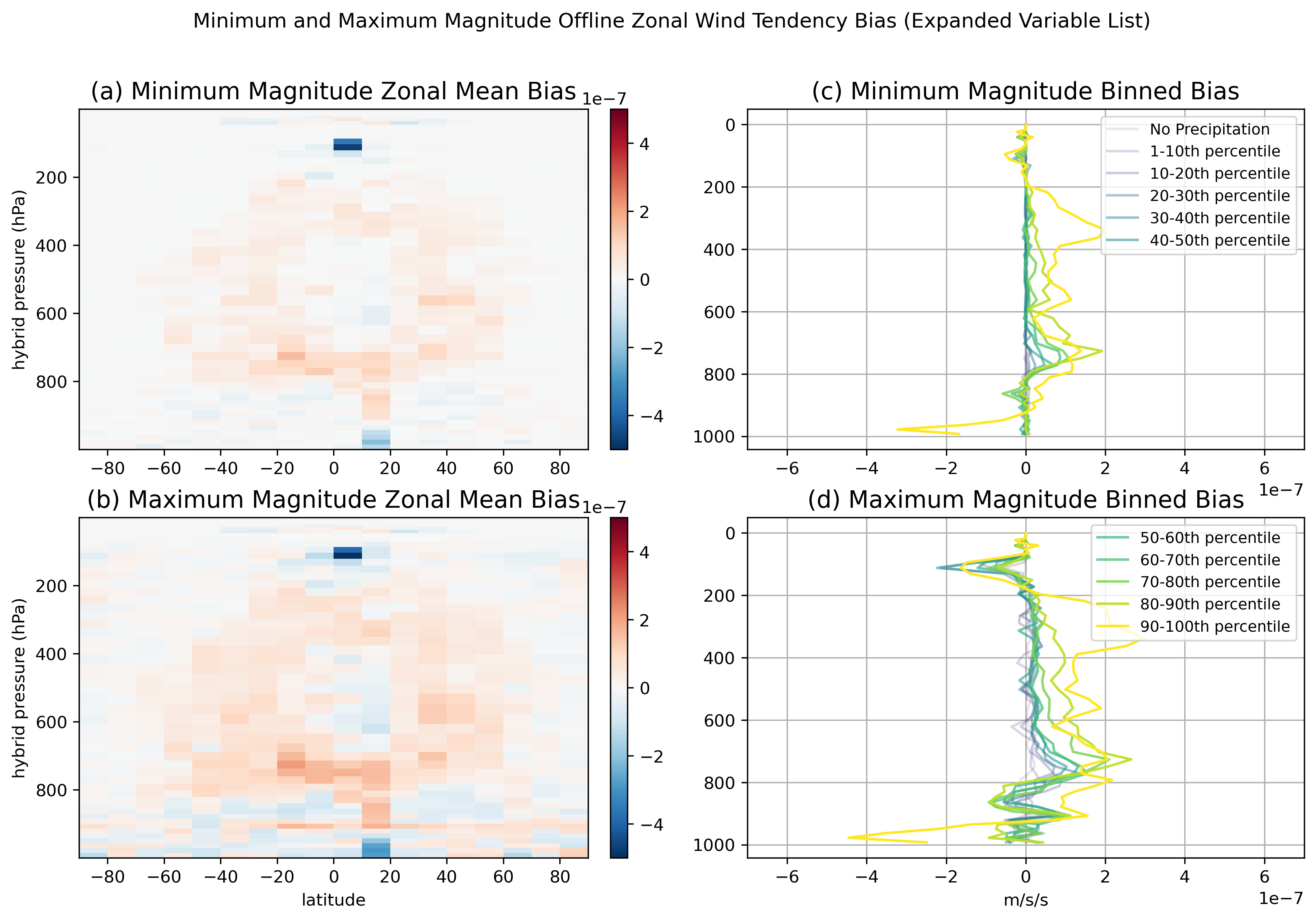}

\caption{Figures S40a and S40b shows minimum and maximum offline zonal mean zonal wind tendency biases across architectures in the expanded variable list configuration after averaging across seeds. Figures S40c and S40d show the vertical profiles of minimum and maximum offline zonal wind tendency biases across architectures (again after averaging across seeds) when binned by precipitation percentile.}
 \label{fig:offline_DUPHYS_bias_minmax_v6}
\end{figure}

\begin{figure}[!htbp]
 \centering
 \includegraphics[width=\textwidth]{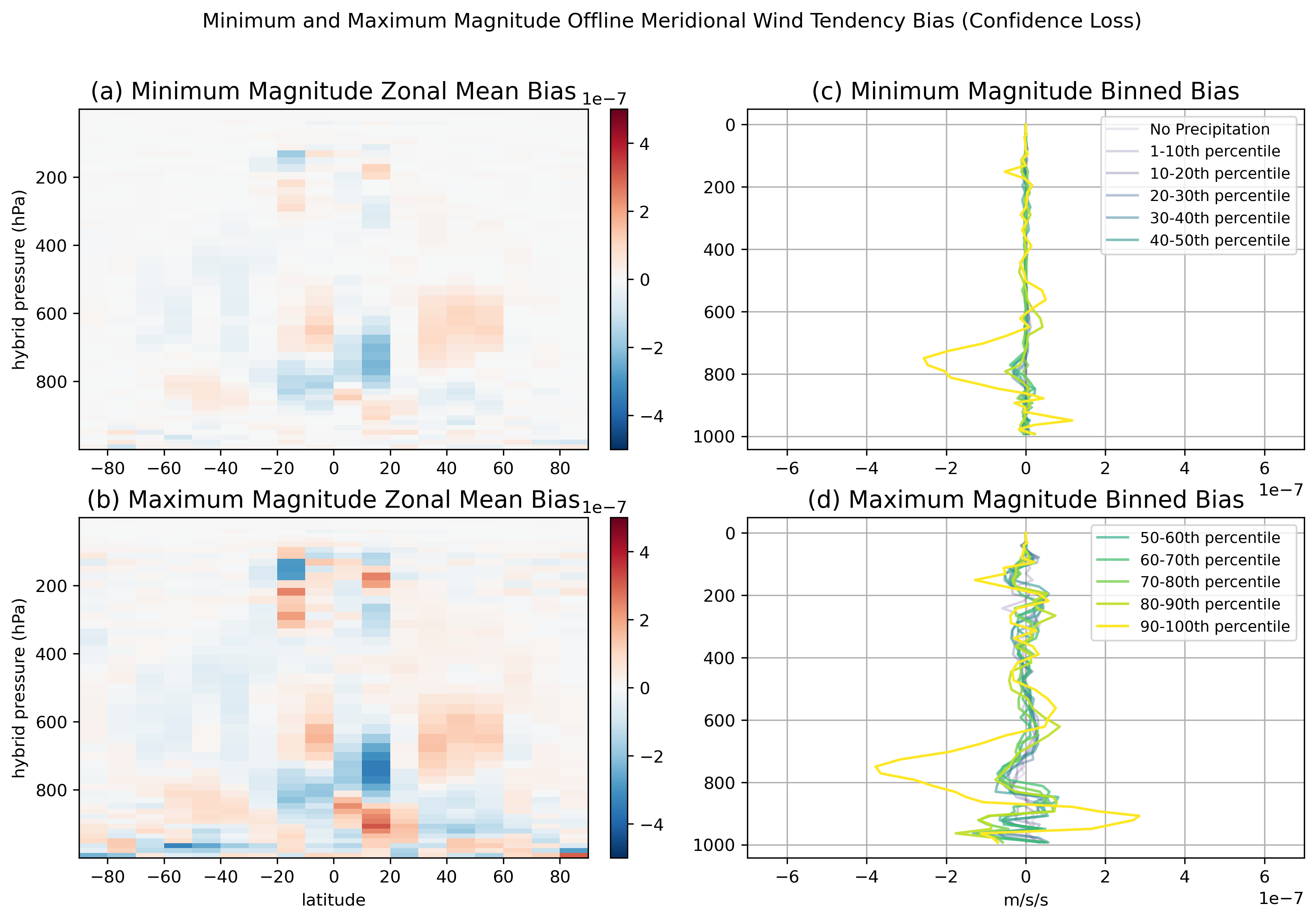}

\caption{Figures S41a and S41b shows minimum and maximum offline zonal mean meridional wind tendency biases across architectures in the confidence loss configuration after averaging across seeds. Figures S41c and S41d show the vertical profiles of minimum and maximum offline meridional wind tendency biases across architectures (again after averaging across seeds) when binned by precipitation percentile.}
 \label{fig:offline_DVPHYS_bias_minmax_conf_loss}
\end{figure}

\begin{figure}[!htbp]
 \centering
 \includegraphics[width=\textwidth]{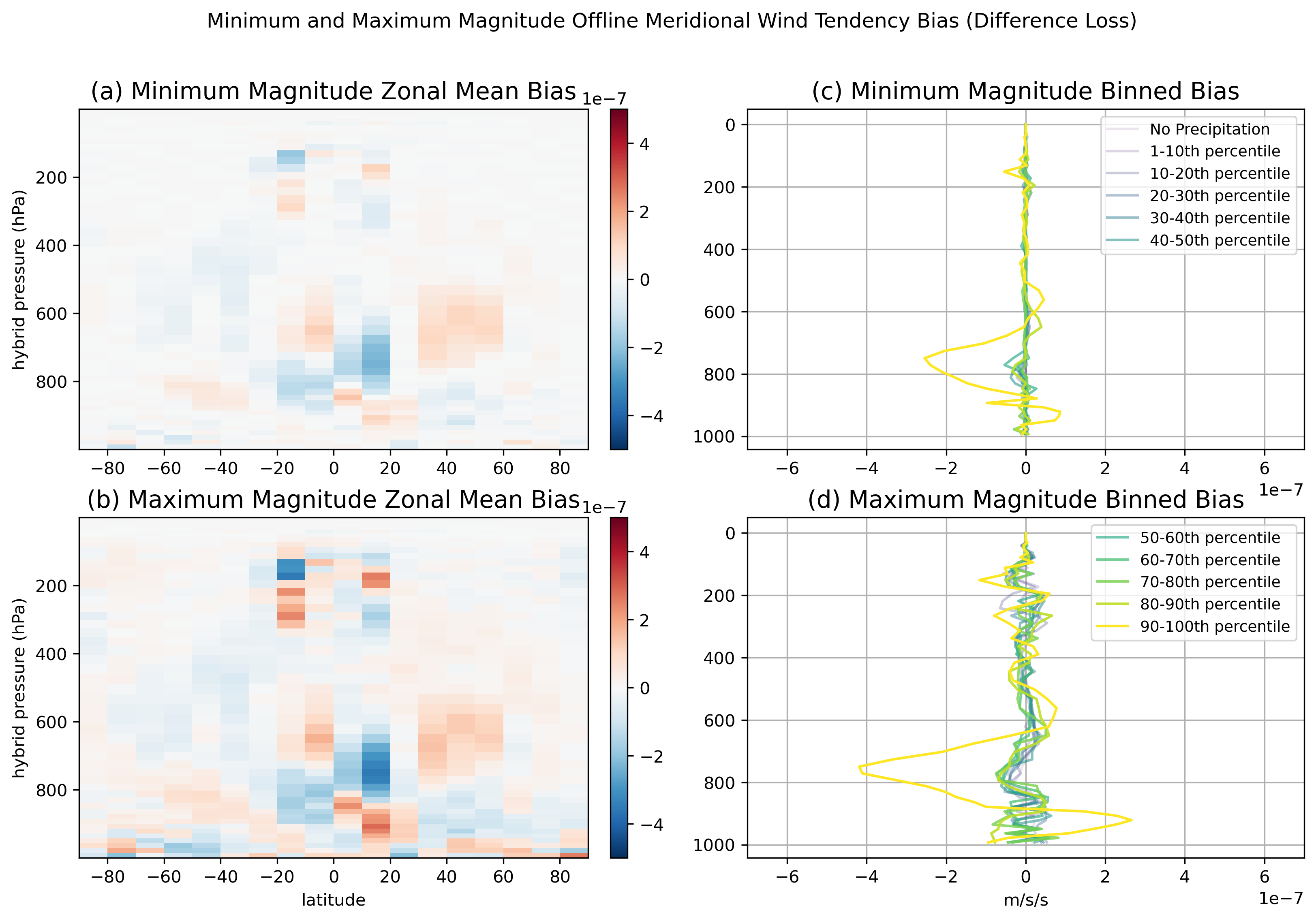}

\caption{Figures S42a and S42b shows minimum and maximum offline zonal mean meridional wind tendency biases across architectures in the difference loss configuration after averaging across seeds. Figures S42c and S42d show the vertical profiles of minimum and maximum offline meridional wind tendency biases across architectures (again after averaging across seeds) when binned by precipitation percentile.}
 \label{fig:offline_DVPHYS_bias_minmax_diff_loss}
\end{figure}

\begin{figure}[!htbp]
 \centering
 \includegraphics[width=\textwidth]{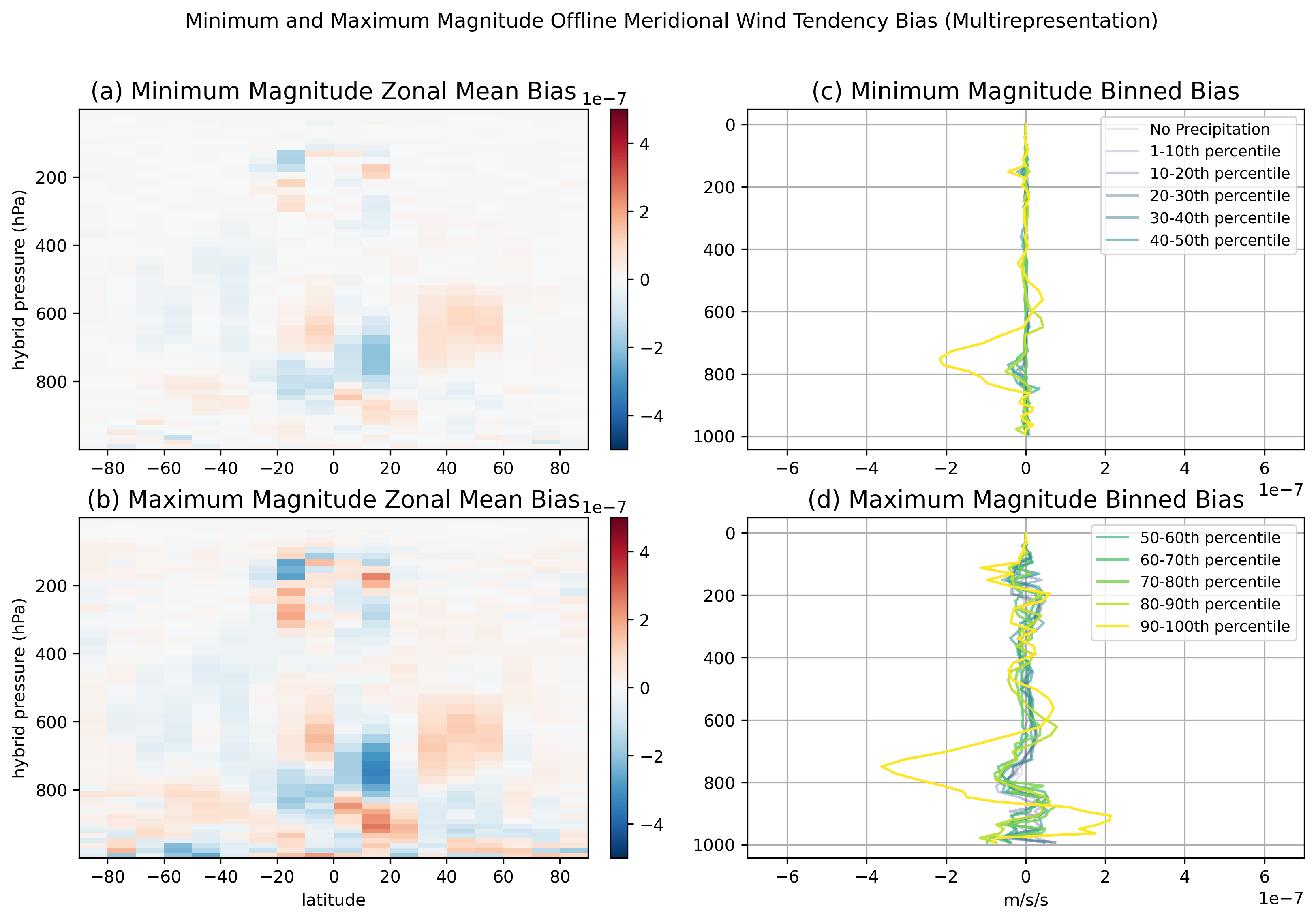}

\caption{Figures S43a and S43b shows minimum and maximum offline zonal mean meridional wind tendency biases across architectures in the multirepresentation configuration after averaging across seeds. Figures S43c and S43d show the vertical profiles of minimum and maximum offline meridional wind tendency biases across architectures (again after averaging across seeds) when binned by precipitation percentile.}
 \label{fig:offline_DVPHYS_bias_minmax_multirep}
\end{figure}

\begin{figure}[!htbp]
 \centering
 \includegraphics[width=\textwidth]{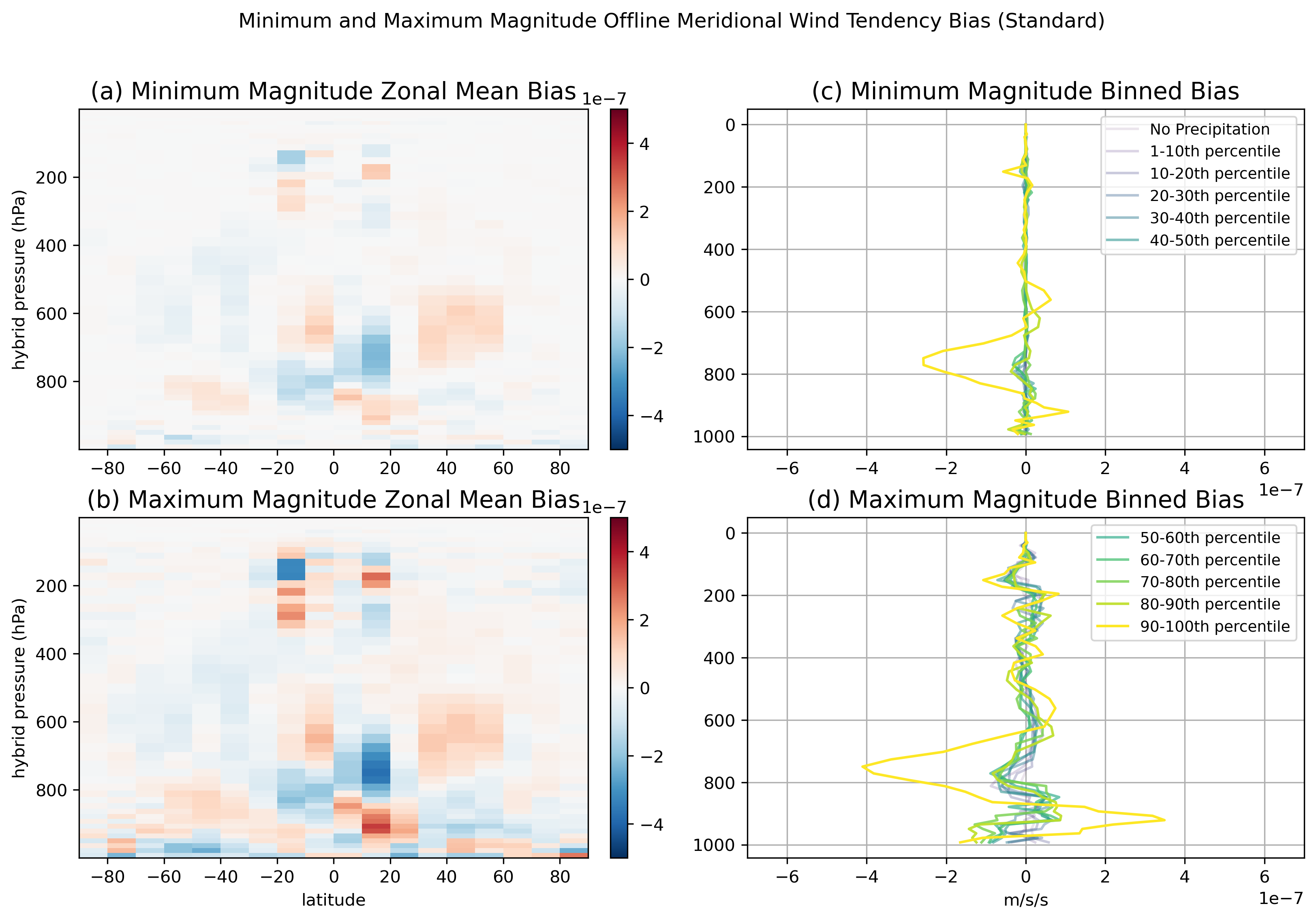}

\caption{Figures S44a and S44b shows minimum and maximum offline zonal mean meridional wind tendency biases across architectures in the standard configuration after averaging across seeds. Figures S44c and S44d show the vertical profiles of minimum and maximum offline meridional wind tendency biases across architectures (again after averaging across seeds) when binned by precipitation percentile.}
 \label{fig:offline_DVPHYS_bias_minmax_standard}
\end{figure}

\begin{figure}[!htbp]
 \centering
 \includegraphics[width=\textwidth]{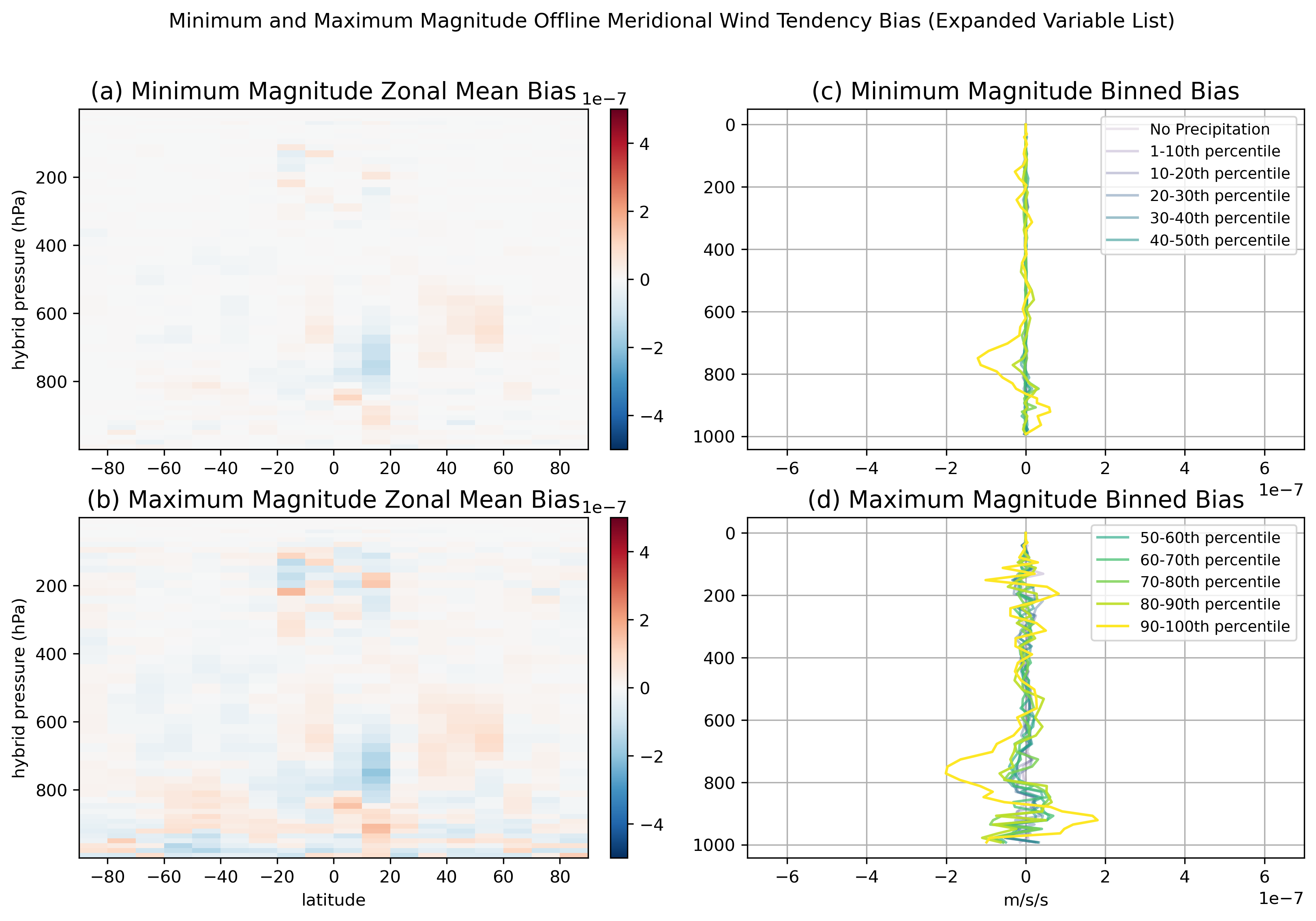}

\caption{Figures S45a and S45b shows minimum and maximum offline zonal mean meridional wind tendency biases across architectures in the expanded variable list configuration after averaging across seeds. Figures S45c and S45d show the vertical profiles of minimum and maximum offline meridional wind tendency biases across architectures (again after averaging across seeds) when binned by precipitation percentile.}
 \label{fig:offline_DVPHYS_bias_minmax_v6}
\end{figure}

\begin{figure}[!htbp]
 \centering
 \includegraphics[width=\textwidth]{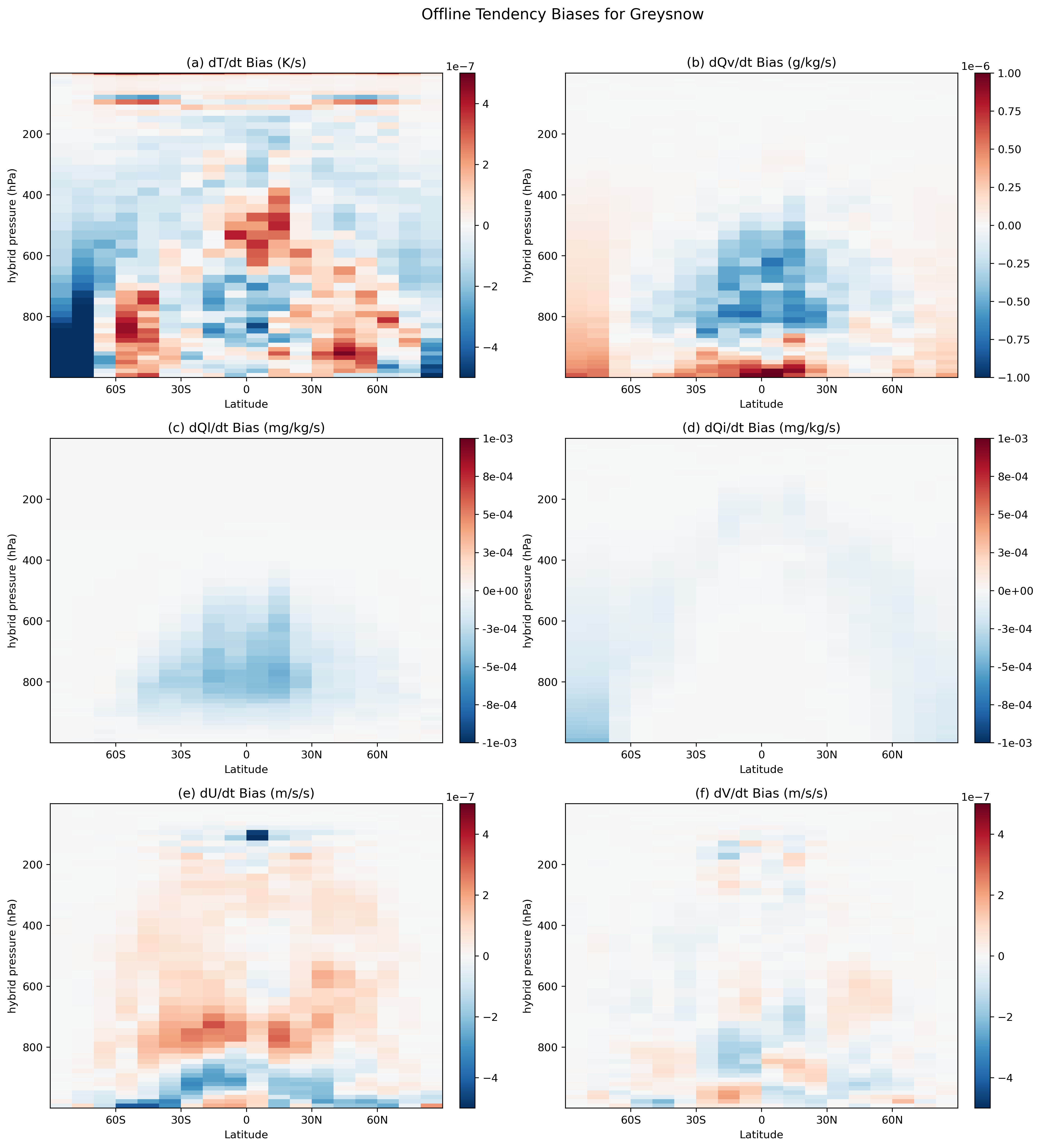}

\caption{This figure shows the zonal means for offline biases from the 1st place winner in the 2024 LEAP ClimSim Kaggle Competition.}
 \label{fig:offline_kaggle_comparison_Adam}
\end{figure}

\begin{figure}[!htbp]
 \centering
 \includegraphics[width=\textwidth]{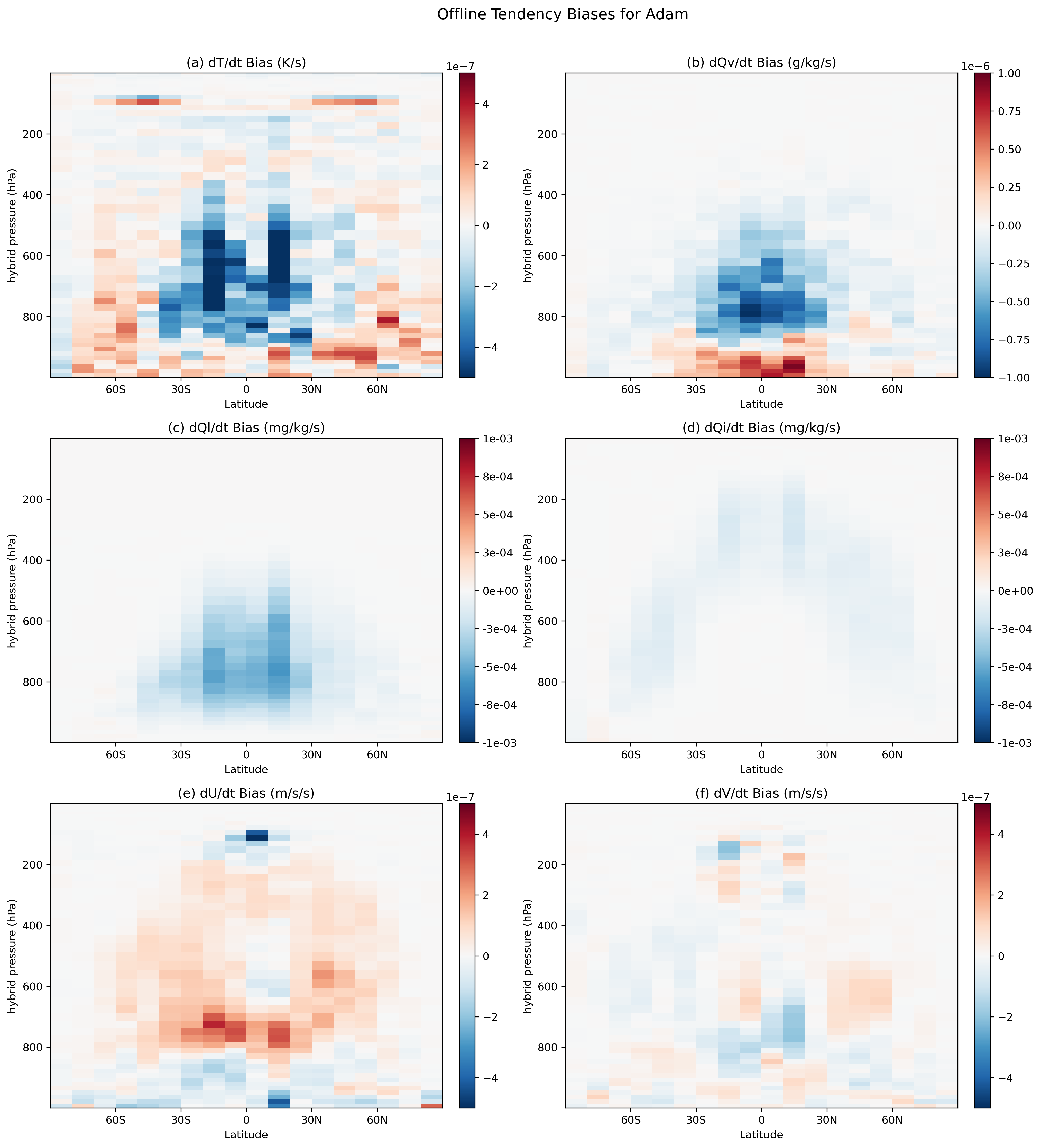}

\caption{This figure shows the zonal means for offline biases from the 2nd place winner in the 2024 LEAP ClimSim Kaggle Competition.}
 \label{fig:offline_kaggle_comparison_Greysnow}
\end{figure}

\end{document}